\documentclass[11pt, english, american]{article}
\usepackage{amsfonts,amsmath,amssymb,mathrsfs}
\usepackage{dsfont}
\usepackage{graphicx}
\usepackage{color}
\usepackage[bookmarksopen=true,colorlinks,linkcolor = black]{hyperref}
\usepackage{hyperref}

\newcommand{\bet}{\beta_1}
\newcommand{\be}{\begin{equation}}
\newcommand{\ee}{\end{equation}}

\newcommand{\landau}{\mbox{\begin{scriptsize}$\mathcal{O}$\end{scriptsize}}}

\newcommand{\llaa}{\left\langle\hspace{-0.14cm}\left\langle}
\newcommand{\rraa}{\right\rangle\hspace{-0.14cm}\right\rangle}
\newcommand{\laa}{\langle\hspace{-0.08cm}\langle}
\newcommand{\raa}{\rangle\hspace{-0.08cm}\rangle}

\newcommand{\LZ}{L^2(\mathbb{R}^2,\mathbb{C})}

\newcommand{\LZN}{L^2(\mathbb{R}^{2N},\mathbb{C})}

\newcommand{\dt}{\frac{\text{d}}{\text{d}t}}
\newcommand{\im}{\text{i}}

\newtheorem{theorem}{Theorem}[section]
\newtheorem{lemma}[theorem]{Lemma}
\newtheorem{notation}[theorem]{Notation}

\newtheorem{corollary}[theorem]  {Corollary}
\newtheorem{remark}[theorem]  {Remark}
\newtheorem{definition}[theorem] {Definition}

\newenvironment{proof}{\emph{Proof:}}{\begin{flushright} $ \Box $ \end{flushright}}

\renewcommand{\phi}{\varphi}

\newcommand{\potdiff}{Z}

\newcommand{\potdiffneu}{\widetilde{Z}_{\beta}}

\newcommand{\asaalt}{\gamma^<_a}
\newcommand{\asalt}{\gamma^<_b}
\newcommand{\asa}{\gamma_a} 
\newcommand{\as}{\gamma_b}
\newcommand{\bs}{\gamma_c}
\newcommand{\cs}{\gamma_d}
\newcommand{\ds}{\gamma_e}

\begin{document}

\title{Derivation of the Time Dependent Gross-Pitaevskii Equation in Two Dimensions}


\author{Maximilian Jeblick\footnote{Mathematisches Institut, Ludwig-Maximilians-Universit\"at M\"unchen, Theresienstra\ss e 39, {80333} M\"unchen, Germany. E-mail: jeblick@math.lmu.de},
 Nikolai Leopold\footnote{Institute of Science and Technology Austria (IST Austria), Am Campus 1, 3400 Klosterneuburg, Austria. E-mail: {\tt nikolai.leopold@ist.ac.at}}\;
 and Peter Pickl\footnote{Duke Kunshan University, Duke Avenue 8, 215316 Kunshan, China.
\newline
 E-mail: {\tt peter.pickl@dukekunshan.edu.cn}} \footnote{Ludwig-Maximilians-Universit\"at M\"unchen, Theresienstra\ss e 39, {80333} M\"unchen, Germany.
 \newline E-mail: {\tt pickl@math.lmu.de}}}

\maketitle

\begin{abstract}
We present a microscopic derivation of the defocusing two-dimensional cubic nonlinear Schr\"odinger equation as a mean field equation starting from an interacting $N$-particle system of Bosons. 
We consider the interaction potential to be given either by $W_\beta(x)=N^{-1+2 \beta}W(N^\beta x)$, for any $\beta>0$, or to be given by $V_N(x)=e^{2N} V(e^N x)$, for some spherical symmetric, nonnegative and compactly supported $W,V \in L^\infty(\mathbb{R}^2,\mathbb{R})$.
In both cases we prove the convergence of the reduced density matrix corresponding to the exact time evolution to the projector onto the solution of the corresponding nonlinear Schr\"odinger equation in trace norm. For the latter potential $V_N$ we show that it is crucial to take the microscopic structure of the condensate into account in order to obtain the correct dynamics.

\end{abstract}

\noindent
\textbf{MSC class:} 35Q55, 35Q41, 81V70   \\
\textbf{Keywords:} Gross-Pitaevskii, NLS, mean-field limit

\newpage
\tableofcontents\newpage


\newpage
\section{Introduction}
This paper deals with the effective dynamics of a two dimensional condensate
of $N$ interacting bosons. 
Fundamentally, the evolution of the system is described by a time-dependent wave-function $\Psi_t \in L^2_{s}(\mathbb{R}^{2N}, \mathbb{C}),
\| \Psi_t \|=1$ (Here and below norms without index $\|\cdot\|$ always denote the $L^2$-norm on the appropriate Hilbert space.).
$L^2_{s}(\mathbb{R}^{2N}, \mathbb{C})$ denotes the set of all $\Psi \in  L^2(\mathbb{R}^{2N}, \mathbb{C})$
 which are symmetric under pairwise permutations of the variables 
$x_1, \dots, x_N \in \mathbb{R}^2$. Assuming that $\Psi_t \in H^2 (\mathbb{R}^{2N},\mathbb{C})$ holds, $\Psi_t$ then solves the  $N$-particle 
Schr\"odinger equation 
\be\label{schroe} \im\partial_t \Psi_t = H_U\Psi_t \ee 
where the  (non-relativistic) Hamiltonian $H_U:H^2 (\mathbb{R}^{2N},\mathbb{C}) \rightarrow L^2 (\mathbb{R}^{2N},\mathbb{C})$ is given by
\be\label{hamiltonian}
 H_U=-\sum_{j=1}^N \Delta_j+\sum_{1\leq  j< k\leq  N} U(x_j-x_k) +\sum _{j=1}^N A_t(x_j)
 \; .
 \ee
In general, even for small particle numbers $N$, \eqref{schroe} cannot be solved neither exactly nor numerically for $\Psi_t$. Nevertheless, for a certain class of scaled potentials $U$ and certain initial conditions $\Psi_0$ it is possible to derive an approximate solution of \eqref{schroe} in the trace class topology of reduced density matrices.
The picture we have in mind is the description of a Bose-Einstein condensate. Initially one starts with the ground state of a trapped, dilute gas and then removes or changes the trap subsequently.
In this paper, we will consider two choices for the interaction potential $U$.
\begin{itemize}
\item
Let  $U(x) = V_N(x)=e^{2N} V(e^N x) $ for a compactly supported, spherically symmetric and nonnegative potential $V\in
L^\infty_c (\mathbb{R}^2, \mathbb{R}) $.
 Below, the exponential scaling of $V_N$ will be explained in detail.
 Note that, in contrast to existing dynamical mean-field results, $\| V_N \|_1= \mathcal{O}(1)$ does not decay like $1/N$.
 \item
Let, for any fixed $\beta>0$, 
 $U(x)= W_\beta(x)= N^{-1+2 \beta} W(N^\beta x)$ for a compactly supported, spherically symmetric and nonnegative potential $W \in L^\infty_c (\mathbb{R}^2, \mathbb{R}) $.
This scaling can be motivated by formally imposing that the total potential energy is of the same order as the total kinetic energy, namely of order $N$,  if  $\Psi_0$ is close to the ground state. 
\end{itemize}
Define the one particle reduced density matrix $\gamma^{(1)}_{\Psi_0}$ of $\Psi_0$ with integral kernel
$$\gamma^{(1)}_{\Psi_0}(x,x')=\int_{\mathbb{R}^{2N-2}} \Psi_{0}^*(x,x_2,\ldots,x_N)\Psi_{0}(x',x_2,\ldots,x_N)d^2x_2\ldots d^2x_N \;. $$
To account for the physical situation of a Bose-Einstein condensate, we assume complete condensation in the limit of large particle number $N$. This amounts to
assume that, for $N \rightarrow \infty$,
 $\gamma^{(1)}_{\Psi_0}  \rightarrow |\phi_0\rangle\langle\phi_0|$ in trace norm for some $\phi_0
\in L^2(\mathbb{R}^2,\mathbb{C}) 
  , \|\phi_0\|=1$.
 Our main goal is to show the persistence of condensation over time. 
  This is of particular interest in experiments if one switches off the trapping potential $A_t$ and monitors the expansion of the condensate.
We prove that the time evolved reduced density matrix $\gamma^{(1)}_{\Psi_t}$
 converges to $ |\phi_t\rangle\langle\phi_t|$ in trace norm as $N \rightarrow \infty$ with convergence rate of order $N^{-\eta}$ for some $\eta>0$. 
 $\phi_t$ then solves  the nonlinear Schr\"odinger equation
\be\label{GP} \im \partial_t
\phi_t=\left(-\Delta +A_t\right) \phi_t+
b_U
|\phi_t|^2\phi_t=:h_{ b_U}^{GP}\phi_t
\ee
with initial datum $\phi_0$. Depending on the interaction potential $U$, we obtain different coupling constants $b_U$.
For $U=W_\beta$, we obtain $b_{W_\beta}= N\|W_\beta\|_1= \|W\|_1$. This result is already expected from a heuristic law of large numbers argument, see below.
In the case $U= V_N$, we have $b_{V_N}= 4\pi$.
We like to remark that it is well known that convergence of $\gamma^{(1)}_{\Psi_t}$
 to $ |\phi_t\rangle\langle\phi_t|$ in trace norm is equivalent to the respective convergence in operator norm since $|\phi_t\rangle\langle\phi_t|$ is a rank-1-projection, see
Remark 1.4. in \cite{rodnianskischlein}. Furthermore, the convergence of the one-particle reduced density matrix
$\gamma^{(1)}_{\Psi_t} \rightarrow| \phi_t \rangle \langle \phi_t |$ in trace norm
 implies convergence of any $k$-particle reduced density matrix $\gamma^{(k)}_{\Psi_t}$ against $| \phi_t^{\otimes k} \rangle \langle \phi_t^{\otimes k} |$ in trace norm as $N \rightarrow \infty$ and $k$ fixed, see for example \cite{knowles}.
\medskip\\
In the case that the time evolution of $\Psi_t$ is generated by $H_{V_N}$
it is interesting to note that the effective evolution equation of $\phi_t$ does not depend on the scattering length $a$. This contrasts the three dimensional case, where the correct mean field coupling is  given by  $ 8 \pi a_{3D}$, $a_{3D}$ denoting the scattering length of the potential in three dimensions. The universal coupling $4 \pi$ in the case of a positive scattering length is known within the physical literature, see e.g. (30) and (A3) in \cite{cherny} (note that $\hbar=1, m= \frac{1}{2}$ in our choice of coordinates). Actually, our dynamical result complements a more general theory describing the ground state properties of dilute Bose gases. 
It was shown in \cite{lsy} that for such a gas with repulsive interaction $V \geq 0$, the ground state energy per particle is to leading order given by either the Gross-Pitaevskii energy functional with coupling parameter $ 8 \pi /| \ln( \overline{\rho} a^ 2 )|$ or a Thomas-Fermi type functional. Here, $\overline{\rho}$ denotes the mean density of the gas, see Equation (1.6) in \cite{lsy} for a precise definition. 
The authors prove further that only if $ N/ |\ln( \overline{\rho} a^ 2 )| =\mathcal{O}(1)$ holds, one obtains the Gross-Pitaevskii regime. This directly implies that scattering length of the interaction potential needs to have an exponential decrease in $N$. In our case, the scattering length of the potential $V_N$ is given by  $ae^{-N}$, $a$ denoting the scattering length of $V$. The mean density of the system we consider is of order one, i.e. $\overline{\rho} = \mathcal{O}(1)$. 
This yields 
$8 \pi N/ |\ln( \overline{\rho} (e^{-N}a)^ 2 )| \approx 4 \pi $ which is in agreement with our findings.
It should be pointed out that there has been some debate about the question whether two dimensional Bose-Einstein condensation can be observed experimentally. This amounts to the question whether condensation takes place for temperatures $T>0$. For an ideal, noninteracting gas in box, the standard grand canonical computation for the critical temperature $T_c$ of a Bose-Einstein condensate shows that there is no condensation for $T>0$. For trapped, noninteracting Bosons in a confining power-law potential, the findings in \cite{Bagnato} however show that in that case $T_c>0$ holds. 
Finally, it was proven in \cite{lieb100bec} that $\gamma^{(1)}_{\Psi}$ converges
 to $ |\phi\rangle\langle\phi|$ in trace norm if $\Psi$ the ground state of $H_{V_N}$
and $\phi$ is the ground state of the Gross-Pitaevskii energy functional, see \eqref{energyfunct}.
The assumptions made in the paper are that  and the external potential $A$ tends to $+ \infty$ as $ |x| \rightarrow \infty$ and the interaction potential $V$ is nonnegative. It is also remarked that one does not observe 100 \% condensation in the ground state of a interacting homogenous system.
The emergence of 100 \% Bose-Einstein condensation as a ground state phenomena thus highly depends on the particular physical system one considers. 
Our approach is the following: Initially, we assume the convergence of $\gamma^{(1)}_{\Psi_0}$ 
 to $ |\phi_0\rangle\langle\phi_0|$. We then show the persistence this condensation for time scales of order one. Our assumption is thus in agreement with the findings in \cite{lieb100bec}.
 We like to remark that the two dimensional Thomas-Fermi regime could be observed experimentally \cite{experiment}.

Next, we want to explain how the different coupling constants $b_U$ are obtained in the dynamical setting.
 For this, we first recall known results from the three dimensional Bose gas. There, one considers the interaction potential to be given by  $V_{\beta}(x)
=N^{-1+3\beta} V(N^{\beta} x)$ for $0 \leq \beta \leq 1$. 
For $0<\beta<1$, one obtains the cubic nonlinear Schr\"odinger equation with coupling constant $\|V\|_1$. This can be seen as a singular mean-field limit, where the full interaction is replaced by its corresponding mean value 
$\int_{\mathbb{R}^3} d^3 y N^{3\beta} V(N^\beta (x-y)) | \phi_t (y) |^2 \rightarrow \|V\|_1 | \phi_t (x) |^2$.  For $\beta=1 $, however, the system develops correlations between the particles which cannot be neglected. As already mentioned, the correct mean field coupling is then given by 
$ 8 \pi a_{3D}$.
This is different for a two dimensional condensate.  
Let us first explain, why the short scale correlation structure is negligible if the potential is given
by $W_{\beta}(x)
=N^{-1+2\beta} W(N^{\beta} x)$ for any $\beta>0$. 
Assuming that the energy of
$\Psi_t$ is comparable to the ground state energy, the wave function will develop short scale correlations between the particles. 
One may heuristically think of $\Psi_t$ of Jastrow-type, i.e.
 $\Psi_t (x_1, \dots,x_N) \approx \prod_{i<j} F(x_i-x_j) \prod_{k=1}^ N \phi_t (x_k) $
\footnote{
One should however note that $\Psi_t$ will not be close to a full product $ \prod_{k=1}^ N \phi_t (x_k) $ in norm. For certain types of interactions, it has been shown rigorously that $\Psi_t$ can be approximated  by a quasifree state satisfying a Bogoloubov-type dynamics, see \cite{SchleinNorm}, \cite{marcin1}, \cite{marcin2} and \cite{picklnorm} for precise statements. }.
The function $F$ accounts for the pair correlations between the particles at short scales of order $N^{-\beta}$. 
It is well known that the correlation function $F$ should be described by the zero energy scattering
state $j_{N,R}$ of the potential $W_{\beta}$, where $j_{N,R}$ satisfies
\begin{equation*}
\begin{cases} 
\left( - \Delta_x + \frac{1}{2} W_\beta(x)  \right) j_{N,R}(x)=0,
\\
j_{N,R}(x)=1 \; \text{for } |x| = R.
  \end{cases}
\end{equation*}
Here, the boundary radius $R$ is chosen of order $N^{-\beta}$. That is, $F(x_i-x_j) \approx  j_{N,R}(x_i-x_j)$ for $|x_i-x_j| =\mathcal{O}
(N^{-\beta})$ and $F(x_i-x_j) =1$ for $|x_i-x_j| \gg \mathcal{O}
(N^{-\beta})$.
Rescaling
to coordinates $y=N^\beta x$, the zero energy scattering state satisfies
\begin{equation}
\label{scaling}
\left(
-\Delta_y+\frac{1}{2}N^{-1}W(y)
\right)
j_{N,N^\beta R}(y)=0\;.
\end{equation}
Due to the factor $N^{-1}$ in front of $W$, the zero energy scattering equation is almost constant, that is $j_{N,R}(x) \approx 1$, for all $|x| \leq R$.
As a consequence, the microscopic structure $F$, induced by the zero energy scattering state, vanishes for any $\beta>0$ and does not effect the dynamics of the reduced density matrix $\gamma^{(1)}_{\Psi_t}$.
Assuming $\gamma^{(1)}_{\Psi_0} \approx|\phi_0\rangle\langle\phi_0|$, one may thus apply a law of large numbers argument and conclude that the interaction on each particle is then approximately given by its mean value
\begin{align*}
 \int_{\mathbb{R}^2} d^2y N W_\beta( x-y) |\phi_t|^2(y)
\rightarrow\|W\|_1 | \phi_t|^2(x)
\;.
\end{align*} 
This yields to the correct coupling in the effective equation \eqref{GP} in the case $U(x)=W_\beta(x)$.

Let us now consider the case for which the dynamics of $\Psi_t$ is generated by the Hamiltonian $H_{V_N}$.
If one would guess the effective coupling of $\phi_t$ to be also given by its mean value w.r.t. the distribution $| \phi_t |^2$, one would end up with the $N$-dependent equation
$\im \partial_t
\phi_t=\left(-\Delta +A_t\right) \phi_t+
N \int_{\mathbb{R}^2} d^2x V (x)
|\phi_t|^2\phi_t$. Note that the coupling constant of the self interaction differs from its correct value by a factor of $\mathcal{O}(N)$.
As in the three dimensional Gross-Pitaevskii regime $\beta=1$, 
 it is now important to take the correlations explicitly into account.
The scaling of the potential yields to
 $j_{N,R}(x)=j_{0,e^{N}R}(e^{N}x)$, which implies that the correlation function will influence the dynamics whenever two particles collide. 
The coupling parameter can then be inferred from the relation
\begin{align*}
 \int_{\mathbb{R}^2}  d^2x V_N( x)j_{N,R}(x)= 
 \frac{4\pi}{ \ln \left(\frac{R}{ae^{-N}} \right)},
\end{align*}
where $a$ denotes the scattering length of the potential $ V$. 
As mentioned, the logarithmic dependence of the integral above on $a$ is special in two dimensions.
Since $ \frac{4\pi}{ \ln \left(\frac{R}{ae^{-N}} \right)} \approx \frac{4 \pi}{N}$ holds for $a>0$, the effective equation for $\phi_t$ will not depend on $a$ anymore. 
Consequently, one obtains as an effective coupling
 \begin{align*}
\int_{\mathbb{R}^2} d^2y N V_N( x-y)j_{N,R}(x-y) |\phi_t|^2(y)
\rightarrow 
4 \pi | \phi_t|^2(x).
\end{align*} 

We like to remark that it is easy to verify that, for any $s>0$, the potential
$V_{sN}(x)= e^{2Ns} V(e^{Ns}x)$ yields to an effective coupling $4 \pi / s$. For the sake of simplicity, we will not consider this slight generalization, although our proof is also valid in this case.

The rigorous derivation of effective evolution equations is well known in the literature, see e.g. \cite{SchleinNorm, benedikter, erdos1, erdos2, erdos3, erdos4, knowles,  marcin1, marcin2,
 pickl2, picklgp3d, pickl1, rodnianskischlein}
 and references therein.
For the two-dimensional case we consider, it has been proven, for $0 <\beta <3/4$ and $W$ nonnegative, that $\gamma^{(1)}_{\Psi_t}$
 converges to $ |\phi_t\rangle\langle\phi_t|$ as $N \rightarrow \infty$ \cite{schlein2d}.
 For $0< \beta < 1/6$, it has been established in \cite{chen2d} that the reduced density matrices converge, assuming that the potential $W$ is attractive, i.e. $W \leq 0$.  This result was later extended to $0<\beta <3/4$, using stability properties of the ground state energy \cite{lewin}.

Another approach which relates more closely to the experimental setup is to consider a three-dimensional gas of Bosons which is strongly confined in one spatial dimension. Then, one obtains an effective two dimensional system in the unconfined  directions. We remark that in this dimensional reduction two limits appear, the length scale in the confined direction and the scaling of the interaction in the unconfined directions.
Results in this direction can be found in \cite{abdallah} and \cite{keler}, see also \cite{teufel}.
It is still an open problem to derive our dynamical result starting from a strongly confined three dimensional system.
For known results regarding the ground state properties of dilute Bose gases, we refer to the monograph \cite{lssy}, which also summarizes the papers \cite{lieb100bec}, \cite{lsy} and\cite{ls}.

Our proof is based on \cite{picklgp3d}, where the emergence of the Gross-Pitaevskii equation was proven by one of us (P.P.) in three dimensions for $\beta=1$. In particular, we adapt some crucial ideas which allow us to control the microscopic structure of $\Psi_t$.

We shall shortly discuss the physical relevance of the different scalings. On the first view, the interactions discussed above do look rather unphysical. It is questionable to assume that the coupling constant and/or the range of the interaction change as the particle number increases.
Nevertheless, one can think of situations,
where for example the support of the interaction is small and the particle number of the system is adjusted accordingly.

The exponential scaling $V_N(x)= e^{2N} V(e^N x)$ is special. In this case it is possible to rescale space- and time-coordinates in such a way that in the new coordinates  the interaction is {\it not} $N$ dependent. Choosing $y=e^Nx$ and $\tau= e^{2N}t$ the Schr\"odinger equation reads
$$
\im\frac{d}{d\tau} \Psi_{e^{-2N}\tau}=\left(-\sum_{j=1}^N \Delta_{y_j}+\sum_{1\leq  j< k\leq  N} V(y_j-y_k) +\sum _{j=1}^N A_{e^{-2N}\tau}(e^{-N}y_j)\right)\Psi_{e^{-2N}\tau}\;.
$$
The latter equation thus corresponds to an extremely dilute gas of bosons with density $\sim e^{-2N}$. In order to observe a nontrivial dynamics, this condensate is then monitored over time scales of order $\tau \sim e^{2N}$.
Since the trapping potential is adjusted according to the density of the gas in the experiment,  the $N$ dependence of  $A_{e^{-2N}\tau}(e^{-N} \cdot)$ is reasonable.


\section{Main result}

We will bound  expressions which are uniformly bounded in $N$ and $t$ by some constant $C$.
Constants
 appearing in a sequence of estimates will not be distinguished, i.e. in $X\leq  CY\leq  CZ$ the constants may differ.

For $U \in \lbrace W_\beta, V_N \rbrace$,  define the energy functional $\mathcal{E}_{U}:H^2(\mathbb{R}^ {2N}, \mathbb{C})\to \mathbb{R}$
$$
\mathcal{E}_{U}(\Psi)=N^{-1}\laa\Psi,H_{U}\Psi\raa\;,$$
where $\laa\cdot,\cdot\raa$ denotes the scalar product on $\LZN$.
Furthermore, define the
Gross-Pitaevskii energy functional $\mathcal{E}_{b_U}^{GP}:H^2(\mathbb{R}^ {2}, \mathbb{C})\to \mathbb{R}$  \begin{align}\label{energyfunct}
\mathcal{E}_{b_U}^{GP}(\phi)=&\langle\nabla\phi,\nabla\phi\rangle+\langle\phi,(A_t+\frac{1}{2} b_U|\phi|^2)\phi\rangle
=\langle\phi,
(h_{b_U}^{GP}-\frac{1}{2} b_U |\phi|^2)\phi\rangle
\end{align}
where $\langle\cdot,\cdot\rangle$ denotes the scalar product on $\LZ$.
Note that both $\mathcal{E}_{U}(\Psi) $  and $\mathcal{E}_{b_U}^{GP}(\phi)$ depend on $t$, due to the time varying external potential $A_t$. 
For the sake of readability, we will not indicate this time dependence explicitly.
We now state our main Theorem:
\begin{theorem}\label{theo}
Let $\Psi_0 \in L^2_{s}(\mathbb{R}^{2N}, \mathbb{C}) \cap H^2(\mathbb{R}^{2N}, \mathbb{C})$ with $\|\Psi_0\|=1$. Let $\phi_0  \in L^{2}(\mathbb{R}^2,\mathbb{C})$ with $\|\phi_0\|=1$ and assume $\lim_{N\to\infty}\gamma^{(1)}_{\Psi_0}=|\phi_0\rangle\langle\phi_0|$ in trace norm. Let the external potential $A_t$, which is defined in \eqref{hamiltonian}, statisfy
$ A_t
 \in L^\infty (\mathbb{R}^2, \mathbb{R})$, 
$ \dot{A}_t 
 \in L^\infty (\mathbb{R}^2, \mathbb{R}) 
$, for all $t \in \mathbb{R}$.
\begin{itemize}
\item[(a)]
For any $\beta>0$, let $W_\beta$ be given by $W_\beta(x)= N^{-1+2 \beta} W(N^{\beta}x)$, for $W \in 
L^\infty_c (\mathbb{R}^2,\mathbb{R}) \;, W\geq0$ and $W$ spherically symmetric.
Let $\Psi_t$ the unique solution to $i \partial_t \Psi_t
= H_{W_\beta} \Psi_t$ with initial datum $\Psi_0$.  
Let $\phi_t$ the unique solution to $i \partial_t \phi_t
= h^{GP}_{ \|W\|_1} \phi_t$ with initial datum $\phi_0$ and assume that $\phi_t \in H^3(\mathbb{R}^2, \mathbb{C})$. 
Let $
\lim_{N \rightarrow \infty}
\left(
\mathcal{E}_{W_\beta}(\Psi_0)
-
\mathcal{E}_{ \|W\|_1}^ {GP}(\phi_0)
\right)
=0
$.
Then, for any $\beta>0$ and for any $t>0$ 
\begin{equation}
\label{convergenls}
\lim_{N\to\infty}\gamma^{(1)}_{\Psi_t}=|\phi_t\rangle\langle\phi_t|
\end{equation} in trace norm.

\item[(b)]
Let $V_N$ be given by $V_N(x)= e^ {2N} V(e^ {N}x)$, for $V \neq 0; V \in 
L^\infty_c (\mathbb{R}^2,\mathbb{R})$, $V \geq 0$ and $V$ spherically symmetric. 
Let $\Psi_t$ the unique solution to $i \partial_t \Psi_t
= H_{V_N} \Psi_t$ with initial datum $\Psi_0$. 
Let $\phi_t$ the unique solution to $i \partial_t \phi_t
= h^{GP}_{4 \pi} \phi_t$ with initial datum $\phi_0$ and assume that $\phi_t \in H^3(\mathbb{R}^2, \mathbb{C})$. Let $
\lim_{N \rightarrow \infty}
\left(
\mathcal{E}_{V_N}(\Psi_0)
-
\mathcal{E}_{4 \pi}^ {GP}(\phi_0)
\right)
=0
$.

Then,  for any $t>0$ 
\begin{equation}
\label{converge}
\lim_{N\to\infty}\gamma^{(1)}_{\Psi_t}=|\phi_t\rangle\langle\phi_t|
\end{equation} in trace norm.

\end{itemize}
\end{theorem}
\textbf{Remark:}
\begin{enumerate}

\item
We expect that for regular enough external potentials $A_t$, the regularity assumption $\phi_t  \in H^3(\mathbb{R}^2, \mathbb{C})$ to follow from
regularity assumptions on the initial datum
$\phi_0$. In particular,
if  $\phi_0  \in 
\Sigma^3(\mathbb{R}^2,\mathbb{C})
=
\lbrace
f \in L^2 (\mathbb{R}^2, \mathbb{C})|
\sum_{ \alpha+ \beta \leq 3} \|x^\alpha \partial_x^\beta f \| < \infty
\rbrace
$ holds,
the bound $\| \phi_t \|_{H^3} < \infty$  has been proven for external potentials which are at most quadratic in space, see \cite{carles} and Lemma  \ref{regularityLemma}. In particular, for $\phi_0 \in \Sigma^3(\mathbb{R}^2, \mathbb{C})$, the bound  $\| \phi_t \|_{H^3} \leq C$  holds if the external potential is not present, i.e. $A_t=0$.

\item 
As already mentioned, the convergence of $\gamma^{(1)}_{\Psi_t}$ to $|\phi_t\rangle\langle\phi_t|$ in trace norm is equivalent to convergence in operator norm, since $|\phi_t\rangle\langle\phi_t|$ is a rank
one projection \cite{rodnianskischlein}. 
Other equivalent definitions of asymptotic 100\% condensation can be found in \cite{michelangeli}.


\item
 In our proof we will give explicit error estimates in terms of the particle number $N$. We shall show that the rate of convergence is of order
$N^{-\delta}$ for some $\delta>0$, assuming that also initially $\gamma^{(1)}_{\Psi_0} \rightarrow|\phi_0\rangle\langle\phi_0|$ converges in trace norm with rate of at least $N^{-\delta}$. 

\item One can relax the conditions on the initial condition and only require $\Psi_0 \in L^2_{s}(\mathbb{R}^{2N}, \mathbb{C}) $ using a standard density argument. 
 
\item 
It has been shown that in the limit $N\to\infty$ the energy-difference $\mathcal{E}_{V_N}(\Psi^{gs})-\mathcal{E}_{4 \pi}^{GP}(\phi^{gs})\to0$, where $\Psi^{gs}$ is the ground state of a trapped Bose gas and $\phi^{gs}$  the ground state of the
respective Gross-Pitaevskii energy functional, see \cite{lsy}, \cite{ls}.

\end{enumerate}

\section{Organization of the proof}\label{secskel}

The method we  use in this paper is introduced in detail in  \cite{pickl1} and was generalized to derive various mean-field equations. 
As we have mentioned, our proof is based on \cite{picklgp3d}, which covers the three-dimensional counterpart of our system.
Heuristically speaking, the method we are going to employ is based on the idea of counting for each time $t$ the relative number of those particles which are
not in the state $\phi_t$. 
It is then possible to show that the rate of particles which leave the condensate is small, if initially almost all particles are in the state $\phi_0$.  
In order to compare the exact dynamic, generated by $H_U$, with the effective dynamic, generated by $h^{GP}_{b_U}$, we define the projectors $p^\phi_j$ and $q^\phi_j$.
\begin{definition}\label{defpro}
Let $\phi\in L^{2}(\mathbb{R}^2,\mathbb{C})$ with $\| \phi  \|=1$.
\begin{enumerate}
\item For any $1\leq  j\leq  N$ the
projectors $p_j^\phi:\LZN\to\LZN$ and $q_j^\phi:\LZN\to\LZN$ are defined as
\begin{align*} p_j^\phi\Psi=\phi(x_j)\int\phi^*(\tilde{x}_j)\Psi(x_1,\ldots, \tilde{x}_j,\dots,x_N)d^2\tilde{x}_j\;\;\;\forall\;\Psi\in\LZN
\end{align*}
and $q_j^\phi=1-p_j^\phi$.

We shall also use, with a slight abuse of notation, the bra-ket notation
$p_j^\phi=|\phi(x_j)\rangle\langle\phi(x_j)|$.
\item 
For any $0\leq  k\leq  N$ we define the set 
$$
\mathcal{S}_k=\left\lbrace (s_1,s_2,\ldots,s_N)\in\{0,1\}^N\;;\;
\sum_{j=1}^N s_j=k\right\rbrace
$$ 
and the orthogonal projector $P_{k}^\phi:\LZN \rightarrow \LZN$ as
$$P_{k}^\phi=\sum_{\vec a\in\mathcal{S}_k}\prod_{j=1}^N\big(p_{j}^{\phi}\big)^{1-s_j} \big(q_{j}^{\phi}\big)^{s_j}\;.$$
For
negative $k$ and $k>N$ we set $P_{k}^\phi=0$.
\item
For any function $m:\mathbb{N}_0\to\mathbb{R}^+_0$ we define the operator $\widehat{m}^{\phi}:\LZN\to\LZN$ as
\be\label{hut}\widehat{m}^{\phi}=\sum_{j=0}^N m(j)P_j^\phi\;.\ee
We also need the shifted operators
$\widehat{m}^{\phi}_d:\LZN\to\LZN$ given by
$$\widehat{m}^{\phi}_d=\sum_{j=-d}^{N-d} m(j+d)P_j^\phi\;.$$
\end{enumerate}
\end{definition}

Following a general strategy, which is described in detail in \cite{pickl1}, we define a functional $\alpha:\LZN\times\LZ\to\mathbb{R}^+_0$ such that
\begin{enumerate}
 \item

$\dt\alpha(\Psi_t,\phi_t)$ can be estimated by $\alpha(\Psi_t,\phi_t)+\landau(1)$.
Using a Gr\"onwall type estimate, it then follows that
$\alpha(\Psi_t,\phi_t) \leq C_1 e^{C_2 t}(\alpha(\Psi_0,\phi_0) + \landau(1))$, for some constants $C_1,C_2>0$.

\item $\alpha(\Psi,\phi)\to0$
implies convergence of the reduced one particle density matrix of $\Psi$ to
$|\phi\rangle\langle\phi|$ in trace norm.
\end{enumerate}
In the case $\beta=0$ it was shown that the choice $$\alpha(\Psi,\phi)=\llaa\Psi,
\left(\widehat{n}^{\phi}\right)^j\Psi\rraa \;,$$ 
where
$n(k)=\sqrt{k/N}$ and $\laa\cdot\raa$ is scalar product on $\LZN$  fulfills these requirements, for arbitrary $j>0$, see for example \cite{pickl1} and \cite{knowles}. For the more involved scaling we consider, it is however necessary to adjust this definition in order to obtain a Gr\"onwall type estimate.

Our proof is organized as follows:
\begin{enumerate}

\item 
In Section \ref{secpre} we recall some important properties of the operator $\widehat{m}$.

\item For the most difficult scaling given by the potential $V_N$, it is crucial to take the interaction-induced correlations between the particles into account. In Section \ref{secmic} we
provide some estimates on the zero-energy scattering state. Furthermore, we explain how the effective coupling parameter $b_{V_N}$ can be inferred from the microscopic structure.

\item
In Section \ref{secpro} we prove our main Theorem stated above. We first consider the potential $W_\beta$ and define a counting measure which allows us to establish a Gr\"onwall estimate for  all $\beta>0$. We will explain in detail how one arrives at this Gr\"onwall estimate.

Afterwards, the counting measure is adjusted to the case $V_N$, taking the microscopic structure $j_{N,R}$ of the wave function into account. We then establish  a Gr\"onwall estimate and finally prove the Theorem for $V_N$. 

The needed estimates in Section \ref{secpro} are then proven in Section \ref{rigestimates}.
\end{enumerate}

\section{Preliminaries}\label{secpre}

We will first fix the notation we are going to employ during the rest of the paper.
\begin{notation}
\begin{enumerate}
 \item
Throughout the paper hats $\;\widehat{\cdot}\;$ will always be
used in the sense of Definition \ref{defpro} (c). The label $n$ will always be used for the function $n(k)=\sqrt{k/N}$.
\item
For better readability, we will often omit the upper index $\phi$ on $p_j$,
$q_j$, $P_j$, $P_{j,k}$ and $\widehat{\cdot}$. It will be placed exclusively formulas where the $\phi$-dependence is crucial.

\item
The operator norm, defined for any linear operator $f:\LZN\to\LZN$, will be denoted by
$$\|f\|_{\text{op}}=\sup_{\psi \in \LZN, \|\Psi\|=1}\|f\Psi\|\;.$$

\item
We will
denote by $\mathcal{K}(\phi_t, A_t)$ a generic polynomial with finite degree  in \\
$\|\phi_t\|_\infty, \|\nabla \phi_t\|_\infty, \|\nabla \phi_t\|, \| \Delta \phi_t \|, \|A_t\|_\infty,
\int_0^t ds \| \dot A_s \|_\infty
$ and $ \| \dot A_t \|_\infty$. Note, in particular, that for a generic constant $C$ the inequality
$C \leq \mathcal{K}(\phi_t, A_t)$ holds.
 The exact form of $\mathcal{K}(\phi_t, A_t)$ which appears in the final bounds can be reconstructed, collecting all contributions from the different estimates.
 \item
We will denote for any multiplication operator 
 $ F:L^2(\mathbb{R}^2, \mathbb{C}) \rightarrow L^2(\mathbb{R}^2, \mathbb{C}) $
 the corresponding operator 
 $$
 \mathds{1}^{\otimes (k-1)} \otimes F \otimes  \mathds{1}^{\otimes (N-k)} :
 L^2(\mathbb{R}^{2N}, \mathbb{C}) \rightarrow L^2(\mathbb{R}^{2N}, \mathbb{C}) 
 $$
 acting on the $N$-particle Hilbert space
 by $F(x_k)$. In particular, we will use, for any $ \Psi,\Omega \in  L^2(\mathbb{R}^{2N}, \mathbb{C})$ the notation
 $$
 \laa \Omega,  \mathds{1}^{\otimes (k-1)} \otimes F \otimes  \mathds{1}^{\otimes (N-k)}\Psi \raa
 =
 \laa \Omega, F(x_k) \Psi \raa 
 \;.
 $$
 In analogy, for any two-particle multiplication operator $K:L^2(\mathbb{R}^2, \mathbb{C}) ^{\otimes 2}\rightarrow L^2(\mathbb{R}^2, \mathbb{C})^{\otimes 2} $, we denote the operator acting on any $ \Psi \in  L^2(\mathbb{R}^{2N}, \mathbb{C})$ by multiplication
 in the variable $x_i$ and $x_j$ by $K(x_i,x_j)$. In particular, we denote
 $$
  \laa \Omega, K(x_i,x_j) \Psi \raa 
  =
\int_{\mathbb{R}^{2N}} 
K(x_i, x_j)
 \Omega^*(x_1,\ldots,x_N) \Psi  (x_1,\ldots,x_N) d^2x_1 \dots d^2x_N   \;.
 $$
\end{enumerate}
\end{notation}
First we prove some properties of the projectors $p_j,q_j$, which were defined in Definition \ref{defpro}.
\begin{lemma}\label{kombinatorik}
\begin{enumerate}
\item For any weights $m,r:\mathbb{N}_0\to\mathbb{R}^+_0$ the commutation relations
$$\widehat{m}\widehat{r}\,=\widehat{mr}=\widehat{r}\,\widehat{m}\;\;\;\;\;\;\;\widehat{m}p_j=p_j\widehat{m}\;\;\;\;\;\;\;\widehat{m}q_j=q_j\widehat{m}\;\;\;\;\;\;\;\widehat{m}P_{k}=P_{k}\widehat{m}$$
hold.
\item Let $n:\mathbb{N}_0\to\mathbb{R}^+_0$ be given by $n(k)=\sqrt{k/N}$.
Then, the square of $\widehat{n}$
equals the relative particle number operator of particles not in the
state $\phi$, i.e.
\be\label{partnumber}\left(\widehat{n}\right)^2=N^{-1}\sum_{j=1}^Nq_j\;.\ee
\item For any weight $m:\mathbb{N}_0\to\mathbb{R}^+_0$ and any function $f \in L^\infty\left(\mathbb{R}^4,\mathbb{R}\right)$ and any
$j,k=0,1,2$  $$\widehat{m} Q_j f(x_1,x_2)Q_k=
Q_j f(x_1,x_2)\widehat{m}_{j-k}Q_k\;,$$ where
$Q_0=p_1 p_2$, $Q_1\in\{p_1q_2,q_1p_2\}$ and
$Q_2=q_1q_2$.
Furthermore, for $j,k\in\{0,1\}$
$$\widehat{m} \widetilde Q_j \nabla_1 \widetilde Q_k=
\widetilde Q_j \nabla_1\widehat{m}_{j-k}\widetilde Q_k\;,$$
where
$\widetilde Q_0=p_1$ and $\widetilde Q_1=q_1$.

\item For any weight $m:\mathbb{N}_0\to\mathbb{R}^+_0$ and any function $f \in L^\infty\left(\mathbb{R}^4,\mathbb{C}\right)$
$$[f(x_1,x_2),\widehat{m}]=\left[f(x_1,x_2),p_1p_2(\widehat{m}-\widehat{m}_2)+(p_1q_2+q_1p_2)(\widehat{m}-\widehat{m}_1)\right]
\;.
$$

\item Let $f\in L^1\left(\mathbb{R}^2,\mathbb{C}\right)$, $g\in L^2\left(\mathbb{R}^2,\mathbb{C}\right)$. Then,
\begin{align}\label{kombeqa}\|p_j f(x_j-x_k)p_j\|_{\text{op}}
\leq&  \|f\|_1\|\phi\|_\infty^2\;,
\\ \label{kombeqb}
\|p_jg^*(x_j-x_k)\|_{\text{op}}=&
\|g(x_j-x_k)p_j\|_{\text{op}}\leq  \|g\|\;\|\phi\|_\infty
\\\label{kombeqc}
\| |\phi(x_j) \rangle \langle \nabla_j \phi(x_j)| h^* (x_j-x_k)\|_{\text{op}}=&
\|h(x_j-x_k)\nabla_j p_j\|_{\text{op}}\leq  \|h\|\|\nabla\phi\|_{\infty}
\;.\end{align}

\end{enumerate}
\end{lemma}
\begin{proof}\begin{enumerate}
 \item
follows immediately from Definition \ref{defpro}, using that $p_j$
and $q_j$ are orthogonal projectors.

\item Note that $\cup_{k=0}^N\mathcal{S}_k=\{0,1\}^N$, so $1=\sum_{k=0}^N P_k$. Using also
$(q_j)^2=q_j$ and $q_j p_j=0$ we get \begin{align*}
\sum_{j=1}^Nq_j=\sum_{j=1}^Nq_j\sum_{k=0}^N
P_k= \sum_{k=0}^N\sum_{j=1}^Nq_j
P_k=\sum_{k=0}^Nk P_k=N\widehat{n^2}=N\widehat{n}^2\;.\end{align*}

\item
Using the definitions above we have  \begin{align*} \widehat{m}
Q_j f(x_1,x_2)Q_k
=&\sum_{l=0}^N m(l)P_l Q_jf(x_1,x_2)Q_k
\;.
\end{align*}
The number of projectors $q_j$ in $P_l Q_j$ in the coordinates $j=3,\ldots,N$ is equal to $l-j$. The $p_j$ and $q_j$ with $j=3,\ldots,N$ commute with $Q_jf(x_1,x_2)Q_k$. Thus
$P_l Q_jf(x_1,x_2)Q_k= Q_jf(x_1,x_2)Q_kP_{l-j+k}$ and
\begin{align*}
\widehat{m}
Q_j f(x_1,x_2)Q_k=& \sum_{l=0}^N  m(l) Q_jf(x_1,x_2)Q_kP_{l-j+k}
\\&\hspace{-3cm}= \sum_{\widetilde l=k-j}^{N+k-j}  Q_jf(x_1,x_2)m(\widetilde l+j-k)P_{\widetilde l} Q_k
=Q_j f(x_1,x_2)\widehat{m}_{j-k}Q_k\;.
 \end{align*}
Similarly one gets the second formula.
\item First note that
\begin{align}\label{multsev}
&\hspace{-1cm}[f(x_1,x_2),\widehat{m}]-\left[f(x_1,x_2),p_1p_2(\widehat{m}-\widehat{m}_2)+p_1q_2(\widehat{m}-\widehat{m}_1)
+q_1p_2(\widehat{m}-\widehat{m}_1)\right]%
\nonumber\\=&[f(x_1,x_2),q_1q_2\widehat{m}]+\left[f(x_1,x_2),p_1p_2\widehat{m}_2+p_1q_2\widehat{m}_1
+q_1p_2\widehat{m}_1\right]%
\;.
\end{align}
We will show that the right hand side is zero.
Multiplying the right hand side with $p_1p_2$ from the left and using (c) one gets
\begin{align*}
&p_1p_2f(x_1,x_2)q_1q_2\widehat{m}+p_1p_2f(x_1,x_2)p_1p_2\widehat{m}_2-p_1p_2\widehat{m}_2f(x_1,x_2)
\\&+p_1p_2f(x_1,x_2)p_1q_2\widehat{m}_1+p_1p_2f(x_1,x_2)q_1p_2\widehat{m}_1%
\\=&p_1p_2\widehat{m}_2f(x_1,x_2)q_1q_2+p_1p_2\widehat{m}_2 f(x_1,x_2)p_1p_2-p_1p_2\widehat{m}_2f(x_1,x_2)
\\&+p_1p_2\widehat{m}_2 f(x_1,x_2)p_1q_2+p_1p_2\widehat{m}_2 f(x_1,x_2)q_1p_2%
\\&=0\;.
\end{align*}
Multiplying (\ref{multsev}) with $p_1q_2$ from the left one gets
\begin{align*}
&p_1q_2f(x_1,x_2)q_1q_2\widehat{m}+p_1q_2f(x_1,x_2)p_1p_2\widehat{m}_2+
p_1q_2f(x_1,x_2)p_1q_2\widehat{m}_1\\&+p_1q_2f(x_1,x_2)q_1p_2\widehat{m}_1-p_1q_2\widehat{m}_1f(x_1,x_2)
\;.
\end{align*}
Using (c) the latter is zero. Also multiplying with $q_1p_2$ yields zero due to symmetry in interchanging
$x_1$ with $x_2$.
Multiplying (\ref{multsev}) with $q_1q_2$ from the left one gets
\begin{align*}
&q_1q_2f(x_1,x_2)\widehat{m}q_1q_2-q_1q_2\widehat{m}f(x_1,x_2)+
q_1q_2f(x_1,x_2)p_1p_2\widehat{m}_2+\\&q_1q_2f(x_1,x_2)p_1q_2\widehat{m}_1+q_1q_2f(x_1,x_2)q_1p_2\widehat{m}_1
\end{align*}
which is again zero and so is (\ref{multsev}).
\item 
First note that, for bounded operators $A, B$,
$\| A B \|_{\text{op}}=\|  B^* A^* \|_{\text{op}}$ holds, where $A^*$ is the adjoint operator of $A$.
To show (\ref{kombeqa}), note that
\begin{align} p_j f(x_j-x_k)p_j=p_j (f\star|\phi|^2)(x_k)\;.
\label{faltungorigin}
\end{align}
It follows that $$\|p_j f(x_j-x_k)p_j\|_{\text{op}}\leq  \|f\|_1\|\phi\|_\infty^2\;.$$

For (\ref{kombeqb}) we write
\begin{align*}
\|g(x_j-x_k)p_j\|_{\text{op}}^2=&\sup_{\|\Psi\|=1}\|g(x_j-x_k)p_j\Psi\|^2=
\\=&\sup_{\|\Psi\|=1}\laa \Psi,p_j  |g(x_j-x_k)|^2 p_j\Psi\raa\\\leq& \|p_j  |g(x_j-x_k)|^2p_j\|_{\text{op}}\;.
\end{align*}
With (\ref{kombeqa}) we get (\ref{kombeqb}).
For \eqref{kombeqc} we use
\begin{align*}
\| g(x_j-x_k) \nabla_j p_j \|_{\text{op}}^ 2 
=&
\sup_{\|\Psi\|=1}
\laa \Psi,
p_j
(|g|^ 2 * |\nabla \phi|^2)(x_k)
\Psi \raa
\leq
\||g|^ 2 * |\nabla \phi|^2\|_{\infty}
 \\
 \leq
 &
 \|g\|^2 \| \nabla \phi \|_\infty^2
\end{align*}

\end{enumerate}
\end{proof}
Within our estimates we will encounter wave functions where some of the symmetry is broken (at this point the
reader should exemplarily think of the wave function
$V_\beta(x_1-x_2)\Psi$ which is not symmetric under exchange of the variables $x_1$ and $x_3$, for example). This leads to the following
definition
\begin{definition}
For any finite set $\mathcal{M}\subset\{1,2,\ldots,N\}$, define the space $\mathcal{H}_{\mathcal{M}}\subset\LZN$ as the set of
functions which are symmetric in all variables in $\mathcal{M}$
\begin{align*}\Psi\in \mathcal{H}_{\mathcal{M}}\Leftrightarrow& \Psi(x_1,\ldots,x_j,\ldots,x_k,\ldots,x_N)=\Psi(x_1,\ldots,x_k,\ldots,x_j,\ldots,x_N)\\&\text{ for all } j,k\in\mathcal{M}\;.
\end{align*}

\end{definition}
Based on the combinatorics of the $p_j$ and $q_j$, we obtain the following
\begin{lemma}\label{kombinatorikb}
For any $f:\mathbb{N}_0\to\mathbb{R}^+_0$ and any finite set $\mathcal{M}_a\subset \{1,2,\ldots,N\}$ with $1\in\mathcal{M}_a$
and any finite set $\mathcal{M}_b\subset\{1,2,\ldots,N\}$ with $1,2\in\mathcal{M}_b$
\begin{align}\label{komb1}
\left\| \widehat{f} q_1\Psi\right\|^2\leq& \frac{N}{|\mathcal{M}_a|}
\|\widehat{f}\widehat{n}\Psi\|^2\;\;\;\;\;\;\;\;\;\;\text{ for any }\Psi\in\mathcal{H}_{\mathcal{M}_a}
,
\\
\label{komb2} \left\| \widehat{f}
q_1 q_2\Psi\right\|^2\leq& \frac{N^2}{|\mathcal{M}_b|(|\mathcal{M}_b|-1)}
\|\widehat{f}(\widehat{n})^2\Psi\|^2\;\;\;\;\;\text{ for any }\Psi\in\mathcal{H}_{\mathcal{M}_b}\;.\end{align}
\end{lemma}
\begin{proof}
Let $\Psi\in\mathcal{H}_{\mathcal{M}_a}$ for some finite set $1\in\mathcal{M}_a\subset\{1,2,\ldots,N\}$.
By Lemma \ref{kombinatorik} (b),  (\ref{komb1}) can be estimated as
\begin{align*}\|\widehat{f}\widehat{n}\Psi\|^2
=&\laa\Psi,(\widehat{f})^{2}(\widehat{n})^2\Psi\raa
=N^{-1}\sum_{k=1}^N\laa\Psi,(\widehat{f})^{2}q_k\Psi\raa
\\\geq&  N^{-1}\sum_{k\in\mathcal{M}_a}\laa\Psi,(\widehat{f})^{2}q_k\Psi\raa
=\frac{|\mathcal{M}_a|}{N}\laa\Psi,(\widehat{f})^{2}q_1\Psi\raa
\\=&\frac{|\mathcal{M}_a|}{N}\|\widehat{f}
q_1\Psi\|^2\;.
 \end{align*}
Similarly, we obtain for $\Psi\in\mathcal{H}_{\mathcal M_b}$
 \begin{align*} \|\widehat{f}
(\widehat{n})^2\Psi\|^2 =&\laa\Psi,(\widehat{f}
)^2(\widehat{n})^4\Psi\raa
\geq N^{-2}\sum_{j,k\in\mathcal{M}_b}\laa\Psi,(\widehat{f} )^2q_j
q_k\Psi\raa
\\=&\frac{|\mathcal{M}_b|(|\mathcal{M}_b|-1)}{N^2}\laa\Psi,(\widehat{f} )^2q_1
q_2\Psi\raa+\frac{|\mathcal{M}_b|}{N^2}\laa\Psi,(\widehat{f}
)^2q_1\Psi\raa
\\\geq& \frac{|\mathcal{M}_b|(|\mathcal{M}_b|-1)}{N^2} \|\widehat{f} q_1q_2\Psi\|^2
 \end{align*}
which concludes the Lemma.
\end{proof}
\begin{corollary}\label{kombinatorikc}
For any weight $m:\mathbb{N}_0\to\mathbb{R}^+_0$ \begin{align}\label{coreins}
\|\nabla_2 \widehat{m}q_2\Psi\|&\leq 2\|\widehat{m}\|_{\text{op}}\|\nabla_2q_2\Psi\| ,
\\\label{corzwei}
\|\nabla_2 \widehat{m}q_1q_2\Psi\|&\leq C\|\widehat{m}\widehat{n}\|_{\text{op}}\|\nabla_2q_2\Psi\|\;.
\end{align}
\end{corollary}
\begin{proof}
Using $p_2+q_2=1$ and triangle inequality,
\begin{align}
\label{azeile}\|\nabla_2 \widehat{m}q_2\Psi\|&\leq\|p_2\nabla_2 \widehat{m}q_2\Psi\|+\|q_2\nabla_2 \widehat{m}q_2\Psi\| ,\\
\label{bzeile}\|\nabla_2 \widehat{m}q_1q_2\Psi\|&\leq\|p_2\nabla_2 \widehat{m}q_1q_2\Psi\|+\|q_2\nabla_2 \widehat{m}q_1q_2\Psi\| \;.
\end{align}
With Lemma \ref{kombinatorik} (c)  we get
\begin{align*}
(\ref{azeile})&=\|\widehat{m}_{1}p_2\nabla_2q_2\Psi\|+\|\widehat{m}q_2\nabla_2q_2\Psi\|
\leq(\|\widehat{m}_{1}\|_{\text{op}}+\|\widehat{m}\|_{\text{op}})\|\nabla_2q_2\Psi\|\;.
\end{align*}
Note that the wave function $p_2\nabla_2q_2\Psi$ is symmetric under the exchange of any two variables but $x_2$. Thus we can use Lemma \ref{kombinatorikb} to get
\begin{align*}
(\ref{bzeile})&=\|q_1\widehat{m}_{1}p_2\nabla_2q_2\Psi\|+\|q_1\widehat{m}q_2\nabla_2q_2\Psi\|
\\&\leq\frac{N}{N-1}(\|\widehat{m}_{1}\widehat{n}\|_{\text{op}}+\|\widehat{m}\widehat{n}\|_{\text{op}})\|\nabla_2q_2\Psi\|\;.
\end{align*}
Since $\sqrt{k}\leq\sqrt{k+1}$ for $k\geq 0$ it follows that the latter is bounded by
$$ C(\|\widehat{m}_{1}\widehat{n}_{1}\|_{\text{op}}+\|\widehat{m}\widehat{n}\|_{\text{op}})\|\nabla_2q_2\Psi\|\;.$$
Using that $\|\widehat{r}\|_{\text{op}}=\sup_{0\leq k \leq N}\{r(k)\}=\|\widehat{r}_d\|_{\text{op}}$
for any $d\in\mathbb{N}$ and any weight $r$, the Corollary follows.
\end{proof}

\begin{lemma}\label{trick}
Let $\Omega,\chi\in \mathcal{H}_{\mathcal{M}}$ for some $\mathcal{M}$, let $1\notin\mathcal{M}$ and $2,3\in\mathcal{M}$.
Let $O_{j,k}$ be an operator acting on the $j^{th}$ and $k^{th}$  coordinate. Then
\begin{align*}
|\laa\Omega,O_{1,2}\chi\raa|&\leq \|\Omega\|^2+\left|\laa O_{1,2}\chi,O_{1,3}\chi\raa\right|+(|\mathcal{M}|)^{-1}\|O_{1,2}\chi\|^2
%
\;.
\end{align*}
\end{lemma}
\begin{proof}
Using symmetry and Cauchy Schwarz
\begin{align*}
|\laa\Omega,O_{1,2}\chi\raa|=&|\mathcal{M}|^{-1}|\laa\Omega,\sum_{j\in\mathcal{M}} O_{1,j}\chi\raa|\leq|\mathcal{M}|^{-1}\|\Omega\|\;\|\sum_{j\in\mathcal{M}} O_{1,j}\chi\|
\end{align*}
For the second factor we can write
\begin{align*}
\|\sum_{j\in\mathcal{M}} O_{1,j}\chi\|^2&=\laa\sum_{j\in\mathcal{M}} O_{1,j}\chi,\sum_{k\in\mathcal{M}} O_{1,k}\chi\raa
\\\leq&\sum_{j\in\mathcal{M}}|\laa O_{1,j}\chi,O_{1,j}\chi\raa|+|\sum_{j\neq k\in\mathcal{M}}\laa O_{1,j}\chi,O_{1,k}\chi\raa|
\\\leq&|\mathcal{M}||\laa O_{1,2}\chi,O_{1,2}\chi\raa|+|\mathcal{M}|(|\mathcal{M}|-1)|\laa O_{1,2}\chi,O_{1,3}\chi\raa|
\end{align*}
Since $ab\leq 1/2a^2+1/2b^2$ and $(a+b)^2\leq 2a^2+2b^2$ holds for any real numbers $a,b$, the Lemma follows.
\end{proof}

In our estimates, we need the regularity conditions 
\begin{align*}
\| \nabla \phi_t \|_\infty < \infty,
\qquad
\|  \phi_t \|_\infty < \infty,
\qquad
\| \nabla \phi_t \| < \infty,
\qquad
\| \Delta\phi_t \| < \infty \;.
\end{align*}
That is, we need $\phi_t \in H^2(\mathbb{R}^2,\mathbb{C}) \cap W^{1,\infty}(\mathbb{R}^2,\mathbb{C})$.
Then, $ \| \Delta|\phi_t| ^2 \|, \| \Delta|\phi_t| ^2 \|_1$ and $ \| \phi_t^ 2 \| $, which also appear in our estimates, can be bounded by
\begin{align*}
 \Delta|\phi_t|^ 2
 =&
 \phi^*_t \Delta \phi_t
  +
 \phi_t  \Delta\phi^*_t
   +
  2   (\nabla \phi^*_t) \cdot (\nabla \phi_t)
  \\
 \| \Delta|\phi_t| ^2 \|
 \leq &
2
 \| \Delta \phi_t \| \| \phi_t \| _\infty
 +
 2
 \| \nabla \phi_t \|  \| \nabla \phi_t \| _\infty
   \\
 \| \Delta|\phi_t| ^2 \|_1 \leq&  4 \| \Delta \phi_t \|
 \\
 \| \phi_t^ 2 \| \leq &\| \phi_t \|_\infty \| \phi_t \|
 \;.
\end{align*}
Recall the Sobolev embedding Theorem, which implies in particular
$H^k(\mathbb{R}^2,\mathbb{C})= W^{k,2}(\mathbb{R}^ 2,\mathbb{C}) \subset C^{k-2}(\mathbb{R}^2,\mathbb{C})$. 
If $\phi \in C^1(\mathbb{R}^2,\mathbb{C}) \cap H^1(\mathbb{R}^2,\mathbb{C})$, then $\phi \in W^{1,\infty}(\mathbb{R}^2,\mathbb{C})$ follows since both $\phi$ and $\nabla \phi$ have to decay at infinity.
 Thus, $\phi_t \in H^3(\mathbb{R}^2,\mathbb{C})$ implies $\phi_t \in H^2(\mathbb{R}^2,\mathbb{C}) \cap W^{1,\infty}(\mathbb{R}^2, \mathbb{C})$, which suffices for our estimates. Since $\phi_t$ obeys a defocusing nonlinear Schr\"odinger equation, we expect the regularity of the solution $\phi_t$ to follow from the regularity of the initial datum $\phi_0$. For a certain class of external potentials $A_t$ this has been proven in \cite{carles}:
\begin{lemma} \label{regularityLemma}
Let $\phi_0  \in 
\Sigma^k(\mathbb{R}^2,\mathbb{C})
=
\lbrace
f \in L^2 (\mathbb{R}^2, \mathbb{C})|
\sum_{ \alpha+ \beta \leq k} \|x^\alpha \partial_x^\beta f \| < \infty
\rbrace
$, for $k\geq2$. 
Let, for $b>0$,
$\phi_t$ the unique solution to
\begin{align*}
i \partial_t \phi_t = (-\Delta+A_t+ b |\phi_t|^2) \phi_t \;.
\end{align*}
Let $A_{\cdot} \in L^\infty_{\text{loc}}(\mathbb{R}_t \times \mathbb{R}^2_x, \mathbb{C})$ real valued and smooth with respect to the space variable: for (almost) all $t \in \mathbb{R}$,
the map $x \mapsto A_t(x)$ is $C^\infty$. Moreover, $A_t$ is at most quadratic in space, uniformly w.r.t. time $t$:
\begin{align*}
\forall \alpha \in \mathbb{N}^2, | \alpha | \geq 2, \qquad \partial_x^\alpha A_{\cdot}
\in L^\infty(\mathbb{R}_t \times \mathbb{R}^d_x, \mathbb{C}).
\end{align*}
In addition, $t \mapsto \sup_{|x| \leq 1} |A_t(x)|$ belongs to $L^{\infty}(\mathbb{R}, \mathbb{C})$.
Then 
\begin{itemize}
\item[(a)]
$ \phi_t \in \Sigma^k(\mathbb{R}^2,\mathbb{C})$, which implies $ \phi_t \in H^k(\mathbb{R}^2,\mathbb{C})$.
\item[(b)]
$\| \phi_t\|=\| \phi_0 \|$.
\item[(c)]  
Let $ \phi_0 \in \Sigma^3(\mathbb{R}^2,\mathbb{C})$. Assume in addition that
$\| A_t \|_\infty < \infty$ and $\| \dot{A}_t \|_\infty < \infty$.
Then, for any fixed $t \geq 0$, 
$ \mathcal{K}(\phi_t, A_t) < \infty$ follows.

\end{itemize}
\end{lemma}
\begin{proof}
Part (a) is Corollary 1.4. in \cite{carles}. We like to remark that $\| \phi_t \| _{H^k} \leq C$ holds, if $A_t=0$, see Section 1.2. in \cite{carles}.
The conditions on $A_t$ are for example satisfied if
 $A_t \in C^\infty_c (\mathbb{R}^2, \mathbb{R})$ for all $t \in \mathbb{R}$, $A_t(x)=0$, for all $|t| \geq T$.
Part (b) can be verified directly, using the existence of global in time solutions. 
Part (c) follows from (a) and the embedding
$H^3(\mathbb{R}^2,\mathbb{C}) \subset  H^2(\mathbb{R}^2,\mathbb{C}) \cap W^{1,\infty}(\mathbb{R}^2, \mathbb{C}) $.

\end{proof}

\section{Microscopic structure in 2 dimensions}
\label{secmic}

\subsection{The scattering state}

In this section we analyze the microscopic structure which is induced by $V_N$. In particular, we explain why the dynamical properties of the system are determined by the low energy scattering regime.

\begin{definition}
Let $V \in L^\infty_c(\mathbb{R}^2, \mathbb{R})$, $V(x) \geq 0$, $V$ spherically symmetric and let $V_N$ be given by $V_N(x) = e^{2N} V(e^N x)$.
For any $R \geq \text{diam}(\text{supp} (V_N))$,
we define the zero energy scattering state $j_{N,R}$ by
\begin{equation}
\label{eq: defj}
\begin{cases} 
\left( - \Delta_x + \frac{1}{2}e^{2N} V(e^N x)  \right) j_{N,R}(x)=0 ,
\\
j_{N,R}(x)=1 \; \text{for } |x| = R \;.
  \end{cases}
\end{equation}

\end{definition}


Next, we want to recall some important properties of the scattering state $j_{N,R}$, see also Appendix C of \cite{lssy}.

\begin{lemma} \label{scattlemma fuer v}
Let $V \in L_c^\infty( \mathbb{R}^2, \mathbb{R})$, $V(x) \geq 0$ and spherically symmetric.
Define $I_{R} = \int_{\mathbb{R}^2} d^2x V_N(x) j_{N,R}(x)$.
For the scattering state defined previously the following relations hold:
\begin{enumerate}
\item[(a)] There exists a nonnegative number $a$, called scattering length of the potential $ V$, such that
\begin{align*}
I_{R}=
  \frac{4 \pi }{\ln \left(\frac{e^N R}{a}\right)}
\end{align*}
(in the case $a=0$ we have $I_R=0$).
The scattering length $a$ does not depend on $R$ and fulfills $a \leq \text{diam}(\text{supp}(V))$.
Furthermore, $I_R \geq 0$ holds.
\item[(b)]
$j_{N,R}$ is a nonnegative function which is spherically symmetric in $|x|$. For $|x| \geq \text{diam}(\text{supp}(V_N))$, $j_{N,R}$ is given by
 $$
 j_{N,R} (x)
= 1 +  \frac{1}{\ln \left(\frac{e^N R}{a}\right)} \ln \left( \frac{|x|}{R} \right). $$
\end{enumerate}

\end{lemma}

\begin{proof}
\begin{enumerate}
\item[(a)+(b)] 

Rescaling $x \rightarrow e^{N} x =y$, we obtain, setting $\tilde{R}=e^{N} R$ and $s_{\tilde{R}}(y)= j_{0, e^{N}R}(y)$, the unscaled scattering equation
\begin{equation}
\begin{cases} 
\left( - \Delta_y + \frac{1}{2}V(y)  \right) s_{\tilde{R}}(y)=0 ,
\\
s_{\tilde{R}}(y)=1 \; \text{for } |y| =\tilde{R} \;.
  \end{cases} \;
\end{equation}
Since we assume $V$ to be nonnegative, one can define the scattering state $s_{\tilde{R}}$ by a variational principle. Theorem C.1 in \cite{lssy} then implies that $s_{\tilde{R}}$ is a nonnegative, spherically symmetric function in $|y|$. 
It is then easy to verify that for $\text{diam (supp} (V))  \leq |y| $ there exists a number $A\in \mathbb{R}$ such that 
\begin{equation} 
\label{jasymp}
s_{\tilde{R}}(y) =  1 + \frac{A}{4 \pi} \ln \left( \frac{|y|}{\tilde{R}} \right)
\;.
\end{equation}

Next, we show that $A= \int_{\mathbb{R}^2}d^2y V(y) s_{\tilde{R}} (y)$. This can be seen by noting that, for $r > \text{diam (supp} (V))   $,
\begin{align*}
 \int_{\mathbb{R}^2}d^2y V(y) s_{\tilde{R}}(y)
 =&
 2 \int_{B_r(0)}d^2y \Delta s_{\tilde{R}}(y)
  =
  2  \int_{\partial B_r(0)} \nabla s_{\tilde{R}}(y) \cdot ds
\nonumber \\   
    =&
   \frac{A}{2 \pi}   \int_{\partial B_r(0)} \nabla  \ln (|y|)\cdot ds
   =
   \frac{A}{2 \pi}   \int_{0}^{2 \pi} \frac{1}{r} r d\varphi
   \\
   =&
A\;.
\end{align*}

By Theorem C.1 in \cite{lssy}, there exists a number $a \geq 0$, not depending on $\tilde{R}$, such that for all $|y| \geq  \text{diam (supp} (V))$
\begin{align*}
s_{\tilde{R}}(y)= \frac{\ln(|y|/a) }{\ln(\tilde{R}/a)}
\;.
\end{align*}
Comparing this with \eqref{jasymp}, we obtain
\begin{align*}
 \int_{\mathbb{R}^2} V(y) s_{\tilde{R}}(y) dy^2=
  \frac{ 4 \pi}{\ln \left(\frac{\tilde{R}}{a}\right)}
 \;.
\end{align*}
Since $s_{\tilde{R}}$ is nonnegative, it furthermore follows that $a \leq  \text{diam (supp} (V))$.
This directly implies $A \geq 0$. By scaling, we obtain
\begin{align*}
I_R= \int_{\mathbb{R}^2} V_N(y) j_{N,R}(y) dy^2
=
 \int_{\mathbb{R}^2} V(y) s_{\tilde{R}}(y) dy^2
 =
 \frac{ 4 \pi}{ \ln \left(\frac{e^N R}{a}\right)}
 \;.
\end{align*}

\end{enumerate}
\end{proof}

Assuming that the energy per particle $\mathcal{E}_{V_N}(\Psi)$ is of order one, the wave function $\Psi$ will have a microscopic structure near the interactions $V_N$, given by $j_{N,R}$. 
The interaction among two particles is then determined by
$   \frac{4 \pi}{N+ \ln \left(\frac{R}{a}\right)} \approx \frac{4 \pi}{N}$. Keeping in mind that each particle interacts with all other $N-1$ particles, we obtain the effective Gross-Pitaevskii equation, for $\varphi_t \in H^2( \mathbb{R}^2, \mathbb{C})$
\begin{align*}
i \partial_t \varphi_t(x) = (- \Delta+A_t+ 4 \pi | \varphi_t(x)|^2) \varphi_t(x).
\end{align*}
Thus, choosing $V_N(x)= e^{2N} V(e^Nx)$ leads in our setting to an effective one-particle equation which is determined by the low energy scattering behavior of the particles. We remark that, for any $s>0$, the potential
$e^{2Ns} V(e^{Ns}x)$ yields to the coupling $4 \pi/ s$.

\subsection{Properties of the scattering state}
Note that the potential $V_N$ is strongly peaked within an exponentially small region. 
In order to control the short scale structure of $\Psi_t$, we define, with a slight abuse of notation, a
potential $M_{\beta}$ with softer scaling behavior in such a way that the potential $V_N -M_\beta$
has scattering length zero.
This allows us to ``replace'' $V_N$  by $M_\beta$, which has better scaling behavior and is easier to control.
In particular, $\|M_\beta\| \leq C N^{-1+\beta}$ can be controlled for $\beta$ sufficiently small, while $\|V_N\|=\mathcal{O}(e^N)$ cannot be bounded by any finite polynomial in $N$. 
The potential $M_\beta$ is \textit{not} of the exact scaling $N^{-1+2\beta} M(N^{\beta} x)$. 
However, it is in the set $\mathcal{V}_{\beta}$, which we will define now.

\begin{definition}
For any $\beta>0$, we define the set of potentials $\mathcal{V}_{\beta}$ as
\begin{align*}
&
\mathcal{V}_{\beta}=
\Big\lbrace
U \in L^2(\mathbb{R}^2,\mathbb{R}) |
U(x) \geq 0 \; \forall x \in \mathbb{R}^2
,
\|U\|_1 \leq CN^{-1} 
,
\|U\| \leq CN^{-1+ \beta},
\\
& 
\|U\| _\infty \leq C N^{-1+2 \beta}
,
U(x)=0 \; \forall |x| \geq C N^{-\beta}
,\;
U\text{ is spherically symmetric}
\Big\rbrace .
\end{align*}
Note that $N^{-1+2\beta}W(N^\beta x) \in \mathcal{V}_\beta$ holds, if $W$ is positive, spherically symmetric and compactly supported.
\end{definition}
All relevant estimates in this paper are formulated for $W_\beta \in \mathcal{V}_\beta$.

\begin{definition}\label{microscopic}
Let $V \in L_c^\infty( \mathbb{R}^2, \mathbb{R})$, $V(x) \geq 0$ and spherically symmetric.
For any $\beta >0$ and any $R_\beta \geq N^{- \beta}$ we define the potential $M_\beta$ via
\begin{align}
\label{eq: defW}
M_\beta(x)
=
 \begin{cases} 
4 \pi N^{-1+2 \beta} & \text{if } N^{- \beta} < |x| \leq R_\beta  \\
   0      & \text{else } 
  \end{cases}
  \;.
\end{align}
Furthermore, we define the zero energy scattering state $f_\beta$ of the potential
$ \frac{1}{2} (V_N-M_\beta)$, that is
\begin{align}
\label{eq: deff}
 \begin{cases} 
\left( - \Delta_x + \frac{1}{2} \left(V_N(x)-M_\beta(x) \right)  \right) f_{\beta}(x)=0
\\
f_{\beta}(x)=1 \; \text{for } |x| = R_\beta 
  \end{cases}
  \;.
\end{align}
\end{definition}
Note that $M_\beta$ and $f_\beta$ depend on $R_\beta$. 
We choose $R_\beta$ such that the scattering length of the potential 
$(V_N-M_\beta)$ is zero. This is equivalent to the condition 
$\int_{\mathbb{R}} d^2x (V_N(x)-M_\beta(x)) f_\beta(x) =0$.

\begin{lemma}\label{defAlemma}
For the scattering state $f_\beta$, defined by \eqref{eq: deff}, the following relations hold:
\begin{enumerate}
\item[(a)] There exists a minimal value $R_\beta <\infty$ such that $\int_{\mathbb{R}^2} d^2x (V_N(x)-M_\beta(x)) f_\beta(x) =0$. 
 \end{enumerate}
For the rest of the paper we assume that $R_\beta$ is chosen such that (a) holds.
  \begin{enumerate}
\item[(b)] There exists $K_{\beta} \in \mathbb{R}, \; K_{\beta}> 0$ such that
$K_{\beta} f_{\beta}(x) = j_{N,R_\beta}(x) \; \forall |x| \leq N^{-\beta}$.

\item[(c)] For $N$ sufficiently large the supports of $V_N$ and $M_{\beta}$ do not overlap.

\item[(d)] 
 $f_{\beta}$ is a nonnegative, monotone nondecreasing function in $|x|$.
\item[(e)]
\begin{align}
f_{\beta}(x)=1 \; \text{for } |x| \geq R_\beta \;.
\end{align}
\item[(f)] 
\begin{align}
1 \geq K_\beta \ge
1+ \frac{1}{N+ \ln \left(\frac{R_\beta}{a}\right)} \ln \left( \frac{N^{-\beta}}{R_\beta} \right)
\; .
\end{align}
\item[(g)]
$ R_\beta \leq C N^{-\beta}$.
\end{enumerate}
For any fixed $0<\beta$, $N$ sufficiently large such that $V_N$ and $M_{\beta}$ do not overlap, we obtain
\begin{enumerate}
\item[(h)] 
\begin{align*}
&| N \| V_{N}f_{\beta} \|_1 - 4 \pi | 
=
| N \| M_{\beta}f_{\beta} \|_1 - 4 \pi | \leq C \frac{\ln(N)}{N} 
  \;.
\end{align*}

\item[(i)]
Define 
\begin{align*}
g_\beta(x) = 1 - f_\beta(x)
\;.
\end{align*}
Then,
\begin{align*}
\|g_{\beta}\|_1&\leq  C N^{-1-2\beta}\ln N \;,\hspace{1cm}\|g_{\beta}\|\leq  C N^{-1-\beta}\ln N
\;,\hspace{1cm}
\|g_{\beta}\|_\infty \leq 1
 \;.
\end{align*}
\item[(j)] 
\begin{align*}
| N \| M_{\beta}\|_1 - 4 \pi | \leq C \frac{\ln(N)}{N} 
  \;.
\end{align*}
\item[(k)]
\begin{align*}
M_{\beta} \in \mathcal{V}_\beta
\;,
M_{\beta} f_\beta \in \mathcal{V}_\beta
\;.
\end{align*}
\end{enumerate}

\end{lemma}

\begin{proof}
\begin{enumerate}
\item[(a)]
In the following, we will sometimes denote, 
with a slight abuse of notation, $f_{\beta}(x)=f_{\beta}(r)$
and $j_{N,R}(x)=j_{N,R}(r)$
 for $r=|x|$ (for this, recall that $f_{\beta}$ and $j_{N,R}$ are radially symmetric). 
 We further denote by $f'_{\beta}(r)$ the derivative of $f_{\beta}$ with respect to

We first show by contradiction that there exists a $x_0 \in \mathbb{R}^2,\; |x_0|\leq N^{-\beta}$, such that $f_\beta(x_0) \neq 0$.
For this, assume  that 
$f_{\beta} (x) =0$ for all $|x| \leq N^{-\beta}$. 
Since $f_{\beta}$ is continuous, there exists a maximal value $r_0 \geq N^{-\beta}$ such that the scattering
equation \eqref{eq: deff} is equivalent to
\begin{align}
\label{nosolution}
 \begin{cases} 
\left( - \Delta_x - \frac{1}{2} M_{\beta}(x) \right) f_{\beta}(x)=0,
\\
f_{\beta}(x)=1 \; \text{for } |x| = R_{\beta},
\\
f_{\beta}(x)=0 \; \text{for } |x| \leq r_0
\;.
  \end{cases}
\end{align}
\\

Using \eqref{eq: deff} and Gauss'-theorem, we further obtain
\begin{align}
\label{Gauss}
f'_{\beta_1, 1}(r) =
\frac{1}{ 4 \pi r }
\int_{B_r(0)} d^2 x (V_N(x)-M_\beta(x)) f_\beta(x) 
\;.
\end{align}
 \eqref{nosolution} and \eqref{Gauss} then imply for $r > r_0$
 \begin{align*}
&
 \left| f'_{\beta}(r)
\right| 
  =
\frac{1}{ 4 \pi r }
\left|
\int_{B_r(0)} d^2 x M_{\beta}(x) f_{\beta}(x) 
\right|
=
\frac{2 \pi  
N^{-1+2 \beta}
}{  r }
\left|
\int_{r_0}^r dr' r'f_{\beta}(r') 
\right|
\\
\leq&
\frac{2 \pi 
N^{-1+2 \beta}
}{ r}
\left|
\int_{r_0}^r dr' r'
(r'-r_0)  
\sup_{r_0 \leq s \leq r}
|f'_{\beta}(s)|
\right| .
 \end{align*}
Taking the supreme over the interval 
$[r_0,r]$, the inequality above then implies that
there exists a constant $C(r,r_0) \neq 0$, $ \lim\limits_{r \rightarrow r_0} C(r,r_0)=0$ such that
 $
 \sup\limits_{r_0 \leq s \leq r}
|f'_{\beta}(s)|
 \leq
C(r,r_0)
N^{-1+3 \beta_1}
\sup\limits_{r_0 \leq s \leq r}
|f'_{\beta}(s)|
 $. Thus, for $r$ close enough to $r_0$, the inequality above can only hold if 
 $
f'_{\beta}(s)=0 
 $ for $s \in [r_0,r]$, yielding a contradiction to the choice of $r_0$.

Consequently, there exists a $x_0 \in \mathbb{R}^2, |x_0| \leq N^{-\beta}$, such that
$f_{\beta}(x_0) \neq 0$.
We can thus define 
\begin{equation*}
h(x) = f_{\beta}(x)\frac{j_{N,R}(x_0)}{f_{\beta}(x_0)}
\end{equation*}
on the compact set $\overline{B_{x_0}(0)}$.
One easily sees  that $h(x)= j_{N,R}(x)$ on $\partial \overline{B_{x_0}(0)}$ and satisfies the zero energy scattering equation~\eqref{eq: defj} for $ x \in \overline{B_{N^{-\beta}}(0)}$.
Note that the scattering equations \eqref{eq: defj} and \eqref{eq: deff} have a unique solution on any compact set.
It then follows that 
$h(x) = j_{N, R}(x) \; \forall x \in \overline{B_{N^{-\beta}}(0)}$.
Since $j_{N, R} (N^{-\beta}) \neq 0$, we then obtain
$f_{\beta}(N^{-\beta_1}) \neq 0$.
Applying Theorem C.1 in \cite{lssy} once more, it then follows that either $f_\beta$ or $-f_\beta$ is a nonnegative, monotone nondecrasing function in $|x|$
 for all $|x| \leq N^{-\beta}$.

Recall that $W_{\beta}$ and hence $f_{\beta}(x) $ depend on $R_\beta
\in [N^{-\beta}, \infty[
$. 
 For conceptual clarity, we
 denote $W_{\beta}^{(R_{\beta})}(x) = W_{\beta}(x) $ and
$f_{\beta}^{(R_{\beta})}(x)= f_{\beta}(x)$ for the rest of the proof of part (a).
For $\beta$ fixed, consider the function
\begin{align}
& s: [N^{-\beta} ,\infty [ 
\rightarrow  \mathbb{R} \\
&R_{\beta}
\mapsto  
\int_{B_{R_{\beta}}(0)} d^2 x
 (V_N(x)-W^{(R_{\beta})}_{\beta}(x)) f^{(R_{\beta})}_{\beta} (x).
\end{align}
We show  by contradiction that the function $s$ has at least one zero.
Assume $s \neq 0$ were to hold.
 We can assume w.l.o.g. $s >0$. 
It  then
follows from Gauss'-theorem that 
$
f'^{(R_{\beta})}_{\beta}(R_{\beta})> 0
$ for all  
$R_\beta \geq N^{-\beta}$.
By uniqueness of the solution of the scattering equation \eqref{eq: deff}, for $\tilde{R}_{\beta}<R_{\beta}$ there exists a constant
$ K_{\tilde{R}_{\beta},R_{\beta}} \neq 0$, such that for all
$
|x|\leq \tilde{R}_{\beta}
$
we have
$
f^{(\tilde{R}_{\beta})}_{\beta}(x)= K_{\tilde{R}_{\beta},R_{{\beta}}}f^{(R_{\beta})}_{\beta}(x)
$. Since $ f^{(R_{\beta})}_{\beta} $ and $s$ are continuous, we can further conclude
 $ K_{\tilde{R}_{\beta},R_{{\beta}}}> 0$.
From $ s \neq 0$, it then follows that, for all $r \in [N^{-\beta} ,\infty [  $ and for all $R_\beta \in [N^{-\beta} ,\infty [ $ , $
f'^{(R_{\beta})}_{\beta}(r) \neq 0 $.
Thus, for all $r \in [N^{-\beta} ,\infty [  $ and for all $R_\beta \in [N^{-\beta_1} ,\infty [ $, the function
$f^{(R_{\beta})}_{\beta}(r)$ doesn't change sign. 
From Lemma \ref{scattlemma fuer v}, the assumption $s(N^{-\beta})>0$ and  $ K_{\tilde{R}_{\beta},R_{{\beta}}} > 0$, we obtain,
 for all $r \in [0, N^{-\beta} ] $ and for all $R_{\beta} \in [N^{-\beta} ,\infty [ $, that
$f^{(R_{\beta})}_{\beta} (r) \geq 0$ holds.
This, however, implies $\lim\limits_{R_{\beta} \rightarrow \infty} s(R_{\beta})= - \infty$ yielding to a contradiction.  
By continuity of $s$, there exists thus a minimal value $R_{\beta} \geq N^{-\beta} $ such that
$s(R_{\beta})=0$.

\begin{remark}
As mentioned, we will from now on fix $R_{\beta} \in [N^{- \beta}, \infty[$ as the minimal value such that $s(R_{\beta})=0$.
Furthermore, we may assume $a>0$ and $R_{\beta} >N^{-\beta}$ in the following. For $a=0$, we can choose $R_{\beta}=N^{-\beta}$, such that $f_{\beta}(x)=j_{N,R}(x)$. It is then easy to verify that the Lemma stated is valid.
\end{remark}

\item[(b)]
From (a), we can conclude that
\begin{equation}
\label{defKbeta}
K_{\beta} = \frac{j_{N,R_\beta}(N^{-\beta})}{f_\beta(N^{-\beta})}.
\end{equation}
Next, we show that the constant $K_\beta$ is positive.
Since $j_{N,R_\beta}(N^{-\beta})$ is positive, it follows from Eq.~\eqref{defKbeta} that  $K_\beta$ and $f_\beta (N^ {-\beta})$ have equal sign. By (a), the sign of $f_\beta$ is constant for $|x|\leq R_\beta$.
Since $j_{N,R_\beta}$ and $V_N$ are nonnegative functions,
we obtain by Gauss-theorem and the scattering equation \eqref{eq: deff}
\begin{align}
\label{eq: sign of derivative of f}
\text{sgn} \left(\frac{\partial f_\beta}{\partial r}|_{r= N^{- \beta}}\right) = \text{sgn}(K_\beta).
\end{align}
Recall that $R_\beta$ is the smallest value such that
  $\frac{\partial f_\beta}{\partial r} \big|_{r=R_\beta} = 0$.
If it were now that $K_\beta$ is negative, we could conclude from \eqref{defKbeta} and \eqref{eq: sign of derivative of f} that
$\frac{\partial f_\beta}{\partial r}|_{r= N^{- \beta}} <0$ and
$ f_\beta ( N^{- \beta}) <0$. 
Since $R_\beta$ is by definition the smallest value where $\frac{\partial f_\beta}{\partial r} = 0$, we were able to conclude from the continuity of the derivative that  $\frac{\partial f_\beta}{\partial r} <0$ for all $r < R_\beta$ and hence $f(R_\beta) <0$. However, this were in contradiction to the boundary condition of the zero energy scattering state (see \eqref{eq: deff}) and thus $K_\beta > 0$ follows.

\item[(c)] This directly follows from $e^{-N} < C N^{- \beta}$ for $N$ sufficiently large.

\item[(d)]
From the proof of property (b), we see that $f_\beta$ and its derivative is positive at $N^{- \beta}$. 
From \eqref{Gauss}, we obtain $f'_\beta(r) = 0 $ for all  $r > R_\beta$. Due to continuity $f'_{\beta}(r)  >0 $ for all  $r< R_\beta$. Since $f_\beta$ is continuous, positive at  $N^{- \beta}$, and its derivative is a nonnegative function, it follows that $f_\beta$ is a nonnegative, monotone nondecreasing function in $|x|$.

\item[(e)]
By definition of $R_\beta$, it follows that $\tilde{I}=\int_{\mathbb{R}^2}d^2x (V_N(x)-W_\beta(x)) f_\beta(x) =0$. Therefore, for all $|x|\geq R_\beta$, $f_\beta$ solves $-\Delta f_\beta(x)=0$, which has the solution
\begin{align*}
f_\beta(x)= 1 + \frac{\tilde{I}}{4 \pi} \ln \left( \frac{|x|}{R_\beta} \right)=1
\end{align*}

\item[(f)]
Since $f_\beta$ is a positive monotone nondecreasing function in $|x|$, we obtain
\begin{align*}
1 \geq f_\beta( N^{- \beta}) =j_{N,R_\beta}( N^{- \beta}) /K_\beta
=
\left(
1+   \frac{1}{N+ \ln \left(\frac{R_\beta}{a}\right)} \ln \left( \frac{N^{-\beta}}{R_\beta} \right)
\right)
/ K_\beta
\end{align*}
We obtain the lower bound
\begin{align*}
K_\beta \ge
1+  \frac{1}{N+ \ln \left(\frac{R_\beta}{a}\right)}\ln \left( \frac{N^{-\beta}}{R_\beta} \right).
\end{align*}
For the upper bound we first prove that $f_{\beta}(x) \geq j_{N,R_\beta}(x)$ holds for all $|x | \leq R_\beta$.
Using the scatting equations \eqref{jasymp} and \eqref{eq: deff} we obtain
\begin{align*}
\Delta_x
(
f_\beta(x)-j_{N,R_\beta}(x)
)
=
\frac{1}{2}
V_N(x)
(
f_\beta(x)-j_{N,R_\beta}(x)
)
-
W_\beta(x) f_{\beta}(x)
\end{align*}
as well as
$f_{\beta}(R_\beta)- j_{N,R_\beta}(R_\beta)=0$.
Since $W_\beta(x) f_{\beta}(x) \geq 0$, we obtain that
$\Delta_x
(
f_\beta(x)-j_{N,R_\beta}(x)
) \leq 0$ for $ N^{-\beta} \leq |x|\leq R_\beta$. That is, $f_\beta(x)-j_{N,R_\beta}(x)$ is superharmonic
for $ N^{-\beta} < |x|< R_\beta$. Using the minimum principle, we obtain, using that 
$f_\beta-j_{N,R_\beta}$ is spherically symmetric
\begin{align}
\label{minimum}
\min_{N^{-\beta} \leq |x|\leq R_\beta} ( f_\beta-j_{N,R_\beta})
=
\min_{|x| \in \lbrace N^{-\beta} ,  R_\beta \rbrace} ( f_\beta-j_{N,R_\beta})
\end{align}
If it were now that $\min_{|x| \in \lbrace N^{-\beta} ,  R_\beta \rbrace} ( f_\beta-j_{N,R_\beta})
=f_{\beta}(N^{-\beta})- j_{N,R_\beta}(N^{-\beta})
\leq
f_{\beta}(R_\beta)- j_{N,R_\beta}(R_\beta)=0
$, we  could conclude that $ f_{\beta}(x)- j_{N,R_\beta}(x) \leq 0$ for all
$ N^{-\beta} \leq |x|\leq R_\beta$. Since $f_\beta(x)-j_{N,R_\beta}(x)$ then obeys
\begin{align*}
 \begin{cases} 
-\Delta 
(
f_\beta(x)-j_{N,R_\beta}(x)
)
+
\frac{1}{2}
V_N(x)
(
f_\beta(x)-j_{N,R_\beta}(x)
)
=
0
\;
&\text{for } |x| \leq N^{-\beta} ,
\\
f_\beta(x)-j_{N,R_\beta}(x) \leq 0 \; &\text{for } |x| = N^{-\beta},
  \end{cases} 
\end{align*}
 we could then conclude that $f_\beta(x)-j_{N,R_\beta}(x) \leq 0$
for all $|x| \leq R_\beta$. From this, we obtain that
 $\Delta 
(
f_\beta(x)-j_{N,R_\beta}(x)
) \leq 0$ for $ |x|\leq R_\beta$. That is, $f_\beta(x)-j_{N,R_\beta}(x)$ is superharmonic for all
$ |x|\leq R_\beta$.
Using the minimum principle once again, we then obtain
\begin{align*}
\min_{\overline{B_{R_\beta}(0)}} ( f_\beta-j_{N,R_\beta})
=
 f_\beta(R_\beta)-j_{N,R_\beta}(R_\beta)=0
\end{align*}
which contradicts $ f_{\beta}(x)- j_{N,R_\beta}(x) \leq 0$ for $|x| \leq R_\beta$.
Therefore, we can conclude in \eqref{minimum} that
$
\min_{N^{-\beta} \leq |x|\leq R_\beta} ( f_\beta-j_{N,R_\beta})
=
  f_\beta(R_\beta)-j_{N,R_\beta}(R_\beta)=0
$ holds. Then, it follows that $f_\beta(x)-j_{N,R_\beta}(x) \geq 0$ for all $N^{-\beta} \leq |x| \leq R_\beta$. Using the zero energy scattering equation
$-\Delta 
(
f_\beta(x)-j_{N,R_\beta}(x)
)
+
\frac{1}{2}
V_N(x)
(
f_\beta(x)-j_{N,R_\beta}(x)
)
=
0$ for $|x| \leq N^{-\beta}$, we can, together with $  f_\beta(N^{-\beta})-j_{N,R_\beta}(N^{-\beta}) \geq 0$, conclude that  $ f_{\beta}(x)- j_{N,R_\beta}(x) \geq 0$ for all $|x| \leq R_\beta$. 

As a consequence, we obtain the desired bound
$
K_\beta=
\frac{j_{N,R_\beta}(N^{-\beta})}{f_\beta(N^{-\beta})}\leq 1$.

\item[(g)]
Since $f_\beta$ is a nonnegative, monotone nondecreasing function in $|x|$ with $f_\beta (x)=1$ $\forall |x| \geq R_\beta$, it follows that
\begin{align*}
C f_\beta(N^{-\beta}) 
=&
f_\beta(N^{-\beta}) 
\int_{\mathbb{R}^2} d^2 x V_N(x) 
\geq
\int_{\mathbb{R}^2} d^2 x V_N(x) f_\beta(x) 
\\
=&
\int_{\mathbb{R}^2} d^2 x M_\beta(x) f_\beta(x) 
\geq
f_{\beta} (N^{-\beta})
\int_{\mathbb{R}^2} d^2 x M_\beta (x)
\;.
\end{align*}
Therfore, $\int_{\mathbb{R}^2} d^2 x M_\beta (x) \leq C$ holds, which implies that $R_\beta \leq C N^{1/2- \beta}$.

From
\begin{align*}
\frac{1}{K_\beta} \frac{4 \pi}{N+ \ln \left(\frac{R_\beta}{a}\right)}
=&
\frac{1}{K_\beta}
\int_{\mathbb{R}^2}d^2x V_N(x) j_{N,R_\beta}(x)
=
\int_{\mathbb{R}^2}d^2x V_N(x) f_\beta(x)
\\
=&
\int_{\mathbb{R}^2}d^2x M_\beta(x) f_\beta(x)
= 
8 \pi^2 N^{-1+2 \beta} 
\int_{N^{-\beta}}^{R_\beta} 
dr r f_\beta(r)
\end{align*}
we conclude that
\begin{align*}
\int_{N^{-\beta}}^{R_\beta} 
dr r f_\beta(r)
=
\frac{N^{1- 2 \beta}}{2 \pi K_\beta
\left(
N+ \ln \left(\frac{R_\beta}{a} \right)
\right)
} 
\;.
\end{align*}
Since $f_\beta$ is a nonegative, monotone nondecreasing function in $|x|$, 
\begin{align*}
\frac{1}{2}( R_\beta^2- N^{-2 \beta})
\frac{j_{N,R_\beta}(N^{-\beta})}{K_\beta}
=
\frac{1}{2}( R_\beta^2- N^{-2 \beta})
f_\beta(N^{-\beta})
\leq
\int_{N^{-\beta}}^{R_\beta} 
dr r f_\beta(r)
\end{align*}
which implies
\begin{align*}
R_\beta^2 N^{2 \beta}
\leq
\frac{N}{\pi 
 \left(N+ \ln \left(\frac{R_\beta}{a} \right) \right)
j_{N,R_\beta}(N^{-\beta})}+1
\end{align*}
Using $R_\beta\leq C N^{1/2- \beta}$, it then follows
\begin{align*}
 j_{N,R_\beta}(N^{-\beta})
 =
 1+   \frac{1}{N+ \ln \left(\frac{R_\beta}{a}\right)} \ln \left( \frac{N^{-\beta}}{R_\beta} \right)
 \geq 1- \frac{C}{N}
 \;,
\end{align*}
which implies $R_\beta \leq C N^{-\beta}$.

\item[(h)]
Using
\begin{align*}
\| M_\beta f_\beta\|_1=&
\| V_N f_\beta \|_1 = K_\beta^ {-1} \|V_N j_{N,R_\beta} \|_1
=
K_\beta^{-1}  \frac{4 \pi}{N+ \ln \left(\frac{R_\beta}{a}\right)}
\;,
\end{align*}
we obtain
\begin{align*}
|N\| V_N f_\beta\|_1- 4\pi|
=&
|N\| M_\beta f_\beta\|_1- 4\pi|
=
4 \pi
\left| K_\beta^{-1}  \frac{N}{N+ \ln \left(\frac{R_\beta}{a}\right)}-1 \right|
\\
=&
\frac{4 \pi}{K_\beta}
\left| \frac{
N-NK_\beta+K_\beta
\ln \left(\frac{R_\beta}{a}\right)}{N+ \ln \left(\frac{R_\beta}{a}\right)} \right|
\leq C \frac{\ln(N)}{N}
\;.
\end{align*}

\item[(i)]
Using for $|x| \leq R_\beta$ the inequalities
$j_{N,R_\beta}(x) \geq 1+
 \frac{1}{N+ \ln \left(\frac{R_\beta}{a}\right)}
 \ln \left( \frac{|x|}{R_\beta} \right) $ 
as well as $1\ge f_{\beta}(x) \geq j_{N,R_\beta}(x)$,  it follows for $|x| \leq R_\beta$ 
\begin{align*}
0\leq& g_{\beta}(x) = 1- f_{\beta}(x) \le 1- j_{N,R_\beta}(x)
\leq -
 \frac{1}{N+ \ln \left(\frac{R_\beta}{a}\right)}
 \ln \left( \frac{|x|}{R_\beta} \right) 
 \nonumber
\\
\label{gbound} 
 \leq&
 C N^ {-1}
  | 
  \ln \left(N |x| \right) 
  |
 \;.
\end{align*}

Since $g_{\beta}(x)=0$ for $\vert x\vert > R_{\beta}$, we conclude with $R_\beta \leq C N^{-\beta}$
that
\begin{align*}
\|g_{\beta}\|_1 
\leq &  
 \frac{C}{N}
 \int_{0}^{R_{\beta}}dr r 
| 
 \ln \left( N r\right)
 |
 \leq  C N^{-1-2\beta} \ln N \;,
\end{align*}
as well as
\begin{align*}
\| g_{\beta}\|^2\leq&
\frac{C}{N^2}
 \int_{0}^{R_{\beta}}  dr r
\left( \ln \left( Nr \right) \right)^2   \\
= &  CN^{-4}
 \Big[  r ^2 ( 2 (\ln(r))^2  -  2\ln (r)+1)\Big]_0^{N R_{\beta}}\\
\le &  C N^{-2-2\beta} \left(\ln (N)\right) ^2.
\end{align*}
$ \| g_{\beta}\|_\infty = \|1- f_{\beta}\|_\infty \leq 1$, since $f_\beta$ is a nonnegative, monotone nondecreasing function with $f_\beta(x) \leq 1$.

\item[(j)] Using (h) and (i), we obtain with $\| M_{\beta}\|_1 \leq CN^{-1}$
\begin{align*}
&| N \| M_{\beta}\|_1 - 4 \pi | 
\leq
| N \| M_{\beta}f_\beta\|_1 - 4 \pi | 
+
 N \| M_{\beta}g_\beta\|_1 
 \\
& \leq
 C
\left( 
  \frac{\ln(N)}{N}  +
   \|\mathds{1}_{|\cdot| \geq N^{-\beta}} g_\beta\|_\infty
  \right)
  \;.
\end{align*}
Since $g_\beta(x)$ is a nonnegative, monotone nonincreasing function, it follows with $K_\beta \leq 1$
\begin{align*}
 &  \|\mathds{1}_{|\cdot| \geq N^{-\beta}} g_\beta\|_\infty
   =
   g_\beta(N^{-\beta})=1-f_\beta(N^{-\beta})
   =
   1- \frac{j_{N,R_\beta}(N^{-\beta})}{K_\beta}
   \\
   \leq&
   1-\left( 1 +  \frac{1}{N+ \ln \left(\frac{R_\beta}{a}\right)} \ln \left( \frac{N^{- \beta}}{R_ \beta} \right)  \right)
  \;.
\end{align*}
and (j) follows.
\item[(k)]
$M_\beta \in \mathcal{V}_\beta$ follows directly from $R_\beta \leq C N^{-\beta}$.
Furthermore, $ 0 \leq M_\beta (x) f_\beta (x)  \leq M_\beta(x)$ implies
$M_\beta f_\beta \in \mathcal{V}_\beta$.
\end{enumerate}
\end{proof}

\section{Proof of the Theorem}\label{secpro}

\subsection{Proof for the potential $W_\beta$}\label{secthe}

\subsubsection{Choosing the weight}
As we have already mentioned, we define a functional $\alpha:\LZN\times\LZ\to\mathbb{R}^+_0$ such that
\begin{enumerate}
 \item[(I)]
$\dt\alpha(\Psi_t,\phi_t)$ can be estimated by $\alpha(\Psi_t,\phi_t)+\landau(1)$,
yielding to a  bound of $\alpha(\Psi_t,\phi_t)$ via a Gr\"onwall estimate. 
\item[(II)] $\alpha(\Psi,\phi)\to0$
implies convergence of the reduced one particle density matrix $ \gamma^{(1)}_\psi$ to
$|\phi\rangle\langle\phi|$ in trace norm.
\end{enumerate}
For $\beta>0$, the interaction gets peaked as $N\to\infty$ and one has to use smoothness properties of $\Psi_t$ to be able to control the dynamics of the condensate. 
For small $\beta$ and  many different choices of the weight, one obtains
\begin{align*}
&\alpha(\Psi_t,\phi_t)\leq 
\alpha(\Psi_0,\phi_0)
\\
+&
\int_0^ t ds
\left(
\mathcal{K}(\phi_s, A_s)\left(\alpha(\Psi_s,\phi_s)+\landau(1)+\laa\Psi_s,\widehat{n}^{\phi_s}\Psi_s\raa+
\left|\mathcal{E}_{W_\beta}(\Psi_s)-\mathcal{E}_{N \|W_\beta\|_1}^{GP}(\phi_s)\right|\right)
\right)
\;.
\end{align*}

 This enables us to perform an integral type Gr\"onwall estimate if we choose
$$\alpha(\Psi_t,\phi_t)=\laa\Psi_t,\widehat{n}^{\phi_t}\Psi_t\raa+\left|\mathcal{E}_{W_\beta}(\Psi_t)-\mathcal{E}_{N \|W_\beta\|_1}^{GP}(\phi_t)\right|\;.$$

For large $\beta$, however, it is necessary to adjust the weight function for the following reason:
Taking the time derivative of $\laa\Psi_t,\widehat{n}^{\phi_t}\Psi_t\raa$, terms of the form $\widehat{n}-\widehat{n}_1$ and $\widehat{n}-\widehat{n}_2$ appear.
The bound
$N\|\widehat{n}-\widehat{n}_i\|_{\text{op}}=\mathcal{O}(N^{1/2})\; , i=1,2$ can then be easily verified.
For $\beta >1/2$ it is not possible to obtain a sufficient decay in $N$, see
Lemma \ref{hnorms}, part (b).
For this reason, it is necessary to choose another weight function $\widehat{m}$ in such a way that 
$N\|\widehat{m}-\widehat{m}_i\|_{\text{op}}$ is better to control.

\begin{definition}\label{defm}
For $0<\xi <\frac{1}{2}$ define
$$m(k)=\left\{
         \begin{array}{ll}
           \sqrt{k/N}, & \hbox{for $k\geq N^{1-2\xi}$;} \\
           1/2(N^{-1+\xi}k+N^{-\xi}), & \hbox{else.}
         \end{array}
       \right.
$$
and
$$\alpha^<(\Psi,\phi)=\laa\Psi,\widehat{m}^\phi\Psi\raa+\left|\mathcal{E}_{W_\beta}(\Psi)-\mathcal{E}_{N \|W_\beta\|_1}^{GP}(\phi)\right|\;.$$
\end{definition}
With this definition, we obtain
$N \|\widehat{m}-\widehat{m}_1\|_{\text{op}} \leq C N^{\xi}$, see \eqref{estab}.

\begin{lemma}\label{equiv}
Let $\Psi \in L^2_s(\mathbb{R}^{2N},\mathbb{C})$ and let $\phi \in L^2(\mathbb{R}^2,\mathbb{C})$.
Let $\alpha^<(\Psi,\phi)$ be defined as above. Then,
\begin{align*}\lim_{N\to\infty}\alpha^<(\Psi,\phi)=0 \;\;\Leftrightarrow \hspace{1cm}&\lim_{N\to\infty}\gamma^{(1)}_{\Psi}=|\phi\rangle\langle\phi|\text{ in trace norm}\\\text{ and }&\lim_{N\to\infty}(\mathcal{E}_{W_\beta}(\Psi)-\mathcal{E}_{N \|W_\beta\|_1}^{GP}(\phi))=0\;.
\end{align*}
\end{lemma}
A proof of this Lemma can be found in \cite{picklgp3d}.
Thus, $\alpha(\Psi_t,\phi_t)$ satisfies condition (II). 
To obtain the desired  Gr\"onwall estimate, we will calculate
$\dt \laa\Psi,\widehat{m}^\phi\Psi\raa$ and $ \dt (
\mathcal{E}_{W_\beta}(\Psi_t)-\mathcal{E}_{N \|W_\beta\|_1}^{GP}(\phi_t)
)$.

For this, define
\begin{definition}\label{alphasplit}
Let $W_\beta \in \mathcal{V}_\beta$. Define
\begin{equation}\label{defz}\potdiff^\phi_\beta(x_j,x_k)=W_{\beta}(x_j-x_k)-\frac{N\|W_{\beta}\|_1}{N-1}|\phi|^2(x_j)-\frac{N\|W_{\beta}\|_1}{N-1}|\phi|^2(x_k)
\;.
\end{equation}
Note, for $W_\beta(x)= N^{-1+2 \beta} W(N^\beta x)$, we have
$N \|W_\beta\|_1= \|W\|_1$.
With
$$
    m^a(k)=m(k)-m(k+1) , \;\;\;\;\; m^b(k)=m(k)-m(k+2)
$$
and
$$\widehat{r}=\widehat{m}^bp_{1}p_{2}+\widehat{m}^a(p_{1}q_{2}+q_1p_2) \;,$$
we define the functionals $\gamma^<_{a,b}:\LZN \times L^2(\mathbb{R}^2, \mathbb{C})
\to\mathbb{R}^+_0$ by
\begin{align}\label{fnochdao} \asaalt(\Psi,\phi)%
=&\laa\Psi,\dot A_t(x_1)\Psi\raa-\langle\phi,\dot A_t\phi\rangle\\
\label{fnochda}
\asalt(\Psi,\phi)%
=&N(N-1)\Im\left(\laa\Psi ,\potdiff^\phi_\beta(x_1,x_2) \widehat{r}
\Psi\raa\right) 
\\
\label{asaaltsplit}
=-&2N(N-1)\Im\left(\laa\Psi ,p_{1}q_{2}\widehat{m}^a_{-1}\potdiff^\phi_\beta(x_1,x_2)p_{1}p_{2}
\Psi\raa\right)
\\\nonumber
-&N(N-1)\Im\left(\laa\Psi ,q_{1}q_{2}\widehat{m}^b_{-2}W_\beta(x_1-x_2) p_{1}p_{2}
\Psi\raa\right)
\\\nonumber
-&2N(N-1)\Im\left(\laa\Psi ,q_{1}q_{2}\widehat{m}^a_{-1}\potdiff^\phi_\beta(x_1,x_2)p_{1}q_{2}
\Psi\raa\right)\;.
\end{align}
\end{definition}
\begin{lemma}\label{ableitung}

Let $W_\beta \in \mathcal{V}_\beta$.
Let $\Psi_t$ the unique solution to $i \partial_t \Psi_t
= H_{W_\beta} \Psi_t$ with initial datum $\Psi_0 \in L^2_{s}(\mathbb{R}^{2N}, \mathbb{C}) \cap H^2(\mathbb{R}^{2N}, \mathbb{C}) ,\; 
\|\Psi_0\|=1$. 
Let $\phi_t$ the unique solution to $i \partial_t \phi_t
= h^{GP}_{N \|W_\beta\|_1} \phi_t$ with $\phi_t \in H^3(\mathbb{R}^2,\mathbb{C}) \;,\|\phi_0\|=1$.
Let $\alpha^<(\Psi_t,\phi_t)$ be defined as in Definition \ref{defm}. Then
\begin{align}
\label{lemmaableitungeq}
 \alpha^<(\Psi_t,\phi_t)\leq
  \alpha^<(\Psi_0,\phi_0)
 +
 \int_0^ t ds
 \left(
\left| \gamma_a^<(\Psi_s,\phi_s)
\right|
+
\left|
\gamma_b^<(\Psi_s,\phi_s)
\right|
\right)
\;.
\end{align}

\end{lemma}
\begin{proof} For the proof of the Lemma we restore the upper index $\phi_t$ in order to pay respect to the time dependence of
$\widehat{m}^{\phi_t}$. 
The time derivative of $\phi_t$ is given by (\ref{GP}), i.e. $i\partial_t \phi_t(x_j)=h_{N \|W_\beta\|_1,j}^{GP}\phi_t(x_j)$. Here, $h_{N \|W_\beta\|_1,j}^{GP}$ denotes the  operator $h_{N \|W_\beta\|_1}^{GP}$ acting on the $j^{\text{th}}$ coordinate $x_j$.
We then obtain
\begin{align*}&
\frac{d}{dt}\laa\Psi_t,\widehat{m}^{\phi_t}\Psi_t\raa
\\=&
i\laa H_{W_\beta}\Psi_t
,\widehat{m}^{\phi_t}\;\Psi_t\raa
-
i\laa\Psi_t
,\widehat{m}^{\phi_t}\;H_{W_\beta}\Psi_t\raa
-
i\laa \Psi_t
,[\sum_{j=1}^N h_{N \|W_\beta\|_1,j}^{GP},\widehat{m}^{\phi_t}\;]\Psi_t\raa
\nonumber
\\=&i\laa\Psi_t
,[H_{W_\beta}-\sum_{j=1}^N h_{N \|W_\beta\|_1,j}^{GP},\widehat{m}^{\phi_t}\;]\Psi_t\raa=i\frac{N(N-1)}{2}\laa\Psi_t
,[\potdiff^{\phi_t}_\beta(x_1,x_2),\widehat{m}^{\phi_t}\;]\Psi_t\raa
\;,
\end{align*}
where we used symmetry of $\Psi_t$ in the last step.
Using Lemma \ref{kombinatorik} (d), it follows that the latter equals (dropping the explicit dependence on $\phi_t$ from now on)
\begin{align*}&i\frac{N(N-1)}{2}\laa\Psi_t
,[\potdiff^{\phi_t}_\beta(x_1,x_2),p_1p_2(\widehat{m}-\widehat{m}_2)]\Psi_t\raa
\\
+&i\frac{N(N-1)}{2}\laa\Psi_t
,[\potdiff^{\phi_t}_\beta(x_1,x_2),(p_1q_2+q_1p_2)(\widehat{m}-\widehat{m}_1)]\Psi_t\raa\;.
\end{align*}
Since $\potdiff^{\phi_t}_\beta$ and $p_1p_2(\widehat{m}-\widehat{m}_2)$ as well as $p_1q_2(\widehat{m}-\widehat{m}_1)$ are selfadjoint, we obtain
\begin{align*}
&\frac{d}{dt}\laa\Psi_t,\widehat{m}^{\phi_t}\Psi_t\raa
=-N(N-1)\\&\Im\left(\laa\Psi_t ,(p_{1}p_{2}+p_{1}q_{2}+q_{1}p_{2}+q_{1}q_{2})\potdiff^{\phi_t}_\beta(x_1,x_2)( \widehat{m}^bp_{1}p_{2}+\widehat{m}^a(p_{1}q_{2}+q_1p_2))
\Psi_t\raa\right)\;.
\end{align*}
Note that in view of Lemma \ref{kombinatorik} (c) $\widehat{r}Q_j\potdiff^{\phi_t}_\beta(x_1,x_2) Q_j= Q_j\potdiff^{\phi_t}_\beta(x_1,x_2) Q_j\widehat{r}$ for any $j\in\{0,1,2\}$ and any weight $r$. Therefore, 
\begin{align*}
\Im\left(\laa\Psi_t ,p_{1}p_{2}\potdiff^{\phi_t}_\beta(x_1,x_2) \widehat{m}^bp_{1}p_{2}
\Psi_t\raa\right)&=0\\
\Im\left(\laa\Psi_t ,(p_{1}q_{2}+q_{1}p_{2})\potdiff^{\phi_t}_\beta(x_1,x_2)\widehat{m}^a(p_{1}q_{2}+q_1p_2)
\Psi_t\raa\right)&=0\;.
\end{align*}
Using Symmetry and Lemma \ref{kombinatorik} (c), we obtain the first line  \eqref{fnochda}.
Furthermore,
\begin{align*}
\frac{d}{dt}\laa\Psi_t,\widehat{m}^{\phi_t}\Psi_t\raa%
=&-2N(N-1)\Im\left(\laa\Psi_t , \widehat{m}^b_{-1}p_{1}q_{2}\potdiff^{\phi_t}_\beta(x_1,x_2)p_{1}p_{2}
\Psi_t\raa\right)
\\
-&N(N-1)\Im\left(\laa\Psi_t ,\widehat{m}^b_{-2}q_{1}q_{2}\potdiff^{\phi_t}_\beta(x_1,x_2)p_{1}p_{2}
\Psi_t\raa\right)
\\
-&2N(N-1)\Im\left(\laa\Psi_t ,p_{1}p_{2}\potdiff^{\phi_t}_\beta(x_1,x_2) \widehat{m}^ap_{1}q_{2}
\Psi_t\raa\right)
\\
-&2N(N-1)\Im\left(\laa\Psi_t ,\widehat{m}^a_{-1}q_{1}q_{2}\potdiff^{\phi_t}_\beta(x_1,x_2)p_{1}q_{2}
\Psi_t\raa\right)\;.
\end{align*}
Since $p_1p_2|\phi_t^2|(x_1)q_1q_2=p_1p_2q_2|\phi_t^2|(x_1)q_1=0=p_1p_2|\phi_t^2|(x_2)q_1q_2$, we can replace $\potdiff^{\phi_t}_\beta(x_1,x_2)$ in the second line by $W_\beta(x_1-x_2)$.
The third line equals $2N(N-1)\Im\left(\laa\Psi ,\widehat{m}^a p_{1}q_{2}\potdiff^{\phi_t}_\beta(x_1,x_2) p_{1}p_{2}
\Psi\raa\right)$. Since
$$m(k-1)-m(k+1)-\left(m(k)-m(k+1)\right)=m(k-1)-m(k)$$
it follows that $\widehat{m}^b_{-1}-\widehat{m}^a=\widehat{m}^a_{-1}$ and we get
\begin{align*}
\frac{d}{dt}\laa\Psi_t,\widehat{m}^{\phi_t}\Psi_t\raa%
=&-2N(N-1)\Im\left(\laa\Psi ,p_{1}q_{2}\widehat{m}^a_{-1}\potdiff^{\phi_t}_\beta(x_1,x_2)p_{1}p_{2}
\Psi\raa\right)
\\\nonumber
-&N(N-1)\Im\left(\laa\Psi ,q_{1}q_{2}\widehat{m}^b_{-2}W_\beta(x_1-x_2) p_{1}p_{2}
\Psi\raa\right)
\\\nonumber
-&2N(N-1)\Im\left(\laa\Psi ,q_{1}q_{2}\widehat{m}^a_{-1}\potdiff^{\phi_t}_\beta(x_1,x_2)p_{1}q_{2}
\Psi\raa\right)\;.
\end{align*}

For the second summand of $\alpha^<(\Psi_t,\phi_t)$ we have
\begin{align*}&
\nonumber\frac{d}{dt}\left(\mathcal{E}_{W_\beta}(\Psi_t)-\mathcal{E}_{N \|W_\beta\|_1}^{GP}(\phi_t)\right)
=\laa\Psi_t,\dot A_t(x_1)\Psi_t\raa-\langle\phi_t,\dot A_t\phi_t\rangle
\\\nonumber+& i
\left\langle\phi_t,\left[h_{N \|W_\beta\|_1}^{GP},\left(h_{N \|W_\beta\|_1}^{GP}-\frac{N \|W_\beta\|_1}{2}|\phi_t|^2\right)\right]\phi_t\right\rangle
+
\left\langle\phi_t,\frac{N \|W_\beta\|_1}{2}\left(\frac{d}{dt}|\phi_t|^2\right)\phi_t\right\rangle
\\\nonumber=&\laa\Psi_t,\dot A_t(x_1)\Psi_t\raa-\langle\phi_t,\dot A_t\phi_t\rangle
+i
\left\langle\phi_t,\left[h_{N \|W_\beta\|_1}^{GP},\frac{N \|W_\beta\|_1}{2}|\phi_t|^2\right]\phi_t\right\rangle
\\
-&
i
\left\langle\phi_t,\left[h_{N \|W_\beta\|_1}^{GP},\frac{N \|W_\beta\|_1}{2}|\phi_t|^2\right]\phi_t\right\rangle
=
\gamma_a^<(\Psi_t,\phi_t)
\;.
\end{align*}
The Lemma then follows using that 
$
|f(x)| \leq |f(0)|+ \int_0^x dy |f'(y)|
$ holds for any $f \in C^1 (\mathbb{R}, \mathbb{C})$.
\end{proof}

\subsubsection{Establishing the Gr\"onwall estimate}\label{secest}
\begin{lemma} \label{gammalemma}
Let $W_\beta \in \mathcal{V}_\beta$.
Let $\Psi_t$ the unique solution to $i \partial_t \Psi_t
= H_{W_\beta} \Psi_t$ with initial datum 
$\Psi_0 \in L^2_{s}(\mathbb{R}^{2N}, \mathbb{C}) \cap H^2(\mathbb{R}^{2N}, \mathbb{C}) \;, \|\Psi_0\|=1$.  
Let $\phi_t$ the unique solution to $i \partial_t \phi_t
= h^{GP}_{N \|W_\beta\|_1} \phi_t$ with $\phi_t \in H^{3}(\mathbb{R}^2,\mathbb{C})$.
Let $\mathcal{E}_{W_\beta}(\Psi_0) \leq C$.
Let $\gamma^<_{a}(\Psi_t,\phi_t)$ and $\gamma^<_{b}(\Psi_t,\phi_t)$ be defined as in Definition \eqref{alphasplit}.
Then, there exists an $\eta>0$ such that
\begin{align} 
\label{abc}
\gamma_a^<(\Psi_t,\phi_t) \leq &
C \|\dot A_t\|_\infty ( \laa\Psi_t,\widehat{n}^{\phi_t}\Psi_t\raa + N^{-\frac{1}{2}} ) 
\\
\label{theoremeq1}
\gamma_b^<(\Psi_t,\phi_t) \leq &
\mathcal{K}(\phi_t, A_t)
\left(
 \laa\Psi_t,\widehat{n}^{\phi_t}\Psi_t\raa + N^{-\eta}+
\left|
\mathcal{E}_{W_\beta}(\Psi_t)
-
\mathcal{E}_{N \|W_\beta\|_1}^ {GP}(\phi_t)
\right|
 \right)
\end{align}
\end{lemma}
The proof of this Lemma can be found in Section \ref{contgamma1}.
Note that
$$| \laa\Psi_t,\widehat{n}^{\phi_t}\Psi_t\raa- \laa\Psi_t,\widehat{m}^{\phi_t}\Psi_t\raa|
\leq
\|\widehat{n}^{\phi_t}-\widehat{m}^{\phi_t}\|_{\text{op}}
=
N^{-\xi}
$$
Once we have proven Lemma \ref{gammalemma}, we obtain with Lemma \ref{ableitung}, Gr\"onwall's Lemma and the estimate above that
\begin{align*}
&\alpha^<(\Psi_t,\phi_t)\leq  
e^{
\int_0^t ds \mathcal{K}(\phi_s, A_s)}
\Big(
\alpha^<(\Psi_0,\phi_0)
\\
&+
\int_0^t ds
\mathcal{K}(\phi_s, A_s)
e^{-\int_0^s d \tau \mathcal{K}(\phi_\tau, A_\tau)}
 N^{-\eta}
 \Big) \;.
\end{align*}
Note that under the assumptions $\phi_t \in H^{3}(\mathbb{R}^2,\mathbb{C})$ and $
A_t \in L^\infty(\mathbb{R}^2, \mathbb{C}),\; \dot{A}_t \in L^\infty(\mathbb{R}^2, \mathbb{C})$
there exists a constant $C_t < \infty$, depending on $t$, $\phi_0$ and $A_t$, such that
 $\int_0^t ds
\mathcal{K}(\phi_s, A_s) \leq C_t$, see Lemma \ref{regularityLemma}.
This proves, using Lemma \ref{equiv},  part (a) of  Theorem \ref{theo}. 
If the potential is switched off, one expects that $C_t$ is of order $t$ since in this case $\|\phi_t\|_\infty$
and $\|\nabla \phi_t\|_\infty$ are expected to decay like $t^{-1}$.

We want to explain on a heuristic level why $\gamma^<_{b}(\Psi_t,\phi_t)$ is small.
The principle argument follows the ideas and estimates of \cite{picklgp3d}.
The first line in (\ref{asaaltsplit}) is the most important one. This expression is only small if the correct coupling parameter $N \|W_\beta\|_1$  is used in  the mean-field equation \eqref{GP}. Then,
$$N p_1 W_\beta(x_1-x_2)p_1=N p_1W_\beta\star|\phi|^2(x_2)p_1
\rightarrow p_1 |\phi|^2(x_2) \|W\|_1 p_1 $$
 converges against the mean-field potential, and hence the first expression of (\ref{asaaltsplit}) is small. 

In order to estimate the second and third line of (\ref{asaaltsplit}), one  tries to bound \\$N^2\laa\Psi ,q_{1}q_{2}\widehat{m}^b_{-2}W_\beta(x_1-x_2) p_{1}p_{2}
\Psi\raa$ and $N^2\laa\Psi ,q_{1}q_{2}\widehat{m}^a_{-1}Z^{\varphi}_\beta(x_1-x_2) p_{1}q_{2}
\Psi\raa$ in terms of $\laa \Psi, \widehat{n} \Psi \raa + \mathcal{O}(N^{-\eta})$ for some $\eta>0$.  
For large $\beta$, one needs  to use additional smoothness properties of $\Psi_t$. This explains the appearance of $\left|
\mathcal{E}_{W_\beta}(\Psi_t)
-
\mathcal{E}_{N \|W_\beta\|_1}^ {GP}(\phi_t)
\right|$ on the right hand side of  \eqref{theoremeq1}. The concise estimates are quite involved and can be found in Section \ref{contgamma1}.

\subsection{Proof for the exponential scaling $V_N$}

\subsubsection{Adapting the weight}
For the most involved scaling $V_N$ it is necessary to modify the counting functional $\alpha^ <(\Psi, \varphi)$ in order to obtain the desired Gr\"onwall estimate. $\gamma^<_b (\Psi, \phi)$, which was defined in \eqref{asaaltsplit}, will not be small if we were to replace $W_\beta$ by $V_N$. 
 In particular, $ \|V_N\|=\mathcal{O}(e^{N})$ cannot be bounded by any finite polynomial in $1/N$.
In order to control the dynamics of the condensate, one needs to account for the
microscopic structure which is induced by  $V_N$, as explained in Section \ref{secmic}.
The idea we will employ is the following: 
For the moment, think of the most simple counting functional, namely
 $\laa\Psi_t,q^{\phi_t}_1\Psi_t\raa=1-\laa\Psi_t,p^{\phi_t}_1\Psi_t\raa$. This functional counts the relative number of particles which are not in the state $\phi_t$. Instead of projecting onto $\phi_t$, we now consider the functional
 $$1-\laa\Psi_t,\prod_{j=2}^Nf_{\beta}(x_1-x_j)p^{\phi_t}_1\prod_{j=2}^Nf_{\beta}(x_1-x_j)\Psi_t\raa\;,$$ which takes the short scale correlation structure into account.
 Neglecting all but two-particle interactions, this can be approximated by
\begin{align*}&1-\laa\Psi_t,\left(1-\sum_{j=2}^Ng_{\beta}(x_1-x_j)\right)p^{\phi_t}_1\left(1-\sum_{j=2}^Ng_{\beta}(x_1-x_j)\right)\Psi_t\raa
\\& \approx\laa\Psi_t,q^{\phi_t}_1\Psi_t\raa+2(N-1)\Re\left(\laa\Psi_t
, g_{\beta}(x_{1}-x_{2}) p^{\phi_t}_1
\Psi_t\raa\right)\;.
\end{align*}
 If we now take the time derivative of this new functional, one gets, among other terms,
 $ 2
 (N-1)
 \Im
 \laa \Psi_t,
 [H_{V_N},f_\beta(x_1-x_2) ]p^{\phi_t}_1
 \Psi_t \raa$. The commutator equals $f_{\beta}(x_1-x_2)(V_{N}(x_1-x_2)-M_{\beta}(x_1-x_2)$ plus mixed derivatives and one
sees, that $V_N$ is ``replaced''
by $M_{\beta}$ for the price of new terms that have to be estimated. 
The strategy we are going to employ is thus to estimate the time derivative of the modified functional and to show that we obtain a Gr\"onwall estimate.
Note, that, using Lemma \ref{kombinatorik} (e) with Lemma \ref{defAlemma} (i)
\begin{equation*}
2(N-1)|\Re\left(\llaa\Psi_t
, g_{\beta}(x_{1}-x_{2})  p^{\phi_t}_1
\Psi_t\rraa\right)|
\leq CN
\|\phi_t\|_\infty
\|g_\beta\|
\leq C
\|\phi_t\|_\infty
N^{-\beta} \ln(N)
\end{equation*} 
holds.
Hence, we obtain the a priori estimate
 $$
 \laa\Psi_t,q^{\phi_t}_1\Psi_t\raa
 \leq
 \laa\Psi_t,q^{\phi_t}_1\Psi_t\raa+2(N-1)\Re\left(\laa\Psi_t
, g_{\beta}(x_{1}-x_{2}) p^{\phi_t}_1
\Psi_t\raa\right) 
+
 C
\|\phi_t\|_\infty
N^{-\beta} \ln(N)
\;.
$$
which explains why the new defined functional implies convergence of the reduced density matrix
 $\gamma^{(1)}_{\Psi_t}$ to $|\phi_t \rangle \langle \phi_t|$ in trace norm.
We now adapt the strategy explained above to modify the counting functional
$\alpha^< (\Psi, \phi)$.
\begin{definition}\label{lambda1}
Let $\widehat{r}=\widehat{m}^b p_{1}p_{2}+\widehat{m}^a(p_{1}q_{2}+q_1p_2)$.

Let the functional $\alpha:\LZN\times\LZ\to\mathbb{R}^+_0$ be defined by 
\begin{align} \label{alphafuerV}
\alpha(\Psi,\phi) =&
\laa \Psi, \widehat{m} \Psi \raa
+
\left|
\mathcal{E}_{V_N} (\Psi)
-
\mathcal{E}^{GP}_{4 \pi} (\phi)
\right|
-N(N-1)\Re\left(\laa\Psi
, g_{\beta}(x_{1}-x_{2}) \widehat{r}
\Psi\raa\right)
 \end{align}
and the functional $\gamma:\LZN\times\LZ\to\mathbb{R} $ be defined by
\begin{align} \label{gamma fuer V}
\gamma(\Psi,\phi)=|\asa (\Psi,\phi)|+|\as (\Psi,\phi)|+|\bs (\Psi,\phi)|+|\cs (\Psi,\phi)|+|\ds (\Psi,\phi)|
+
|\gamma_f (\Psi, \phi)|
\;,
\end{align}
where the different summands are:
\begin{enumerate}
\item[(a)] The change in the energy-difference
$$\asa (\Psi,\phi)=\laa\Psi,\dot A_t(x_1)\Psi\raa-\langle\phi,\dot A_t\phi\rangle\;.$$
\item [(b)] The new interaction term
\begin{align*}\as (\Psi,\phi)=&-N(N-1)\Im\left(\laa\Psi ,
\potdiffneu^\phi(x_1,x_2)\widehat{r}\,
\Psi\raa\right)\\&
-N(N-1)\Im\left(\laa\Psi ,g_{\beta}(x_{1}-x_{2})
\widehat{r}\,\mathcal{Z}^\phi(x_1,x_2) \Psi\raa\right),
\end{align*}
where, using $M_\beta$ from Definition \ref{microscopic},
 \begin{align}
 \label{defzz}
 &\potdiffneu^\phi(x_1,x_2)=\left(M_{\beta}(x_1-x_2) - 4 \pi\frac{|\phi|^2(x_1)+|\phi|^2(x_2)}{N-1}\right)f_{\beta}(x_{1}-x_{2})
\\ 
 &
 \nonumber 
 \mathcal{Z}^\phi(x_1,x_2)=
 V_N(x_1-x_2)- 
 \frac{4  \pi}{N-1} |\phi|^ 2(x_1)
 -
  \frac{4 \pi}{N-1} |\phi|^ 2(x_2)
 \;.
\end{align}

 \item [(c)] The mixed derivative term
\begin{align*} \bs (\Psi,\phi)=&-4N(N-1)\laa\Psi
, (\nabla_1g_{\beta}(x_1-x_2))\nabla_1
\widehat{r}\Psi\raa\;.\end{align*}
\item [(d)] Three particle interactions
\begin{align*}\cs (\Psi,\phi)=&2N(N-1)(N-2)\Im\left(\laa\Psi ,g_{\beta}(x_{1}-x_{2})
\left[V_N(x_1-x_3),
\widehat{r}\right] \Psi\raa\right)
\\-&N(N-1)(N-2)\Im\left(\laa\Psi ,g_{\beta}(x_{1}-x_{2})
\left[4 \pi |\phi|^2(x_3),
\widehat{r}\right] \Psi\raa\right)\;.\end{align*}
\item [(e)] Interaction terms of the correction
\begin{align*}&\ds (\Psi,\phi)=\frac{1}{2}N(N-1)(N-2)(N-3)
\Im\left(\laa\Psi ,
g_{\beta}(x_{1}-x_{2})\left[V_N(x_3-x_4),
\widehat{r}\right] \Psi\raa\right)\;.
\end{align*}
\item[(f)] Correction terms of the mean field
\begin{align*}
\gamma_f(\Psi,\phi)=
-2N (N-2)\Im\left(\laa\Psi,g_{\beta}(x_{1}-x_{2})
\left[4 \pi |\phi|^2(x_1),
\widehat{r}\right] \Psi\raa\right)
\;.
\end{align*}
\end{enumerate}
\end{definition}

\begin{lemma}\label{secondadjlemma}
Let $\Psi_t$ the unique solution to $i \partial_t \Psi_t
= H_{V_N} \Psi_t$ with initial datum $\Psi_0 \in L^2_{s}(\mathbb{R}^{2N}, \mathbb{C}) \cap H^2(\mathbb{R}^{2N}, \mathbb{C}) \;, 
\|\Psi_0\|=1$. 
Let $\phi_t$ the unique solution to $i \partial_t \phi_t
= h^{GP}_{4 \pi} \phi_t$ with $\phi_t \in H^3(\mathbb{R}^2,\mathbb{C}) \;,\|\phi_0\|=1$.
Let $\alpha(\Psi_t,\phi_t)$ and $\gamma(\Psi_t,\phi_t)$ be defined as in \eqref{alphafuerV} and \eqref{gamma fuer V}. Then
$$\alpha(\Psi_t,\phi_t)\leq 
\alpha(\Psi_0,\phi_0)
+
\int_0^t ds
\gamma(\Psi_s,\phi_s) \;.$$
\end{lemma}

\begin{proof}
We first calculate
\begin{align*}
\dt &
\left(
\laa \Psi, \widehat{m} \Psi \raa
-N(N-1)\Re\left(\laa\Psi
, g_{\beta}(x_{1}-x_{2}) \widehat{r}
\Psi\raa\right)
\right)
\\
=&
-N(N-1)\Im\left(\laa\Psi_t ,\mathcal{Z}^{\phi_t}(x_1,x_2) \widehat{r}
\Psi_t\raa\right)
\\&-N(N-1)\Re\left(i\laa\Psi_t , g_{\beta}(x_{1}-x_{2})\left[H_{V_N}-\sum_{i=1}^N h^{GP}_{4 \pi,i},
\widehat{r}\right] \Psi_t\raa\right)
\\&-N(N-1)\Re\left(i\laa\Psi_t ,\left[H_{V_N},
g_{\beta}(x_{1}-x_{2})
\right]\widehat{r}
\Psi_t\raa\right)
\;.
\end{align*}
Using symmetry and $\Re(iz)=-\Im(z)$, we obtain
\begin{align*}
\dt &
\left(
\laa \Psi, \widehat{m} \Psi \raa
-N(N-1)\Re\left(\laa\Psi
, g_{\beta}(x_{1}-x_{2}) \widehat{r}
\Psi\raa\right)
\right)
\\
=&
-N(N-1)\Im\left(\laa\Psi_t ,\mathcal{Z}^{\phi_t}(x_1,x_2) \widehat{r}
\Psi_t\raa\right)
\\&+N(N-1)\Im\left(\laa\Psi_t ,g_{\beta}(x_{1}-x_{2})\left[\mathcal{Z}^{\phi_t}(x_1,x_2),
\widehat{r}\right] \Psi_t\raa\right)
\\&+2N(N-1)(N-2)\Im\left(\laa\Psi _t,g_{\beta}(x_{1}-x_{2})
\left[V_N(x_1-x_3),
\widehat{r}\right] \Psi_t\raa\right)
\\&-N(N-1)(N-2)\Im\left(\laa\Psi_t,g_{\beta}(x_{1}-x_{2})
\left[4 \pi |\phi_t|^2(x_3),
\widehat{r}\right] \Psi_t\raa\right)
\\&+\frac{1}{2}N(N-1)(N-2)(N-3)\Im\left(\laa\Psi_t ,
g_{\beta}(x_{1}-x_{2})\left[V_N(x_3-x_4),
\widehat{r}\right] \Psi_t\raa\right)
\\&+N(N-1)\Im\left(\laa\Psi_t ,\left[H_{V_N},
g_{\beta}(x_{1}-x_{2})
\right]\widehat{r}
\Psi_t\raa\right)\;.
\\
&-
2
N(N-2)\Im\left(\laa\Psi_t,g_{\beta}(x_{1}-x_{2})
\left[4 \pi |\phi_t|^2(x_1),
\widehat{r}\right] \Psi_t\raa\right)
\;.
\end{align*}

The third and fourth lines equal $\gamma_d$ (recall that $\Psi$ is symmetric), the fifth line equals  $\gamma_e$ and the seventh line equals $\gamma_f$.
Using that $(1-g_{\beta}(x_{1}-x_{2}))\mathcal{Z}^\phi(x_1,x_2)=\potdiffneu^\phi(x_1,x_2)+(V_{N}(x_1-x_2)-M_{\beta}(x_1-x_2))f_{\beta}(x_{1}-x_{2})$
we get
\begin{align}\label{commu}
\dt &
\left(
\laa \Psi, \widehat{m} \Psi \raa
-N(N-1)\Re\left(\laa\Psi
, g_{\beta}(x_{1}-x_{2}) \widehat{r}
\Psi\raa\right)
\right)
\nonumber
\\
\leq &\gamma_d(\Psi_t,\phi_t)+\gamma_e(\Psi_t,\phi_t)+\gamma_f(\Psi_t,\phi_t)
\nonumber
\\
&-N(N-1)\Im\left(\laa\Psi_t ,\potdiffneu^{\phi_t}(x_1,x_2) \widehat{r}
\Psi_t\raa\right)
\\\nonumber&-N(N-1)\Im\left(\laa\Psi_t ,
(V_{N}(x_1-x_2)-M_{\bet}(x_1-x_2))f_{\beta}(x_1-x_2)\widehat{r}
\Psi_t\raa\right)
\\\nonumber&-N(N-1)\Im\left(\laa\Psi_t ,g_{\beta}(x_{1}-x_{2})
\widehat{r}\mathcal{Z}^{\phi_t}(x_1,x_2) \Psi_t\raa\right)
\\\nonumber&+N(N-1)\Im\left(\laa\Psi_t ,\left[H_{V_N},
g_{\beta}(x_{1}-x_{2})
\right]\widehat{r}
\Psi_t\raa\right)\;.
\end{align}
The first, second and the fourth line give $\gamma_b+\gamma_d+\gamma_e+ \gamma_f$. Using Definition~(\ref{microscopic}) the commutator in the fifth line equals
\begin{align*}[H_{V_N},g_{\beta}(x_1-x_2)]=&-[H_{V_N},f_{\beta}(x_1-x_2)]\\\nonumber=&[\Delta_1+\Delta_2,f_{\beta}(x_1-x_2)]
\\\nonumber=&(\Delta_1+\Delta_2)f_{\beta}(x_1-x_2)\\\nonumber&+(2\nabla_1f_{\beta}(x_1-x_2))\nabla_1+(2\nabla_2f_{\beta}(x_1-x_2))\nabla_2
\\\nonumber=&(V_{N}(x_1-x_2)-M_{\beta}(x_1-x_2))f_{\beta}(x_1-x_2)\\\nonumber&-(2\nabla_1g_{\beta}(x_1-x_2))\nabla_1-(2\nabla_2g_{\beta}(x_1-x_2))\nabla_2\;.
\end{align*}
Using symmetry the third and fifth line in (\ref{commu}) give
$$-4N(N-1)\laa\Psi_t
, (\nabla_1g_{\beta}(x_1-x_2))\nabla_1
\widehat{r}\Psi_t\raa=\gamma_c (\Psi_t,\phi_t)\;.$$
Using 
$$
\frac{d}{dt}\left(\mathcal{E}_{W_\beta}(\Psi_t)-\mathcal{E}_{N \|W_\beta\|_1}^{GP}(\phi_t)\right)
=
\gamma_a (\Psi_t, \phi_t)
\;,
$$
we obtain the desired result.
\end{proof}

\subsubsection{Establishing the Gr\"onwall estimate}
Again, we will bound the time derivative of $\alpha(\Psi_t,\phi_t)$ such that we can employ a Gr\"onwall estimate.
\begin{lemma} \label{gammalemma fuer V}
Let $\Psi_t$ the unique solution to $i \partial_t \Psi_t
= H_{V_N} \Psi_t$ with initial datum 
$\Psi_0 \in L^2_{s}(\mathbb{R}^{2N}, \mathbb{C}) \cap H^2(\mathbb{R}^{2N}, \mathbb{C}) ,\; \|\Psi_0\|=1$.  
Let $\phi_t$ the unique solution to $i \partial_t \phi_t
= h^{GP}_{4 \pi} \phi_t$ with $\phi_t \in H^{3}(\mathbb{R}^2,\mathbb{C})$.
Let $\mathcal{E}_{V_N}(\Psi_0) \leq C$.
Let $\gamma(\Psi_t,\phi_t)$ be defined as in \eqref{gamma fuer V}.
Then, there exists an $\eta>0$ such that
\begin{align} 
\label{theoremeq1 fuer V}
\gamma(\Psi_t,\phi_t) \leq &
 \mathcal{K}(\phi_t, A_t)
\left(
	 \laa\Psi_t,\widehat{n}\Psi_t\raa + N^{-\eta}+
	\left|
	\mathcal{E}_{V_N(\Psi_0)}
-
\mathcal{E}_{b_{V_N}}^ {GP}(\phi_0)
\right|
 \right)
 \;.
\end{align}
\end{lemma}
A prove of the Lemma can be found in Section \ref{gammacontrolsection}.

 The most important estimate is the first part of $\gamma_b$, which can be estimated in the same way as $\gamma_b^<$.
 All other estimates are based on the smallness of the $L^p$-norms of $g_{\beta}$, see Lemma \ref{defAlemma} (i).
 We now show that Lemma \ref{gammalemma fuer V} implies convergence of the  reduced density matrix
 $\gamma^{(1)}_{\Psi_t}$ to $|\phi_t \rangle \langle \phi_t|$ in trace norm.
 Using $\| \widehat{m}^a\|_{\text{op}}
 +
 \| \widehat{m}^b\|_{\text{op}} \leq C N^{-1+\xi}$, see \eqref{estab}, together with Equation \eqref{kombeqb} and Lemma \ref{defAlemma} (i), we obtain
\begin{align*}
& \|g_{\beta}(x_{1}-x_{2}) \widehat{r}\|_{\text{op}}
 \leq
 \|g_{\beta}(x_{1}-x_{2})p_1( 
 \widehat{m}^b p_2+
 \widehat{m}^a q_2)
 \|_{\text{op}}
 +
 \|g_{\beta}(x_{1}-x_{2}) p_2
q_1 
 \widehat{m}^a\|_{\text{op}}
\\
 \leq&
 \mathcal{K}(\phi, A_t)
  \|g_{\beta} \|
 (
 \| \widehat{m}^a\|_{\text{op}}
 +
 \| \widehat{m}^b\|_{\text{op}}
 )
 \leq 
 \mathcal{K}(\phi, A_t) 
 N^{\xi-2- \beta} \ln(N)
 .
\end{align*}
Therefore, we bound
  
\begin{equation}
\label{zweimal}
N(N-1)|\Re\left(\llaa\Psi
, g_{\beta}(x_{1}-x_{2}) \widehat{r}
\Psi\rraa\right) |
\leq  \mathcal{K}(\phi, A_t)
N^{-\beta+\xi} \ln(N).
\end{equation} 
For $\beta$ large enough, (\ref{theoremeq1 fuer V}) implies together with (\ref{zweimal})  that 
$$\gamma(\Psi_t,\phi_t)\leq   \mathcal{K}(\phi_t, A_t)
\left(\alpha(\Psi_t,\phi_t)
+ N^{-\eta}\right)\;,
$$
for some $\eta>0$.
We get with Lemma \ref{secondadjlemma} and Gr\"onwall's Lemma, 
using (\ref{zweimal}) again,
that
\begin{align*}
&\alpha^<(\Psi_t,\phi_t)\leq  
e^{
\int_0^t ds \mathcal{K}(\phi_s, A_s)}
\Big(
\alpha^<(\Psi_0,\phi_0)
\\
&+
\int_0^t ds
\mathcal{K}(\phi_s, A_s)
e^{-\int_0^s d \tau \mathcal{K}(\phi_\tau, A_\tau)}
 N^{-\eta}
 \Big) \;.
\end{align*}

Therefore, we obtain part (b) of  Theorem \ref{theo}.

\section{Rigorous estimates}
\label{rigestimates}
\subsection{Smearing out the potential $W_\beta$}
In Section \ref{secmic} we have defined  the potential $M_\beta$ to control the strongly peaked potential $V_N$. 
We will employ a similar strategy to ''smear out'' the potential $W_\beta$ when $\beta$ is large. For this, we define, for $\beta_1<\beta$, a potential $U_{\beta_1,\beta} \in \mathcal{V}_{\beta_1}$ such that
$\|W_\beta\|_1=\|U_{\beta_1,\beta} \|_1$. Furthermore, define  $h_{\beta_1,\beta}$ by
$ \Delta h_{\beta_1,\beta}=W_{\beta}-U_{\beta_1,\beta}$. 
The function $h_{\beta_1,\beta}$ can be thought as an electrostatic potential which is caused by the charge $ W_{\beta}-U_{\beta_1,\beta}$.
It is then possible to rewrite 
\begin{align*}
&\laa \chi, W_\beta(x_1-x_2) \Omega \raa =
\laa \chi, U_{\beta_1,\beta}(x_1-x_2)  \Omega \raa
\\
-&
\laa \nabla_1 \chi, (\nabla_1 h_{\beta_1,\beta}) (x_1-x_2) \Omega \raa
-
\laa \chi, (\nabla_1 h_{\beta_1,\beta}) (x_1-x_2) \nabla_1 \Omega \raa
\; ,
\end{align*}
for $\chi, \omega \in L^2_s(\mathbb{R}^{2N},\mathbb{C})$. 
 It is easy to verify that $h_{\beta_1,\beta}$ and $ \nabla h_{\beta_1,\beta}$ are faster decaying than the potential $W_\beta$. 
The right hand side of the equation above is hence better to control, if one has additional control of $\nabla_1 \Omega$ and $\nabla_1 \chi$.

\begin{definition}\label{udef}
For any $0\leq\beta_1< \beta$ and any $W_{\beta}\in\mathcal{V}_\beta$ we define
$$U_{\beta_1,\beta}(x)=\left\{
         \begin{array}{ll}
           \frac{4}{\pi}\|W_{\beta}\|_1N^{2\beta_1} & \hbox{for \ $|x|<1/2 N^{-\beta_1}$,} \\
           0 & \hbox{else.}
         \end{array}
       \right.
$$
and
\begin{align} \label{defh} h_{\beta_1,\beta}(x)=  \frac{1}{2\pi} \int_{\mathbb R^2}   \ln|x-y| 
(W_{\beta}(y)-U_{\beta_1,\beta}(y))d^2y
\; .
\end{align}
\end{definition}
\begin{lemma}\label{ulemma}
For any $0\leq  \beta_1<  \beta$ and any $W_{\beta}\in\mathcal{V}_\beta$, we obtain with the above definition
\begin{enumerate}
\item
\begin{align*}
& U_{\beta_1,\beta} \in \mathcal{V}_{\beta_1}\;,
\\
&\Delta h_{\beta_1,\beta}=W_{\beta}-U_{\beta_1,\beta}
.
\end{align*}
\item Pointwise estimates
\begin{align}\label{extraz}
|h_{\beta_1,\beta}(x)|\leq&  CN^{-1} \ln (N) , \hspace{1cm}
h_{\beta_1,\beta}(x) =0 \text{ for } |x| \geq N^{- \beta_1} ,
\\|\nabla h_{\beta_1,\beta}(x)|\leq&  CN^{-1}\left(\vert x\vert ^2+N^{-2\beta}\right)^{-\frac{1}{2}}.\label{extrazb}
\end{align}
\item Norm estimates
\begin{align*}
\|h_{\beta_1,\beta}\|_\infty & \leq CN^{-1} \ln (N),\\
\hspace{1cm}\|h_{\beta_1,\beta}\|_{\lambda}& \leq CN^{-1-\frac{2}{\lambda}\beta_1} \ln (N)\;\text{ for }1\le \lambda \leq\infty ,
\\\|\nabla h_{\beta_1,\beta}\|_{\lambda}&\leq CN^{-1+\beta-\frac{2}{\lambda}\beta_1} \hspace{0.63cm} \text{ for }1\le \lambda \leq\infty.
 \end{align*}
Furthermore, for $\lambda=2$, we obtain the improved bounds
 \begin{align}
\|h_{0,\beta}\| \leq &C N^{-1} \text{ for } \beta>0
\; ,
\\
 \|\nabla h_{\beta_1,\beta}\|\leq& CN^{-1} (\ln (N))^{1/2}
 \;.
\end{align}
 \end{enumerate}
\end{lemma}
\begin{proof} 
\begin{enumerate}
\item
$U_{\beta_1,\beta} \in \mathcal{V}_{\beta_1} $ follows directly from the definition of $U_{\beta_1,\beta}$.
\\
The second statement is a well known result of standard electrostatics (therefore recall that the radially symmetric Greens function of the Laplace operator in two dimensions is given by $-\frac{1}{2\pi} \ln \vert x-y \vert$). $W_\beta$ can be understood as a
given charge density. $-U_{\beta_1,\beta}$ then corresponds to a smeared out charge density of opposite sign such that the ``total charge'' is zero. Hence, the
``potential'' $h_{\beta,\beta_1}$ can be chosen to be zero outside the support of the total charge density.\footnote{To see this, recall that the solution of $\Delta h(r) = \rho(r)$ for radially symmetric and regular enough charge density $\rho$ is given by
\begin{align*}
h(r) = \ln (r) \int_0^r r'\rho(r') dr' + \int_r^{\infty} \ln (r') \rho(r') r' dr' + C,
\end{align*}
where $C\in \mathbb R$. The r.h.s. is zero for $r \not\in \text{supp}(\rho)$ when the total charge vanishes $\int_0^{\infty} r \rho(r) dr=0$ and $C$ is chosen equal to zero.}

\item  
First note that $\vert h_{\beta_1,\beta}(x) \vert=0$ for $\vert x \vert \ge 1/2 N^{-\beta_1}$, which implies the pointwise estimate
\begin{align*}
\vert h_{\beta_1,\beta}(x)|  \leq &
\frac{1}{2 \pi} \int_{B_{1/2 N^{-\beta_1}(0)}} d^2y
\left|
\ln |x-y| W_\beta (y)
\right|
\\
+&
\frac{1}{2 \pi} 
\int_{B_{1/2 N^{-\beta_1}(0)}} d^2y
\left|
\ln |x-y| U_{\beta_1, \beta} (y)
\right| \;.
\end{align*}
We estimate each term separately.
For $ R N^{-\beta} < \vert x \vert$, we obtain
\begin{align*}
\int_{B_{1/2 N^{-\beta_1}(0)}} d^2y |\ln |x-y| |
W_{\beta}(y) \le C \|W_\beta \|_1   \vert \ln (\vert x \vert - RN^{-\beta} )\vert,
\end{align*}
which in turn implies
\begin{align*}
\int_{B_{1/2 N^{-\beta_1}(0)}} d^2y |\ln |x-y| |
W_{\beta}(y)
 \le C \|W_\beta \|_1 \ln N^{\beta}\le C N^{-1} \ln \left( N \right)
\end{align*}
for all $2R N^{-\beta} \le \vert x \vert$.\\
\\
Let next $|x|\leq 2RN^{-\beta}$.
Note that $|x-y| \leq 1$ in the integral above, using $h_{\beta_1, \beta}(x)=0$, whenever
$|x| > 1/2 \beta_1$. This implies
$|\ln|x-y||= - \ln|x-y|$ in the integral.
Thus,
\begin{align*}
&\int_{B_{N^{-\beta_1}(0)}}
  |\ln |x-y| |
W_{\beta}(y)d^2y \nonumber\\
& \le C \| W_{\beta} \|_{\infty} \int_{ B_{RN^{-\beta}(0)}} - \ln  \vert x-y \vert  d^2y \nonumber\\
& \le C N^{-1 + 2 \beta} \int_{ B_{RN^{-\beta}(x)}} - \ln \vert y\vert  d^2y \nonumber\\
& \le C N^{-1 + 2 \beta} \int_{B_{4RN^{-\beta}(0)}} - \ln  \vert y\vert    d^2y \nonumber\\
& = C N^{-1 + 2 \beta}\Big[- \vert y\vert^2 (2\ln \vert y \vert -1) \Big]_0^{4RN^{-\beta}}  \le C N^{-1} \ln  \left( N^{\beta} \right),\nonumber
\end{align*}
Repeating these estimates for $U_{\beta_1, \beta}$ proves the first statement.\\
For the gradient, we estimate the two terms on the r.h.s. of
\begin{align*}|\nabla h_{\beta_1,\beta}(x)|\leq \frac{1}{2\pi }\int \frac{1}{|x-y|} W_{\beta}(y)d^2y + \frac{1}{2\pi }\int \frac{1}{|x-y|} U_{\beta,\beta_1}(y)d^2y
\end{align*}
separately. Let first $2R N^{-\beta} \le \vert x \vert$. Similarly as in the previous argument, one finds
\begin{align*} 
\int \frac{1}{|x-y|} W_{\beta}(y)d^2y & \leq   \int_{B_{R N^{-\beta}}(0)} \frac{1}{\vert x-y\vert} 
W_{\beta}(y)d^2y  \le \frac{\|W_\beta \|_1}{\vert x \vert - RN^{-\beta}}
\end{align*}
for $R N^{-\beta} \le \vert x \vert$, which implies that
\begin{align*}
\int \frac{1}{|x-y|} W_{\beta}(y)d^2y \le \frac{C \| W_{\beta}\|_1}{( \vert x \vert^2 + N^{-2\beta})^{\frac{1}{2}}} \le  \frac{C N^{-1}}{( \vert x \vert^2 + N^{-2\beta})^{\frac{1}{2}}} 
\end{align*}
for all $2R N^{-\beta} \le \vert x \vert$. 
For $\vert x \vert \le 2R N^{-\beta}$, we make use of
\begin{align*}
N^{\beta} \le \frac{C}{\left( \vert x \vert^2 + N^{- 2 \beta} \right)^{1/2}}
\end{align*}
and estimate
\begin{align*}
\int \frac{1}{|x-y|} W_{\beta}(y) d^2y & \le \| W_{\beta} \|_{\infty} \int_{ B_{RN^{-\beta}(0)}} \frac{1}{\vert x-y \vert } d^2y \\
& \le C N^{2\beta-1}\int_{0}^{RN^{-\beta }} d\vert y\vert = C N^{-1+\beta}
\le  \frac{C N^{-1}}{\left( \vert x \vert^2 + N^{- 2 \beta} \right)^{1/2}}  .
\end{align*}
Equivalently, we obtain
\begin{align*}
\int \frac{1}{|x-y|} U_{\beta_1,\beta}(y) d^2y & \le \| U_{\beta_1, \beta} \|_{\infty} \int_{ B_{N^{-\beta_1}(0)}} \frac{1}{\vert x-y \vert } d^2y \\
& = C N^{-1+\beta_1}
\le  \frac{C N^{-1}}{\left( \vert x \vert^2 + N^{- 2 \beta_1} \right)^{1/2}} 
\le  \frac{C N^{-1}}{( \vert x \vert^2 + N^{-2\beta})^{\frac{1}{2}}} ,
\end{align*}
for $\vert x \vert \leq N^{-\beta_1}$.
Since $\nabla h_{\beta_1,\beta}(x) = 0 $ for $\vert x \vert \geq N^{-\beta_1}$,  the second statement of (b) follows.

\item The first part of (c) follows from (b) and the fact that the support of $h_{\beta1,\beta}$ and $\nabla h_{\beta_1,\beta}$ has radius $\leq CN^{-\beta_1}$.
The bounds on the $L^2$-norm can be improved by
\begin{align*}
 \|\nabla h_{\beta_1,\beta}\|_{2}^2\leq&
 C \int_0^{C N^{-\beta_1}} dr r |\nabla h_{\beta_1,\beta}(r)|^2
 \leq
  \frac{C}{N^2} \int_0^{C N^{-\beta_1}} dr \frac{r}{r^2+N^{-2\beta } } 
  \\
  =&
  \frac{C}{N^2} \ln \left(\frac{N^{-2\beta_1 }+N^{-2\beta }}{N^{-2\beta }} \right)
  \leq    \frac{C}{N^ 2} \ln(N)
\end{align*}
Using, for $|x|\geq 2 RN^{-\beta}$, the inequality
\begin{align*}
\vert h_{0,\beta}(x)| &\leq C N^{-1}  \vert \ln (\vert x \vert - RN^{-\beta} )\vert,
\end{align*}
we obtain
\begin{align*}
\|h_{0,\beta}\|_2^2
=&
\int_{\mathbb{R}^2} d^2x
\mathds{1}_{B_{2RN^{-\beta}(0)}}(x)
|h_{0,\beta}(x)|^2
+
\int_{\mathbb{R}^2} d^2x
\mathds{1}_{B^c_{2RN^{-\beta}(0)}}(x)
|h_{0,\beta}(x)|^2
\\
\leq &
\|h_{0\beta}\|_\infty^2
|B_{2RN^{-\beta}(0)}|
+
C N^{-2}
\int_{2 RN^{-\beta}}^1 dr
r
 \vert \ln (r - RN^{-\beta} )\vert^2
 \\
 \leq &
C
\left(
 N^{-2-2\beta} (\ln(N))^2
+
N^{-2} 
\int_{ RN^{-\beta}}^1 dr
(r+ RN^{-\beta} )
( \ln (r))^2
 \right)
 \;.
\end{align*}
Using
\begin{align*}
&\int_{ RN^{-\beta}}^1 dr
(r+ RN^{-\beta} )
( \ln (r))^2
\\
 =&
 \left(
 \frac{1}{4}r^2
  (2 (\ln(r))^2-2\ln(r)+1)
 +
 R N^{-\beta}
r( (\ln(r))^2-2 \ln(r)+2)
\right)\Big|^1_{RN^{-\beta}}
\\
\leq&
C \left(1+ N^{-\beta}+ N^{-2 \beta} (\ln(N))^2 \right) \;,
\end{align*}
we obtain, for any $\beta>0$,
\begin{align*}
\|h_{0,\beta}\|_2^2
\leq
C N^{-2}
\left(
1+
N^{-\beta}
+
N^{-2\beta}
(\ln(N))^2
\right)
\leq
C N^{-2} \;.
\end{align*}

\end{enumerate}

\end{proof}

\subsection{Estimates on the cutoff}
In order to smear out singular potentials  as explained in the previous section and to obtain sufficient bounds, it seems at first necessary to show that $\|\nabla_1 q_1\Psi_t\|$  decays in $N$. However,
 this term will in fact not be small for the dynamic generated by $V_N$. There, we rather expect that $\|\nabla_1q_1\Psi_t\| = \mathcal{O}(1)$ holds.
It has been shown in \cite{cherny} and \cite{lieb100bec}  that the interaction energy is purely kinetic in the Gross-Pitaevskii regime, which 
implies that a relevant part of the kinetic energy is concentrated around the scattering centers. 
We must thus cutoff the part which is used to form the microscopic structure.
For this, we define the set $\overline{\mathcal{A}}^{(d)}_j$ which includes all configurations where the distance between particle $x_i$ and particle $x_j \;,j \neq i$ is smaller than $N^ {-d}$. 
It is then possible to prove that the kinetic energy concentrated on the complement of $\overline{\mathcal{A}}_j^{(d)}$, i.e. $\|\mathds{1}_{\mathcal{A}^{(d)}_1}\nabla_1q_1\Psi\|$, is small, see Lemma \ref{energylemma}.

\begin{definition}
\label{hdetail} 
For any $j,k=1, \dots,N$ and $ d>0$ 
let
 \be
 \label{defkleins}a^{(d)}_{j,k}=\{(x_1,x_2,\ldots,x_N)\in
\mathbb{R}^{2N}: |x_j-x_k|<N^{-d}\}
\subseteq \mathbb{R}^{2N}
\ee
$$\overline{\mathcal{A}}^{(d)}_j=\bigcup_{k\neq j}a^{(d)}_{j,k}\;\;\;\;\;\;\;\mathcal{A}^{(d)}_j=\mathbb{R}^{2N}\backslash \overline{\mathcal{A}}_j^{(d)}
\;\;\;\;\;\;\;\overline{\mathcal{B}}^{(d)}_{j}=\bigcup_{k\neq l\neq j}a^{(d)}_{k,l}\;\;\;\;\;\;\;\mathcal{B}^{(d)}_{j}=\mathbb{R}^{2N}\backslash
\overline{\mathcal{B}}^{(d)}_{j}\;.$$
\end{definition}

\begin{lemma}
\label{propo}
Let $\Psi \in L^2_{s}(\mathbb{R}^{2N}, \mathbb{C}) \cap H^1(\mathbb{R}^{2N}, \mathbb{C})$ $\| \Psi\|=1$ and let $\| \nabla _1\Psi \|$ be uniformly bounded in $N$.
Then, for all $j\neq k$ with $1 \leq j,k \leq N$,
\begin{enumerate}
 \item
\begin{align*}
\|\mathds{1}_{\overline{\mathcal{A}}^{(d)}_{j}}p_j\|_{\text{op}}&\leq  C \| \phi\|_{\infty}N^{1/2-d}\;,
\\
\|\mathds{1}_{\overline{\mathcal{A}}^{(d)}_{j}} \nabla_j p_j\|_{\text{op}}&\leq  C\| \nabla\phi\|_{\infty} N^{1/2-d}\;,
\\
\|\mathds{1}_{a^{(d)}_{j,k}}p_j\|_{\text{op}}&\leq  C \|\phi\|_\infty N^{-d}
\;.
\end{align*}

\item
For any $1<p< \infty$
\begin{align*}\|\mathds{1}_{\overline{\mathcal{A}}^{(d)}_{j}}\Psi\|
\leq &
C
N^{\frac{1-2d}{2} \frac{p-1}{p}}
\;,
\end{align*}
which implies that
$$
\|\mathds{1}_{\overline{\mathcal{A}}^{(d)}_{j}}\Psi\| \leq
C
N^{\frac{1}{2}-d+ \epsilon}
$$
for any $\epsilon>0$.
\item
\begin{align*}
\|\mathds{1}_{\overline{\mathcal{B}}^{(d)}_{j}}\Psi\|\leq &
C
N^{1-d+ \epsilon} 
\end{align*}
for any $\epsilon>0$.
\item For any $k\neq j$
$$\|[\mathds{1}_{\overline{\mathcal{A}}^{(d)}_{j}},p_k]\|_{\text{op}}=\|[\mathds{1}_{a^{(d)}_{j,k}},p_k]\|_{\text{op}}=\|[\mathds{1}_{\mathcal{A}^{(d)}_{j}},p_k]\|_{\text{op}}\leq C \|\phi\|_\infty N^{-d}
\;.
$$

\end{enumerate}\end{lemma}
\begin{proof}
\begin{enumerate}
 \item First note that the volume of the sets $a^{(d)}_{j,k}$ introduced in Definition \ref{hdetail} are
$|a^{(d)}_{j,k}|=\pi N^{-2d}$.
 \begin{align*}
 \|\mathds{1}_{\overline{\mathcal{A}}^{(d)}_{j}}p_j\|_{\text{op}}=&
 \|\mathds{1}_{\overline{\mathcal{A}}^{(d)}_{1}}p_1\|_{\text{op}}=
 \|p_1\mathds{1}_{\overline{\mathcal{A}}^{(d)}_{1}}p_1\|_{\text{op}}^{\frac{1}{2}}
 \leq \left(\|\phi\|_\infty^2\|\mathds{1}_{\overline{\mathcal{A}}^{(d)}_{1}}\|_{1,\infty}\right)^{1/2}
\end{align*}
  where we defined
\begin{align*}
\|f \|_{p, \infty} =
\sup_{x_2, \dots, x_N \in \mathbb{R}^2} 
\left(
\int dx_1 |f(x_1, \dots, x_N)|^{p}
\right)^{\frac{1}{p}}
\;.
 \end{align*}
Using $\mathds{1}_{\overline{\mathcal{A}}^{(d)}_{1}} \leq \sum_{k=2}^N \mathds{1}_{a_{1,k}^{(d)}} $
as well as $\left(\mathds{1}_{\overline{\mathcal{A}}^{(d)}_{1}} \right)^p =\mathds{1}_{\overline{\mathcal{A}}^{(d)}_{1}}$,
  we obtain
\begin{align*}
\|\mathds{1}_{\overline{\mathcal{A}}^{(d)}_{1}} \|_{p, \infty}
\leq&
\sup_{x_2, \dots, x_N \in \mathbb{R}^2} 
\left(
\int dx_1 \sum_{k=2}^N \mathds{1}_{a_{1,k}^{(d)}} 
\right)^{\frac{1}{p}}
\leq
\left(
N |a_{1,k}|
\right)^{\frac{1}{p}}
\leq
C
N^{(1-2d) \frac{1}{p}}
\;.
\end{align*}
This implies
\begin{align*}
 \|\mathds{1}_{\overline{\mathcal{A}}^{(d)}_{j}}p_j\|_{\text{op}}
  \leq&
  C \|\phi\|_\infty N^{\frac{1}{2}-d}
  \;.
\end{align*}
The second statement of (a) can be proven similarly. Analogously, we obtain
 \begin{align*}
 \|\mathds{1}_{a^{(d)}_{j,k}}p_j\|_{\text{op}}\leq&\|\phi\|_\infty  |a^{(d)}_{j,k}|^{1/2} \leq  C \|\phi\|_\infty N^{-d}.
 \end{align*}
 \item
Without loss of generality, we can set $j=1$.
Recall the two-dimensional Sobolev inequality, for $\varrho \in H^1(\mathbb{R}^2,\mathbb{C})$,
$
\| \varrho \|_m \leq
 C \| \nabla \varrho\|^{\frac{m-2}{m}} \| \varrho \|^{\frac{2}{m}}
$ holds for any $2<m<\infty$.
Using H\"older and Sobolev for the $x_1$-integration, we get, for $p>1$
\begin{align*} \|
&\mathds{1}_{\overline{\mathcal{A}}^{(d)}_{1}}\Psi\|^2
=\laa\Psi,\mathds{1}_{\overline{\mathcal{A}}^{(d)}_{1}}\Psi\raa
=
\int d^2 x_2\dots d^2x_N \int d^2x_1 |\Psi(x_1, \dots,x_N)|^2\mathds{1}_{\overline{\mathcal{A}}^{(d)}_{1}}(x_1,\dots,x_N)
\\
\leq &
\|\mathds{1}_{\overline{\mathcal{A}}^{(d)}_{1}} \|_{\frac{p}{p-1}, \infty} 
\int d^2 x_2\dots d^2x_N 
\left(
\int d^2x_1  |\Psi(x_1, \dots,x_N)|^{2p}
\right)^{1/p}
 \\
 \leq &
 C N^{(1-2d)\frac{p-1}{p}}
 \int d^2 x_2\dots d^2x_N 
\left(
\int d^2x_1  |\nabla_1 \Psi(x_1, \dots,x_N)|^{2}
\right)^{\frac{p-1}{p}}
\left(
\int d^2 \tilde{x}_1  | \Psi(\tilde{x}_1, \dots,x_N)|^{2}
\right)^{\frac{1}{p}}
\;.
 \end{align*}

Using  H\"older for the $x_2,\dots x_N$-integration with the conjugate pair $r= \frac{p}{p-1} $ and $s =p$, we obtain
\begin{align*}
 \|
\mathds{1}_{\overline{\mathcal{A}}^{(d)}_{1}}\Psi\|^2
\leq &
C N^{(1-2d)\frac{p-1}{p}}
\| \nabla_1 \Psi \|^{2\frac{p-1}{p}}
\| \Psi \|^{\frac{2}{p}} \;.
 \end{align*}
 Using $\|\nabla_1\Psi \|<C$, (b) follows.

 \item We use that $\overline{\mathcal{B}}^{(d)}_{j}\subset\bigcup_{k=1}\overline{\mathcal{A}}^{(d)}_{k}$.
Hence one can find pairwise disjoint sets $ \mathcal{C}_k\subset\overline{\mathcal{A}}^{(d)}_{k}$, $k=1,\ldots,N$
such that $\overline{\mathcal{B}}^{(d)}_{j}\subset\bigcup_{k=1} \mathcal{C}_{k}$. Since the sets $ \mathcal{C}_k$ are pairwise disjoint,
the $\mathds{1}_{ \mathcal{C}_{k}}\Psi $ are pairwise orthogonal and we get
\begin{align*}
\|\mathds{1}_{\overline{\mathcal{B}}^{(d)}_{j}}\Psi \|^2=\sum_{k=1}\|\mathds{1}_{ \mathcal{C}_{k}}\Psi \|^2
\leq \sum_{k=1}^N\|\mathds{1}_{\overline{\mathcal{A}}^{(d)}_{k}}\Psi \|^2 \;.
\end{align*}
\item
\begin{align*}
\|[\mathds{1}_{\overline{\mathcal{A}}^{(d)}_{1}},p_2]\|_{\text{op}} \leq &\|[\mathds{1}_{a_{1,2}},p_2]\|_{\text{op}}
\leq\|\mathds{1}_{a_{1,2}}p_2\|_{\text{op}}+\|p_2\mathds{1}_{a_{1,2}}\|_{\text{op}}
\\\leq &2\|\phi\|_\infty|a_{1,2}|^{\frac{1}{2}}\leq C \|\phi\|_\infty N^{-d}\;.
\end{align*}
\end{enumerate}
\end{proof}


\subsection{Estimates for the functionals $\gamma_a$, $\gamma_a^<$ and $\gamma_b^<$}\label{contgamma1}

\paragraph{Control of $\asa$ and $\asaalt$}

\begin{lemma}\label{opdiff}
For any multiplication operator $B: L^2(\mathbb{R}^2,\mathbb{C}) \rightarrow L^2(\mathbb{R}^2,\mathbb{C})$ and any $\phi\in L^2(\mathbb{R}^ 2, \mathbb{C}) $ and any $\Psi\in L_s^2(\mathbb{R}^{2N},\mathbb{C})$ we have
$$|\laa\Psi , B(x_1) \Psi\raa-\langle\phi,B\phi\rangle|\leq C \|B\|_\infty( \laa\Psi , \widehat{n}^\phi\Psi\raa + N^{-\frac{1}{2}} ) \;.$$
\end{lemma}

\begin{proof}
Using $1=p_1+q_1$,
\begin{align*}
&\laa\Psi, B(x_1)  \Psi\raa-\langle\phi,B\phi\rangle
\\
=&
 \laa\Psi, p_1B (x_1) p_1\Psi\raa+2\Re\laa\Psi , q_1B(x_1)  p_1\Psi\raa+\laa\Psi , q_1B (x_1) q_1\Psi\raa-\langle\phi ,  B\phi\rangle
 \\
 \leq&
 \langle\phi , B\phi\rangle(\|p_1\Psi\|^2-1)+2\Re\laa\Psi, \widehat{n}^{-1/2}q_1B(x_1) p_1\widehat{n}_1^{1/2}\Psi\raa \nonumber\\
  + &\laa\Psi, q_1B(x_1) q_1\Psi\raa ,
\end{align*}
where we used Lemma \ref{kombinatorik} (c). Since $\|p_1\Psi\|^2-1 =\|q_1\Psi\|^2$
it follows that
\begin{align}\label{neededtocomplete}
|\laa\Psi, B(x_1) \Psi\raa-\langle\phi,B\phi\rangle|&\leq C\| B\|_\infty \left(\laa\Psi,\widehat{n}^2\Psi\raa+
\laa\Psi,\widehat{n}_1\Psi\raa + \laa\Psi,\widehat{n}\Psi\raa  \right)\nonumber\\
& \le C \|B\|_\infty( \laa\Psi,\widehat{n}\Psi\raa + N^{-\frac{1}{2}}) \;.
\end{align}
\end{proof}
Using Lemma \ref{opdiff}, setting $B=\dot A_t$, we get 
$$\gamma_a^<(\Psi_t,\phi_t)  =\gamma_a(\Psi_t,\phi_t) \leq C \|\dot A_t\|_\infty ( \laa\Psi_t,\widehat{n}^{\phi_t}\Psi_t\raa + N^{-\frac{1}{2}} ) \;,$$
which yields the first bound \eqref{abc} in Lemma \ref{gammalemma}.

\paragraph{Control of $\asalt$} 
To control $\gamma_b^<$ we will first prove that $\| \nabla_1 \Psi_t \| $ is uniformly bounded in $N$, if initially the energy per particle $\mathcal{E}_{U}(\Psi_0)$ is of order one.

\begin{lemma} \label{kinenergyboundedlemma}
Let $\Psi_0 \in L^2_{s}(\mathbb{R}^{2N}, \mathbb{C}) \cap H^2(\mathbb{R}^{2N}, \mathbb{C})$ with $\|\Psi_0\|=1$. 
For any $U \in L^2(\mathbb{R}^2,\mathbb{R}) ,\; U(x) \geq 0$,
let $\Psi_t$ the unique solution to $i \partial_t \Psi_t
= H_{U} \Psi_t$ with initial datum $\Psi_0$.  
Let $\mathcal{E}_{U}(\Psi_0) \leq C$.
Then
\begin{align*}
\| \nabla_1 \Psi_t \| \leq \mathcal{K}(\phi_t, A_t) .
\end{align*}

\end{lemma}

\begin{proof}
Using 
$
 \frac{d}{dt}  \mathcal{E}_{U}(\Psi_t) \leq \| \dot A_t\|_\infty
$, we obtain $\mathcal{E}_{U}(\Psi_t)  \leq \mathcal{K}(\phi_t, A_t)$. 
This yields
\begin{align*}
\| \nabla_1 \Psi_t \|^2
\leq
\mathcal{K}(\phi_t, A_t)- (N-1) \| \sqrt{U}\Psi_t \|^2
+
\| A_t \|_\infty
\leq \mathcal{K}(\phi_t, A_t) .
\end{align*}

\end{proof}

Next, we control $\widehat{m}^a$ and $\widehat{m}^b$ which were defined in Definition \ref{defm}.
The difference $m(k)-m(k+1)$ and
$m(k)-m(k+2)$ is of leading order given by the derivative of the function $m(k)$ -- $k$ understood as real variable --
with respect to $k$.
The $k$-derivative of $m(k)$ equals
\begin{equation}\label{mprime}m(k)^\prime=\left\{
         \begin{array}{ll}
           1/(2\sqrt{kN}), & \hbox{for $k\geq N^{1-2\xi}$;} \\
           1/2(N^{-1+\xi}), & \hbox{else.}
         \end{array}
       \right.\end{equation}
It is then easy to show
that, for any $j\in\mathbb{Z}$, there exists a $C_j<\infty$ such that 
\begin{align}\label{estab1}\widehat{m}_j^x&\leq C_jN^{-1} \widehat{n}^{-1}\text{ for }x\in\{a,b\}\\
\label{estab}\|\widehat{m}_j^x\|_{\text{op}}&\leq C_jN^{-1+\xi} \text{ for }x\in\{a,b\}
\\\label{estabn}\|\widehat{n}\widehat{m}_j^x\|_{\text{op}}&\leq C_jN^{-1} \text{ for }x\in\{a,b\}
\\\label{rop}\|\widehat{r}\|_{\text{op}}&\leq \|\widehat{m}^a\|_{\text{op}}+\|\widehat{m}^b\|_{\text{op}}
\leq CN^{-1+\xi}\;.
\end{align}
The different terms we have to estimate for $\asalt$ are found in (\ref{asaaltsplit}). In order to facilitate the notation, let $ \widehat{w} \in \lbrace N \widehat{m}^a_{-1},N \widehat{m}^b_{-2} \rbrace$. Then $w(k) < n(k)^{-1}$ and $\|\widehat{w}\|_{\text{op}}\leq C N^{\xi}$ follows. 
\begin{lemma}\label{hnorms}
Let $\beta >0$ and $W_\beta\in\mathcal{V}_\beta$. 
Let $\Psi  \in L^2_{s}(\mathbb{R}^{2N}, \mathbb{C}) \cap H^2(\mathbb{R}^{2N}, \mathbb{C}) \; , \| \Psi \|=1$
and let $\|\nabla_1\Psi\|
\leq \mathcal{K}(\phi, A_t)$.
Let $w(k)<n(k)^{-1}$ and $\| \widehat{w}\|_{\text{op}} \leq C N^{ \xi}$ for some $\xi \geq 0$. Then,
\begin{enumerate}
\item   $$N\left|\laa\Psi p_1p_2
\potdiff^\phi_\beta(x_1,x_2) q_1p_2\widehat{w}\Psi\raa\right|\leq   
\mathcal{K}(\phi, A_t)
 \left( N^{-1}+   N^{- 2 \beta } \ln(N) \right)
\;.
$$

\item 
 \begin{align*} 
 & N|\laa\Psi, p_1p_2
  W_{\beta}(x_1-x_2)
  \widehat{w}q_1q_2\Psi\raa|
  \nonumber
\\ 
 & \leq 
\mathcal{K}(\phi, A_t)
	\left(
		\laa \Psi, \widehat{n} \Psi \raa 
		 +
		  \inf_{\eta>0}
  \inf_{\beta>\beta_1 >0}
  \left(
 			N^{ \eta- 2 \beta_1} \ln(N)^2
 				 +\| \widehat{w}\|_{\text{op}} N^{-1+2 \beta_1}
+
\| \widehat{w}\|_{\text{op}}^2  N^{ - \eta}
\right)  
\right) 
\;.
\end{align*}

\item
\begin{align*} &N|\laa\Psi p_1q_2\potdiff^\phi_\beta(x_1,x_2)\widehat{w}q_1q_2\Psi\raa|
\leq 
\mathcal{K}(\phi, A_t)
\Big(
\laa\Psi,\widehat{n}\Psi\raa+ N^{-1/6} \ln(N)
\\
+&
\inf \left\lbrace 
\left|\mathcal{E}_{V_N}(\Psi)-\mathcal{E}_{4 \pi}^{GP}(\phi)\right|,
\left|\mathcal{E}_{W_\beta}(\Psi)-\mathcal{E}_{N \|W_\beta\|_1}^{GP}(\phi)\right|+ N^{-2 \beta} \ln(N)
\right\rbrace
\Big)
\;.
\end{align*}

\end{enumerate}
\end{lemma}

\begin{proof}
\begin{enumerate}
\item
In view of Lemma \ref{kombinatorikb}, we obtain
\begin{align*}
N \left|\laa\Psi, p_1p_2
\potdiff^\phi_\beta(x_1,x_2) q_1p_2\widehat{w}\Psi\raa\right|
\leq&  N\|p_1p_2
\potdiff^\phi_\beta(x_1,x_2) q_1p_2\|_{\text{op}}\|\widehat{n}\widehat{w}\Psi\|
\\\leq&  CN\|p_1p_2
\potdiff^\phi_\beta(x_1,x_2) q_1p_2\|_{\text{op}}\;.
\end{align*}
$\|p_1p_2
\potdiff^\phi_\beta(x_1,x_2) q_1p_2\|_{\text{op}}$ can be estimated using $p_1q_1=0$ and (\ref{faltungorigin}): 
\begin{align*}
& N
\left\|
	p_1p_2
	\left(
		W_\beta(x_1-x_2)-\frac{N\|W_\beta\|_1}{N-1}|\phi(x_1)|^2-\frac{N\|W_\beta\|_1}{N-1}|		
		\phi(x_2)|^2
	\right) q_1p_2
\right\|_{\text{op}}
\\
&\leq\|p_1p_2 
(NW_\beta(x_1-x_2)-N\|W_\beta\|_1|\phi(x_1)|^2)
p_2\|_{\text{op}}
+C \|\phi\|_\infty^2 N^{-1}
\\
&\leq \| \phi \|_{\infty} \ \|N(W_\beta\star|\phi|^2)-\|NW_\beta\|_1|\phi|^2\|+C \|\phi\|_\infty^2 N^{-1}\;.
\end{align*}
Let $h$ be given by 
$$h(x)=- \frac{1}{2 \pi}\int_{\mathbb{R}^2}d^2y \ln |x-y| NW_\beta(y) + \frac{1}{2\pi}\|NW_\beta\|_1\ln |x|
\;,
 $$
which implies
$$\Delta h(x)=N W_\beta(x) - \|NW_\beta\|_1 \delta(x)\;.$$ 
As above (see Lemma \ref{ulemma}), we obtain
$h(x)=0$ for $x \notin B_{RN^{- \beta}}(0)$, where $R N^{-\beta}$ is the radius of the support of $W_\beta$.
Thus,
\begin{align} \label{roemisch I potetnialestimate}
\|h\|_1
\leq&
\frac{1}{2 \pi}
\int_{\mathbb{R}^2}d^2x \int_{\mathbb{R}^2}  d^2y|\ln |x-y|| \mathds{1}_{B_{RN^{- \beta}}(0)}(x) NW_\beta(y)
\\
-& 
\frac{1}{2 \pi}
 N\|W_\beta\|_1\int_{\mathbb{R}^2}  d^2x  \ln(|x|) \mathds{1}_{B_{RN^{- \beta}}(0)}(x)
\leq 
C N^{-2 \beta} \ln(N)
\end{align}
Integration by parts and Young's inequality give that
\begin{align*}
&\| N(W_\beta\star|\phi|^2)-\|NW_\beta\|_1|\phi|^2\|=\|(\Delta h)\star|\phi|^2\|
\\\leq& \| h\|_1 \|\Delta|\phi|^2\|_2
\leq \mathcal{K}(\varphi, A_t) N^{-2\beta} \ln(N) 
\;.
\end{align*}
Thus, we obtain the bound
\begin{align}
N \left|\laa\Psi, p_1p_2
\potdiff^\phi_\beta(x_1,x_2) q_1p_2\widehat{w}\Psi\raa\right|
\leq
\mathcal{K}(\phi, A_t)
 \left( N^{-1}+   N^{- 2 \beta } \ln(N) \right)
\;,
\end{align}
which then proves (a).

\item
We will first consider  $\beta<1/2$.\\
Using Lemma \ref{kombinatorik} (c) and Lemma \ref{trick} with $O_{1,2}=q_2
W_\beta(x_1-x_2) p_2$, $\Omega=N^{-1/2}(\widehat{w})^{1/2}q_1\Psi$ and $\chi=N^{1/2}p_1(\widehat{w}_{2})^{1/2}\Psi$ we get
 \begin{align*}
&|\laa\Psi ,p_1 p_2 W_\beta(x_1-x_2) q_1 q_2
\widehat{w}\Psi\raa|
\\=&|\laa\Psi ,(\widehat{w})^{1/2}q_1 q_2
W_\beta(x_1-x_2)p_1 p_2(\widehat{w}_{2})^{1/2}\Psi\raa|
%
%
\\&\leq N^{-1}\left\|(\widehat{w})^{1/2}q_1\Psi\right\|^2+N\big|\laa q_2(\widehat{w}_{2})^{1/2}\,\Psi,p_1\sqrt{W_\beta}(x_1-x_2)
p_3
\sqrt{W_\beta}(x_1-x_3)\\&\hspace{3.7cm}\sqrt{W_\beta}(x_1-x_2)p_2\sqrt{W_\beta}(x_1-x_3)p_1
q_3(\widehat{w}_{2})^{1/2}\,\Psi\raa\big|
\\&+N(N-1)^{-1}\|q_2
W_\beta(x_1-x_2) p_2p_1(\widehat{w}_{2})^{1/2}\Psi\|^2&
\\&\leq N^{-1}\left\|(\widehat{w})^{1/2}q_1\Psi\right\|^2+N\|\sqrt{W_\beta}(x_1-x_2)p_1\|_{\text{op}}^4\;\|
q_2(\widehat{w}_{2})^{1/2}\,\Psi\|^2
\\&+2 N(N-1)^{-1}\|p_1q_2(\widehat{w}_{1})^{1/2}
W_\beta(x_1-x_2) p_2p_1\Psi\|^2
\\&+2 N(N-1)^{-1}\|q_1q_2 (\widehat{w})^{1/2}
W_\beta(x_1-x_2) p_2p_1\Psi\|^2
\;.
 \end{align*}
 With  Lemma \ref{kombinatorik} (e) we get the bound
 \begin{align*}
 \leq&  N^{-1}\|(\widehat{w})^{1/2}\widehat{n}\Psi\|^2+N\|\phi  \|_\infty^4\|W_\beta\|_1^2\;\|
\widehat{n}(\widehat{w}_{2})^{1/2}\,\Psi\|^2
\\&+2 N(N-1)^{-1}\|W_\beta\|^2\|\phi  \|_\infty^2
\left(
\| \widehat{w}_{1}\|_{\text{op}}
+
\| \widehat{w}\|_{\text{op}}
\right)
\;.
 \end{align*}
Note, that $\|W_\beta\|_1\leq CN^{-1}$, $\|W_\beta\|^2\leq CN^{-2+2\beta}$.
Furthermore, using $\widehat n<\widehat n_2$, we have under the conditions on $\widehat{w}$
\be
\label{einfuegen}
\|(\widehat{w})^{1/2}\widehat{n}_2\Psi\|
\leq
\|(\widehat{w}_2)^{1/2}\widehat{n}_2\Psi\|\leq
\|(\widehat{n}_2)^{1/2}\Psi\|
\leq
\sqrt{\langle\Psi,\widehat{n}\Psi\rangle}+2N^{-\frac{1}{2}}
\;.
\ee
In total, we obtain
\begin{align*}
N|\laa\Psi ,p_1 p_2 W_\beta(x_1-x_2) q_1 q_2
\widehat{w}\Psi\raa|\leq 
 \mathcal{K}(\phi, A_t)
\left(\laa\Psi,\widehat{n}\Psi\raa+ \| \widehat{w}\|_{\text{op}} N^{-1+2 \beta}\right)
\end{align*}
and we get (b) for the case $\beta<1/2$.

\item[b)] for $1/2\leq \beta$:
We use $U_{\beta_1,\beta}$ from Definition \ref{udef} for some $0<\beta_1<1/2$.
 We then obtain
 \begin{align}
 \nonumber
&   N\laa\Psi, p_1p_2
 W_{\beta}(x_1-x_2)
  \widehat{w}q_1q_2\Psi\raa
\\ 
\label{eins}
  =&
    N\laa\Psi, p_1p_2
U_{\beta_1,\beta}(x_1-x_2)
  \widehat{w}q_1q_2\Psi\raa
  \\
  \label{zwei}
  +&
    N\laa\Psi, p_1p_2
  \left(W_{\beta}(x_1-x_2)-U_{\beta_1,\beta}(x_1-x_2)\right)
  \widehat{w}q_1q_2\Psi\raa
 \end{align}
Term \eqref{eins} has been controlled above.
So we are left to control \eqref{zwei}.

Let  $\Delta h_{\beta_1,\beta}=W_{\beta}-U_{\beta_1,\beta}$. Integrating by parts and using that\\ $\nabla_1 h_{\beta_1,\beta}(x_1-x_2)=-\nabla_2 h_{\beta_1,\beta}(x_1-x_2)$ gives
 \begin{align}
\nonumber
&N\left|\laa\Psi, p_1p_2\left(W_{\beta}(x_1-x_2)-U_{\beta_1,\beta}(x_1-x_2)\right)\widehat{w}q_1q_2\Psi\raa\right|
\\\label{csplit1}&
\leq N\left|\laa\nabla_1p_1\Psi, p_2\nabla_2 h_{\beta_1,\beta}(x_1-x_2)\widehat{w} q_1q_2\Psi\raa\right|
\\\label{csplit2}
&+N\left|\laa \Psi, p_1p_2\nabla_2 h_{\beta_1,\beta}(x_1-x_2) \nabla_1\widehat{w}q_1q_2\Psi\raa\right|
\;.
\end{align}
Let $ t_1 \in \lbrace p_1, \nabla_1 p_1 \rbrace$ and let
 $\Gamma \in \lbrace \widehat{w}q_1\Psi, \nabla_1 \widehat{w}q_1\Psi \rbrace$.

For both \eqref{csplit1} and \eqref{csplit2}, we use 
 Lemma \ref{trick} with 
 $O_{1,2}=N^{1+ \eta/2} q_2\nabla_2 h_{\beta_1,\beta}(x_1-x_2)p_2$, $\chi=t_1\Psi$ and
 $\Omega= N^{- \eta/2} \Gamma$. This yields
 
\begin{align}
\label{gammaterm}
&\eqref{csplit1}+\eqref{csplit2}
\leq 
2 
\sup_{t_1 \in  \lbrace p_1, \nabla_1 p_1 \rbrace,  \Gamma \in \lbrace \widehat{w}q_1\Psi, \nabla_1 \widehat{w}q_1\Psi \rbrace }
\Big(
N^{-\eta} \| \Gamma \|^2
\\
\label{diagonalterm}
+&
\frac{N^{2+\eta}}{N-1}\|q_2\nabla_2 h_{\beta_1,\beta}(x_1-x_2)t_1p_2\Psi\|^2
 \\
\label{nebendiagonale}
+&
N^{2+\eta}\left|\laa \Psi,t_1p_2 q_3 \nabla_2 h_{\beta_1,\beta}(x_1-x_2)\nabla_3 h_{\beta_1,\beta}
(x_1-x_3)t_1q_2 p_3\Psi\raa\right|
\Big)
\;.
 \end{align}
The first term can be bounded using Corrolary \ref{kombinatorikc} by
\begin{align*}
N^{-\eta}\|\nabla_1 \widehat{w}q_1\Psi\|^2
\leq &
N^{-\eta} \| \widehat{w} \|^2_{\text{op}} \|\nabla_1 q_1 \Psi\|^2
\\
N^{-\eta}\| \widehat{w}q_1\Psi\|^2
\leq&
C N^{-\eta}
\;.
\end{align*}
Thus $\eqref{gammaterm} \leq \mathcal{K}(\phi, A_t)
N^{-\eta} \| \widehat{w} \|^2_{\text{op}} $ using that $\|\nabla_1 q_1 \Psi\| \leq \mathcal{K}(\phi, A_t)$.
By $\|t_1\Psi\|^2 \leq \mathcal{K}(\phi, A_t)$, we obtain
\begin{align*}
 (\ref{diagonalterm})
 \leq&
 \mathcal{K}(\phi, A_t)
 \frac{N^{2+\eta}}{N-1}\|\nabla_2 h_{\beta_1,\beta}(x_1-x_2)p_2\|_{\text{op}}^2
\leq 
 \mathcal{K}(\phi, A_t)
\frac{N^{2+\eta}}{N-1}
\|\phi\|_\infty ^2
\|\nabla h_{\beta_1,\beta}\|^2
\\
\leq&
 \mathcal{K}(\phi, A_t)
N^{\eta-1}\ln(N)
 \;,
 \end{align*}
where we used Lemma \ref{ulemma} in the last step.

Next, we estimate 
\begin{align*}
(\ref{nebendiagonale})\leq& N^{2+\eta}\| p_2
\nabla_2 h_{\beta_1,\beta}(x_1-x_2)t_1 q_2
\Psi\|^2\
\nonumber
\\
\leq&
2N^{2+\eta}\| p_2
 h_{\beta_1,\beta}(x_1-x_2)t_1 \nabla_2q_2
\Psi\|^2
\nonumber
\\+&
2N^{2+\eta}\||\varphi(x_2)\rangle \langle \nabla \varphi(x_2)|
 h_{\beta_1,\beta}(x_1-x_2)t_1 q_2
\Psi\|^2\
\nonumber
\\
\leq &
2N^{2+\eta}\| p_2
 h_{\beta_1,\beta}(x_1-x_2)
 \|_{\text{op}}^2
 \|
t_1 
  \nabla_2 q_2
\Psi\|^2\
\nonumber
\\
+&
2N^{2+\eta}
\| |\varphi(x_2)\rangle \langle \nabla \varphi(x_2)|
 h_{\beta_1,\beta}(x_1-x_2) \|_{\text{op}}^2
  \| t_1 q_2
\Psi\|^2\
\nonumber
\\
\leq &
 \mathcal{K}(\phi, A_t)
 N^{2+\eta}
\| h_{\beta_1,\beta} \|^2
\nonumber
\\ \leq&
 \mathcal{K}(\phi, A_t)
N^{\eta- 2\beta_1} \ln(N)^2
\;.
\end{align*}
Thus, for all $\eta \in \mathbb{R}$
\begin{align*}
&  N\laa\Psi, p_1p_2
  \left(W_{\beta}(x_1-x_2)-U_{\beta_1,\beta}(x_1-x_2)\right)
  \widehat{w}q_1q_2\Psi\raa
  \nonumber
\\ 
  \leq &
  \mathcal{K}(\phi,A_t)
  \left(\| \widehat{w}\|_{\text{op}}^2  N^{ - \eta} + N^{ \eta-1} \ln(N) +N^{ \eta- 2 \beta_1} \ln(N)^2 \right).
\end{align*}
Combining both estimates for  $\beta<1/2$ and $\beta \geq 1/2$, we obtain,
using $ N^{ \eta-1} \ln(N) < N^{ \eta- 2 \beta_1} \ln(N)$, 
 \begin{align*} 
 & N\laa\Psi, p_1p_2
  W_{\beta}(x_1-x_2)
  \widehat{w}q_1q_2\Psi\raa
  \nonumber
\\ 
 & \leq 
\mathcal{K}(\phi,A_t)
	\left(
		\laa \Psi, \widehat{n} \Psi \raa 
		 +
		  \inf_{\eta>0}
  \inf_{\beta_1 >0}
  \left(
 			N^{ \eta- 2 \beta_1} \ln(N)^2
 				 +N^{-1+2 \beta_1}
+
\| \widehat{w}\|_{\text{op}}^2  N^{ - \eta}
\right)  
\right) 
   \;.
\end{align*}

 and we get (b)  in full generality.

\item
We first estimate, noting that $q_1 p_2 |\phi|^2(x_1) q_1q_2=0$, 
\begin{align*}
&N
\left|\laa\Psi, q_1p_2 
\frac{N \|W_\beta\|_1}{N-1} |\phi|^2(x_2)
\widehat{w}q_1q_2\Psi\raa\right|
\leq
C
\|\phi\|_\infty^2
\|\widehat{w} q_2\|_{\text{op}}
\|q_1 \Psi \|^2
\\
&\leq
 \mathcal{K}(\phi, A_t)
\laa \Psi, \widehat{n} \Psi \raa
\;.
\end{align*}
Is is left to estimate $N
\left|\laa\Psi, q_1p_2W_\beta(x_1-x_2)\widehat{w}q_1q_2\Psi\raa\right|$.
Let $U_{0,\beta}$ be given as in Definition \ref{udef}. Using Lemma \ref{kombinatorik} (c) and integrating by parts we get
\begin{align}\nonumber
N&\left|\laa\Psi, q_1p_2V_\beta(x_1-x_2)\widehat{w}q_1q_2\Psi\raa\right|\\\leq&\nonumber
N\left|\laa\Psi,q_1p_2U_{0,\beta}(x_1-x_2)q_1q_2 \widehat{w}\Psi\raa\right|+
N\left|\laa\Psi,q_1p_2(\Delta_1
h_{0,\beta}(x_1-x_2))q_1q_2\widehat{w}\Psi\raa\right|
\\\leq&\nonumber
\|U_{0,\beta}\|_\infty
N\|q_1\Psi\|\;\|\widehat{w}q_1q_2\Psi\|
\\\nonumber%
&+N\left|\laa\nabla_1q_1 p_2\Psi,(\nabla_1
h_{0,\beta}(x_1-x_2))\widehat{w}q_1q_2\Psi\raa\right|
\\&+\nonumber
N\left|\laa\Psi, \widehat{w}_1q_1p_2(\nabla_1
h_{0,\beta}(x_1-x_2))\nabla_1q_1q_2\Psi\raa\right|
\\\leq&\label{s1}
N
 \|U_{0,\beta}\|_\infty
\|q_1\Psi\|\;\|\widehat{w}q_1q_2\Psi\|
\\\label{s2}
&+N\left|\laa\mathds{1}_{\mathcal{A}^{(d)}_{1}}\nabla_1q_1\Psi,p_2(\nabla_1
h_{0,\beta}(x_1-x_2))\widehat{w}q_1q_2\Psi\raa\right|
\\\label{s3}
&+N\left|\laa\nabla_1q_1\Psi,\mathds{1}_{\overline{\mathcal{A}}^{(d)}_{1}}p_2(\nabla_1
h_{0,\beta}(x_1-x_2))q_1q_2\widehat{w}\Psi\raa\right|
\\&+\label{s4}
N\left|\laa\Psi, \widehat{w}_1q_1p_2(\nabla_1
h_{0,\beta}(x_1-x_2))q_2\mathds{1}_{\mathcal{A}^{(d)}_{1}}\nabla_1q_1\Psi\raa\right|
\\&+\label{s5}
N\left|\laa\Psi, \widehat{w}_1q_1p_2(\nabla_1
h_{0,\beta}(x_1-x_2))q_2\mathds{1}_{\overline{\mathcal{A}}^{(d)}_{1}}\nabla_1q_1\Psi\raa\right|
\;.
\end{align}
Lemma \ref{kombinatorikb}  and Lemma \ref{ulemma} (a) yields the bound
$$(\ref{s1})\leq
C
\laa\Psi,\widehat{n}\Psi\raa
\;.
$$
For (\ref{s3}) and (\ref{s5}) we use Cauchy Schwarz and then Sobolev inequality as in Lemma \ref{propo} to get, for any $p > 1$,
\begin{align*}
&(\ref{s3})+(\ref{s5})\leq
N\left\|\nabla_1q_1\Psi\right\|\left\|\mathds{1}_{\overline{\mathcal{A}}^{(d)}_{1}}p_2(\nabla_1
h_{0,\beta}(x_1-x_2))q_1q_2\widehat{w}\Psi\right\|
\\&+N\left\|\nabla_1q_1\Psi\right\|\left\|\mathds{1}_{\overline{\mathcal{A}}^{(d)}_{1}}q_2(\nabla_1
h_{0,\beta}(x_1-x_2))q_1p_2\widehat{w}_1\Psi\right\|
\\\leq& CN\|\nabla_1q_1\Psi\|
\;
N^{\frac{1-2d}{2} \frac{p-1}{p}}
\|\nabla_1 p_2(\nabla_1
h_{0,\beta}(x_1-x_2))q_1q_2\widehat{w}\Psi\|^{\frac{p-1}{p} }
\|p_2(\nabla_1
h_{0,\beta}(x_1-x_2))q_1q_2\widehat{w}\Psi\|^{1/p}
\\+&CN\|\nabla_1q_1\Psi\|\;
N^{\frac{1-2d}{2} \frac{p-1}{p}}
\|\nabla_1 q_2(\nabla_1
h_{0,\beta}(x_1-x_2))q_1p_2\widehat{w}_1\Psi\|^{ \frac{p-1}{p} }
\| q_2(\nabla_1
h_{0,\beta}(x_1-x_2))q_1p_2\widehat{w}_1\Psi\|^{1/p}
\;.
\end{align*}
Using Lemma \ref{kombinatorik}, Lemma \ref{kombinatorikb}, Corollary \ref{kombinatorikc} and Lemma \ref{ulemma}, we obtain
\begin{align*}\|\nabla_1 p_2(\nabla_1
h_{0,\beta}(x_1-x_2))q_1q_2\widehat{w}\Psi\|\leq&
\|p_2(\Delta_1
h_{0,\beta}(x_1-x_2))q_1q_2\widehat{w}\Psi\|
\\&+
\| p_2(\nabla_1
h_{0,\beta}(x_1-x_2))\nabla_1q_1q_2\widehat{w}\Psi\|
\\&\hspace{-3cm}\leq  C \left(\|p_2(W_\beta-U_{0,\beta})(x_1-x_2)\|_{\text{op}}+\|p_2\nabla_1
h_{0,\beta}(x_1-x_2))\|_{\text{op}}\right)
\\&\hspace{-3cm}\leq  C
\|\phi\|_\infty
\left(
 N^{-1+\beta}
+
N^{-1} (\ln(N))^{1/2}
\right)
\; ,
\end{align*}
and similarly
\begin{align*}\|\nabla_1q_2(\nabla_1
h_{0,\beta}(x_1-x_2))q_1p_2\widehat{w}_1\Psi\|\leq&
\|q_2(\Delta_1
h_{0,\beta}(x_1-x_2))q_1p_2\widehat{w}_1\Psi\|
\\&+
\| q_2(\nabla_1
h_{0,\beta}(x_1-x_2))\nabla_1q_1p_2\widehat{w}_1\Psi\|
\\&\hspace{-3cm}\leq C \left(\|p_2(W_\beta-U_{0,\beta})(x_1-x_2)\|_{\text{op}}+
\| \widehat{w}_1\|_{\text{op}}
\|p_2\nabla_1
h_{0,\beta}(x_1-x_2))\|_{\text{op}}\right)
\\&\hspace{-3cm}\leq
C
\|\phi\|_\infty
\left(
 N^{-1+\beta}
+
\| \widehat{w}\|_{\text{op}}
N^{-1} (\ln(N))^{1/2}
\right)
\;.
\end{align*}
Moreover, we estimate
\begin{align*}
&\|
p_2
(\nabla_1
h_{0,\beta}(x_1-x_2))q_1q_2\widehat{w}\Psi\|
\leq
C\|\phi\|_\infty \|\nabla_1
h_{0,\beta}\|_2 \leq C\|\phi\|_\infty N^{-1} (\ln(N))^{1/2}
\\
&
\|
q_2
(\nabla_1
h_{0,\beta}(x_1-x_2))q_1p_2\widehat{w}\Psi\|
\leq
C\|\phi\|_\infty \|\nabla_1
h_{0,\beta}\|_2 \leq C \|\phi\|_\infty N^{-1} (\ln(N))^{1/2}
\;.
\end{align*}
Hence, we obtain, for any $p >1$, 
\begin{align*}
(\ref{s3})+(\ref{s5})&
\leq
C \|\phi\|_\infty
N^{1+\frac{1-2d}{2} \frac{p-1}{p}}
\left(
 N^{-1+\beta}
+
\| \widehat{w}\|_{\text{op}}
N^{-1} (\ln(N))^{1/2}
\right)^{\frac{p-1}{p}}
\left(
N^{-1} (\ln(N))^{1/2}
\right)^{1/p}
\;.
\end{align*}
For $d$ large enough, the right hand side can be bounded by $N^{-1}$, that is
\begin{align*}
\eqref{s3}+\eqref{s5}&
\leq C 
 \|\phi\|_\infty
N^{-1}
\;.
\end{align*}

For (\ref{s2}) we use that $\nabla_2
h_{0,\beta}(x_1-x_2)=-\nabla_1
h_{0,\beta}(x_1-x_2)$, Cauchy Schwarz and $ab\leq a^2+b^2$ and get
\begin{equation}\label{s2ref}(\ref{s2})\leq \|\mathds{1}_{\mathcal{A}^{(d)}_{1}}\nabla_1q_1\Psi\|^2+N^2\|p_2(\nabla_2
h_{0,\beta}(x_1-x_2))\widehat{w}q_1q_2\Psi\|^2\;.
\end{equation}
$\|\mathds{1}_{\mathcal{A}^{(d)}_{1}}\nabla_1q_1
\Psi \|^2 $ can be bounded using Lemma \ref{energylemma}.

Integration by parts and Lemma \ref{kombinatorik} (c) as well as $(a+b)^2\leq 2a^2+2b^2$ gives for the second summand 
\begin{align}\nonumber
&N^2\|p_1(\nabla_1h_{0,\beta}(x_1-x_2))q_1q_2\widehat{w}\Psi\|^2\leq 2N^2\|p_1h_{0,\beta}(x_1-x_2)\nabla_1q_1q_2\widehat{w}\Psi\|^2
\\
\nonumber+&2N^2\||\phi(x_1)\rangle\langle\nabla_1\phi(x_1)|h_{0,\beta}(x_1-x_2)q_1q_2\widehat{w}\Psi\|^2
\\\label{linea}\leq&2N^2\|p_1h_{0,\beta}(x_1-x_2)q_2(p_1\widehat{w}_1+q_1\widehat{w})\mathds{1}_{\mathcal{A}^{(d)}_{1}}\nabla_1q_1\Psi\|^2
\\\label{lineb}+&2N^2\|p_1h_{0,\beta}(x_1-x_2)q_2p_1\widehat{w}_1\mathds{1}_{\overline{\mathcal{A}}^{(d)}_{1}}\nabla_1q_1\Psi\|^2
\\\label{linec}+&2N^2\|p_1h_{0,\beta}(x_1-x_2)q_2q_1\widehat{w}\mathds{1}_{\overline{\mathcal{A}}^{(d)}_{1}}\nabla_1q_1\Psi\|^2
\\\label{lined}+&2N^2\||\phi(x_1)\rangle\langle\nabla_1\phi(x_1)|h_{0,\beta}(x_1-x_2)q_1q_2\widehat{w}\Psi\|^2
\;.
\end{align}
For (\ref{linea}) we use Lemma \ref{kombinatorikb}, Lemma \ref{kombinatorik} (e) with Lemma \ref{ulemma} (c) and then Lemma \ref{energylemma}.
\begin{align*}
 (\ref{linea})&\leq CN ^2\|p_1h_{0,\beta}(x_1-x_2)\|_{\text{op}}^2\|\mathds{1}_{\mathcal{A}^{(d)}_{1}}\nabla_1q_1\Psi\|^2
\\& \leq 
\mathcal{K}(\phi,A_t)
  \Big( \llaa\Psi,\widehat{n}^{\phi}\Psi\rraa+N^{-1/6} \ln(N)
\\
&+
\inf \left\lbrace 
\left|\mathcal{E}_{V_N}(\Psi)-\mathcal{E}_{4 \pi}^{GP}(\phi)\right|,
\left|\mathcal{E}_{W_\beta}(\Psi)-\mathcal{E}_{N \|W_\beta\|_1}^{GP}(\phi)\right|+ N^{-2 \beta} \ln(N)
\right\rbrace
\Big)
\;.
\end{align*}

Let $s_1 \in \lbrace p_1, q_1 \rbrace$ and let $
\widehat{d} \in \lbrace \widehat{w}, \widehat{w}_1 \rbrace$. 
Note that $ \| \widehat{d} \|_{\text{op}}= \| \widehat{w} \|_{\text{op}}$.
Then,
\eqref{lineb} and (\ref{linec}) can be estimated as
\begin{align*}
&
\eqref{lineb},
\eqref{linec}
\leq
2N^2
\| \nabla_1 q_1 \Psi \|
\|
\mathds{1}_{\overline{\mathcal{A}}^{(d)}_{1}} \widehat{d}  s_1 q_2 h_{0,\beta}(x_1-x_2)
p_1h_{0,\beta}(x_1-x_2)q_2s_1\widehat{d} \mathds{1}_{\overline{\mathcal{A}}^{(d)}_{1}}\nabla_1q_1\Psi\|
\\
& 
\leq
C
N^{2+ \frac{1-2d}{2}\frac{p-1}{p}}
\| \nabla_1 q_1 \Psi \|
\|
\nabla_1\widehat{d} s_1 q_2 h_{0,\beta}(x_1-x_2)
p_1h_{0,\beta}(x_1-x_2)q_2s_1\widehat{d}\mathds{1}_{\overline{\mathcal{A}}^{(d)}_{1}}\nabla_1q_1\Psi
\|^{\frac{p-1}{p}}
\\
&\times 
\|
\widehat{d} s_1 q_2 h_{0,\beta}(x_1-x_2)
p_1h_{0,\beta}(x_1-x_2)q_2s_1\widehat{d}\mathds{1}_{\overline{\mathcal{A}}^{(d)}_{1}}\nabla_1q_1\Psi\|
^{\frac{1}{p}}
\\
&
\leq 
C N^{2+ \frac{1-2d}{2}\frac{p-1}{p}}
\| \nabla_1 q_1 \Psi \|
\|  \widehat{w}\|_{\text{op}}
\|p_1  h_{0,\beta}(x_1-x_2)\|_{\text{op}}
\|\mathds{1}_{\overline{\mathcal{A}}^{(d)}_{1}}\nabla_1q_1\Psi \|
\\
&\times 
\|\nabla_1\widehat{d} s_1 q_2 h_{0,\beta}(x_1-x_2)
p_1\|_{\text{op}}^{\frac{p-1}{p}}
\|\widehat{d} s_1 q_2 h_{0,\beta}(x_1-x_2)
p_1\|_{\text{op}}^{\frac{1}{p}}
\\
& \leq
\mathcal{K}(\phi,A_t)
 N^{1+\frac{1-2d}{2}\frac{p-1}{p}} \|  \widehat{w}\|_{\text{op}}^2
\|\nabla_1 s_1 h_{0,\beta}(x_1-x_2)
p_1\|_{\text{op}}^{\frac{p-1}{p}}
\|h_{0,\beta}(x_1-x_2)
p_1\|_{\text{op}}^{\frac{1}{p}}
\\
&
\leq
\mathcal{K}(\phi,A_t)
 N^{1+\frac{1-2d}{2}\frac{p-1}{p}}
 \|  \widehat{w}\|_{\text{op}}^2
\left(
\|\nabla \phi\|
\| \nabla_1 h_{0,\beta}\|+ \|h_{0,\beta}\| 
\right)^{\frac{p-1}{p}}
\|h_{0,\beta}\| ^{\frac{1}{p}}
\\
& \leq \mathcal{K}(\phi,A_t)  \|  \widehat{w}\|_{\text{op}}^2
(1 +\ln(N))^{ \frac{p-1}{2p}}
N^{\frac{1-2d}{2}\frac{p-1}{p}}
\;.
\end{align*}
Here, we used, for $s_1 \in \lbrace p_1, 1- p_1 \rbrace$,
\begin{align*}
&\|\nabla_1 s_1 h_{0,\beta}(x_1-x_2)
p_1\|_{\text{op}}
\leq
\|\nabla_1 p_1 h_{0,\beta}(x_1-x_2)
p_1\|_{\text{op}}
+\|\nabla_1 h_{0,\beta}(x_1-x_2)
p_1\|_{\text{op}}
\\
\leq &
\| \varphi \|_\infty
\left(
\| \nabla \varphi \|
 \| h_{0,\beta}\| 
+
\| \nabla h_{0,\beta}\| 
\right)
\end{align*}
and then applied Lemma \ref{kombinatorik} (e).

For $d$ large enough, we obtain 
\begin{align*}
\eqref{lineb}+\eqref{linec}
\leq
 \mathcal{K}(\phi, A_t)
N^{-2}
\;.
\end{align*}

Line (\ref{lined}) can be bounded by
\begin{align*}
(\ref{lined})\leq&N^2\|h_{0,\beta}(x_1-x_2)\nabla_1p_1\|_{\text{op}}^2\;\|q_1q_2\widehat{w}\Psi\|^2
\leq
N^2\|h_{0,\beta}\|^2 \| \nabla \phi \|_\infty^2  \| q_1 \widehat{w} \|_{\text{op}}^2
\|q_1 \Psi \|^2
\\ 
\leq& C
 \| \nabla \phi \|_\infty^2
\laa\Psi\widehat{n}\Psi\raa
\;.
\end{align*}

For (\ref{s4}) we use Lemma \ref{trick} with $\Omega=\mathds{1}_{\mathcal{A}^{(d)}_{1}}\nabla_1q_1\Psi$,\\
$O_{1,2}=Nq_2(\nabla_2
h_{0,\beta}(x_1-x_2))p_2$ and $\chi=\widehat{w}q_1\Psi$.
\begin{align} 
(\ref{s4})&\leq \|\mathds{1}_{\mathcal{A}^{(d)}_{1}}\nabla_1q_1\Psi\|^2\label{ehklar}\\
&+
\label{ding3}
2 N\|q_2(\nabla_2
h_{0,\beta}(x_1-x_2))\widehat{w}q_1p_2\Psi\|^2
\\\label{ding4}&+N^2\big|\laa \Psi,q_1q_3\widehat{w}(\nabla_2
h_{0,\beta}(x_1-x_2))
p_2p_3(\nabla_3
h_{0,\beta}(x_1-x_3))\widehat{w}q_1q_2\Psi\raa\big|
\;.
\end{align}
Line \eqref{ding3} is bounded by
\begin{align*}
\eqref{ding3}
\leq &
C
 \|  \phi \|_\infty^2
N 
\|(\nabla_2 h_{0,\beta}(x_1-x_2))p_2 \|_{\text{op}}^2
\|\widehat{w}q_1\|_{\text{op}}^2
\\
\leq &
 C  \|  \phi \|_\infty^2 N 
\|\nabla_2 h_{0,\beta}(x_1-x_2) \|^2
\leq
C  \| \phi \|_\infty^2 N^{-1} \ln(N)
\;.
\end{align*}
(\ref{ehklar})+(\ref{ding4}) is bounded by $$
\|\mathds{1}_{\mathcal{A}^{(d)}_{1}}\nabla_1q_1\Psi\|^2+N^2\| p_2 (\nabla_2
h_{0,\beta}(x_1-x_2))\widehat{w}q_1 q_2\Psi\|^2\;.$$
Both terms have been controlled above (see (\ref{s2ref})). In total, we obtain
\begin{align*} &N|\laa\Psi p_1q_2\potdiff^\phi_\beta(x_1,x_2)\widehat{w}q_1q_2\Psi\raa|
\leq 
\mathcal{K}(\phi,A_t)
\Big(
\laa\Psi,\widehat{n}\Psi\raa+ N^{-1/6} \ln(N)
\\
+&
\inf \left\lbrace 
\left|\mathcal{E}_{V_N}(\Psi)-\mathcal{E}_{4 \pi}^{GP}(\phi)\right|,
\left|\mathcal{E}_{W_\beta}(\Psi)-\mathcal{E}_{N \|W_\beta\|_1}^{GP}(\phi)\right|+ N^{-2 \beta} \ln(N)
\right\rbrace
\Big)
\;.
\end{align*}

\end{enumerate}
\end{proof}
Using this Lemma, it follows that there exists an $\eta>0$ such that
\begin{equation*}
\gamma_b^< (\Psi_t,\phi_t) \leq 
\mathcal{K}(\phi_t,A_t)
\left(
 \laa\Psi_t,\widehat{n}^{\phi_t}\Psi_t\raa + N^{-\eta}+
\left|
\mathcal{E}_{W_\beta}(\Psi_0)
-
\mathcal{E}_{N \|W_\beta\|_1}^ {GP}(\phi_0)
\right|
 \right)\;.
 \end{equation*}
This proves Lemma \ref{gammalemma}.

\subsection{Estimates for the functional $\gamma$}
\label{gammacontrolsection}
For the most involved scaling which is induced by $V_N$, we need to control $ \|p_1 V_N \Psi\|$.
\begin{lemma}
Let $\Psi \in L_s^2 (\mathbb{R}^{2N}, \mathbb{C})$ and let
$\mathcal{E}_{V_N}(\Psi) \leq C$. Then
\begin{align} \label{p1 VN abschaetzung}
\|p_1 V_N \Psi\| \leq   \mathcal{K}(\phi, A_t) N^{-\frac{1}{2}}
\;.
\end{align}
\end{lemma}
\begin{proof}
We estimate
\begin{align*}
&\|p_1 V_N(x_1-x_2) \Psi\|= \|p_1 \mathds{1}_{\text{supp}(V_N)}(x_1-x_2) V_N(x_1-x_2) \Psi\|
\\
\leq & \|p_1 \mathds{1}_{\text{supp}(V_N)}(x_1-x_2) \|_{\text{op}} \| V_N(x_1-x_2) \Psi\|
\;.
\end{align*}
We have
\begin{align*}
\|p_1 \mathds{1}_{\text{supp}(V_N)} (x_1-x_2)\|_{\text{op}}^2\leq \| \varphi \|_\infty^2 
\|\mathds{1}_{\text{supp}(V_N)} \|_1 \leq C  \| \varphi \|_\infty^2  e^{-2N}
\;.
\end{align*}
Using
\begin{align*}
C \geq \mathcal{E}_{V_N}(\Psi)= \| \nabla \Psi \|^2 + (N-1) \| \sqrt{V_N}(x_1-x_2) \Psi \|^2
+
\laa \Psi, A_t(x_1) \Psi \raa
\end{align*}
as well as
\begin{align*}
\| V_N(x_1-x_2) \Psi\|^2 =& \| \sqrt{V_N} (x_1-x_2)\sqrt{ V_N}(x_1-x_2) \Psi\|^2
\leq
\| \sqrt{V_N}\|_{\infty}^2 \| \sqrt{V_N}(x_1-x_2) \Psi \|^2
\nonumber
\\
\leq&
C e^{2N} \frac{\mathcal{E}_{V_N}(\Psi)+ \| A_t \|_\infty }{N} 
\leq
 C  (1+ \| A_t \|_\infty) \frac{e^{2N}}{N}
 \;,
\end{align*}
we obtain
\begin{align*} 
\|p_1 V_N \Psi\| \leq 
 \mathcal{K}(\phi, A_t) N^{-\frac{1}{2}} \;.
\end{align*}
\end{proof}

\paragraph{Control of $\as$}
Recall that
\begin{align*}\as (\Psi,\phi)&=-N(N-1)\Im\left(\laa\Psi ,
\potdiffneu^{\phi}(x_1,x_2)\widehat{r}\,
\Psi\raa\right)\\&
-N(N-1)\Im\left(\laa\Psi ,g_{\beta}(x_{1}-x_{2})
\widehat{r}\,\mathcal{Z}^\phi (x_1,x_2) \Psi\raa\right)\;.
\end{align*}
Estimate \eqref{p1 VN abschaetzung} yields to the bound
$\|p_1 \mathcal{Z}^\phi (x_1,x_2)\Psi\| \leq   \mathcal{K}(\phi, A_t)  N^{-1/2}$. 
Therefore, the second line of $\gamma_b$ is controlled by
\begin{align*} 
&N^2\|g_{\beta}(x_{1}-x_{2})
p_{1}\|_{\text{op}}\|\widehat{r}\|_{\text{op}}\|p_1\mathcal{Z}^\phi (x_1,x_2)\Psi\|
\\
\leq & \mathcal{K}(\phi, A_t)N^{3/2}\|g_{\beta}\| \|\widehat{r}\|_{\text{op}}
\leq   \mathcal{K}(\phi, A_t) N^{\xi-1/2-\beta/2} \ln(N) \;.
\end{align*}
The first line of $\as$ can be bounded with (\ref{defzz}) and $f_{\beta}=1-g_{\beta}$ by
\begin{align} 
\nonumber
&N(N-1)|\Im\left(\laa\Psi ,
\potdiffneu^{\phi}(x_1,x_2)\widehat{r}\,
\Psi\raa\right)|
\\
\leq&
\label{oans}
N^2 |
\Im
\left(
\laa\Psi ,
\left(
	M_{\beta}(x_1-x_2)f_{\beta}(x_{1}-x_{2})-
	\frac{N}{N-1}
		\left(
		\| M_{\beta}f_{\beta}  \|_1
		|\phi(x_1)|^2 
		+\|  M_{\beta}f_{\beta} \|_1
		|\phi(x_2)|^2
	\right)
\right)\widehat{r}
\Psi\raa
\right)
|
\\&+
\label{zwoa}
\frac{N^2}{N-1}  |\laa\Psi ,
\left(
\| N M_{\beta}f_{\beta} \|_1- 4\pi\right)\left(  |\phi(x_1)|^2+|\phi(x_2)|^2\right)\widehat{r}
\Psi\raa|
\\&+
\label{gsuffa}
\frac{N^2}{N-1}|\laa\Psi ,
  \left(4\pi|\phi(x_1)|^2+4\pi|\phi(x_2)|^2\right)g_{\beta}(x_{1}-x_{2})
\widehat{r}\Psi\raa|
\;.
\end{align}
Since $M_{\beta}f_{\beta} \in \mathcal{V}_\beta$,
 \eqref{oans} is of the same form as $\asalt (\Psi,\varphi)$. 
Using Lemma \ref{defAlemma} (h), the second term is controlled by
\begin{align*}
\eqref{zwoa}
\leq
C
 \| \varphi \|_\infty^2 
N  
\left(
N\|M_\beta f_\beta\|_1-4 \pi
\right)
\|\hat{r}\|_{\text{op}}
\leq
C  \| \varphi \|_\infty^2 
N^{-1+ \xi}  \ln(N)
\;.
\end{align*}
The last term is controlled by
$$
\eqref{gsuffa} \leq
C N\|\phi\|_\infty^2\|g_{\beta}(x_{1}-x_{2})
p_{1}\|_{\text{op}}\|\widehat{r}\|_{\text{op}}\leq
C\|\phi\|_\infty^3N^{-1-\beta+\xi} \ln(N)
\;.
$$
and we get
$$|\as (\Psi,\phi)|\leq  
\mathcal{K}(\phi, A_t)
\left(
\laa \Psi, \widehat{m} \Psi \raa + |\mathcal{E}_{V_N}(\Psi)
- \mathcal{E}^{GP}_{4 \pi}(\phi) |
 +N^{-\eta}\right)$$
for some $\eta>0$.

\paragraph{Control of $\bs$}
Recall that \begin{align*} \bs (\Psi,\phi)=&-4N(N-1)\laa\Psi
, (\nabla_1g_{\beta}(x_1-x_2))\nabla_1
\widehat{r}\Psi\raa\;.\end{align*}

Using $\widehat{r}=(p_2+q_2)\widehat{r}=p_2\widehat{r}+p_1q_2\widehat{m}^a$  and $\nabla_1g_{\beta}(x_1-x_2)=-\nabla_2g_{\beta}(x_1-x_2)$, integration  by parts yields to
\begin{align}\label{extra1} 
|\bs (\Psi,\phi)|\leq&
4N^2|\laa\Psi
, g_{\beta}(x_1-x_2)\nabla_1\nabla_2
(
p_2\widehat{r}+p_1q_2\widehat{m}^a
)
\Psi\raa|
\\\label{extra2}&+
4N^2|\laa\nabla_2\Psi
, g_{\beta}(x_1-x_2)\nabla_1
p_2\widehat{r}\Psi\raa|
\\\label{extra3}&+4N^2|\laa\nabla_2\Psi
, g_{\beta}(x_1-x_2)\nabla_1
p_1q_2\widehat{m}^a\Psi\raa|
\;.\end{align}
We begin with 
\begin{align*}
(\ref{extra1})\leq &C N^2
\|g_\beta\| 
 \| \nabla \varphi \|_\infty
\left(
 \| \nabla_1 \widehat{r} \psi \|
 +
 \| \nabla_ 2 q_2 \widehat{m}^a\Psi \|
 \right)
 \\
 \leq &
 C
 N^{1-\beta} \ln(N)
  \| \nabla \varphi \|_\infty 
  \left(
\| \nabla_1 \widehat{r} \psi \|
 +
 \| \nabla_ 2 q_2 \widehat{m}^a\Psi \|
 \right)
 \;.
\end{align*}
Let $s_1,t_1\in \lbrace p_1,q_1 \rbrace , \; \;
s_2,t_2\in \lbrace p_2,q_2 \rbrace$. Inserting the identity $1=(p_1+q_1)(p_2+q_2)$, we obtain,
for $a \in \lbrace -2,-1,0,1,2 \rbrace$, 
\begin{align*}
\|\nabla_1\widehat{r}\Psi\|
\leq& C
\sup_{s_1,s_2,t_1,t_2, a}
 \|\widehat{r}_a s_1 s_2 \nabla_1 t_1 t_2 \Psi \|
\leq C
\sup_{t_1, a}
 \|\widehat{r}_a\|_{\text{op}} \|\nabla_1 t_1 \Psi \|
 \\
 \leq &
 C N^{-1 + \xi}
 \;.
\end{align*}
In analogy $ \| \nabla_ 2 q_2 \widehat{m}^a\Psi \| \leq C \| \widehat{m}^a \|_{\text{op}} \leq C N^{-1 + \xi}$.
This yields the bound
\begin{align*}
(\ref{extra1})\leq
\mathcal{K}(\phi, A_t)
N^{-\beta+ \xi} \ln( N)
\;.
\end{align*}

Furthermore,  (\ref{extra2}) is bounded by
\begin{align}
(\ref{extra2})\leq&4N^2\|\nabla_2\Psi\|\;
\|g_{\beta}\|\;\| \nabla \phi\|_\infty\|\nabla_1\widehat{r}\Psi\|
\leq C
\|\nabla \phi\|_\infty\
 N^{\xi-\beta} \ln(N) \;.
\end{align}
Similarly, we obtain
\begin{align*}(\ref{extra3})\leq&
4N^2\|\nabla_2\Psi\|\;
\|g_{\beta}\|\;\|\nabla \phi\|_\infty\| q_2 \widehat{m}^a  \Psi\|
\leq C \| \nabla \phi\|_\infty\  N^{ \xi-\beta} \ln(N) \;.
\end{align*}
It follows that $ |\bs (\Psi,\phi)|\leq 
\mathcal{K}(\phi, A_t)
	 N^{  \xi-\beta} 
 	\ln(N)
$.

\paragraph{Control of $\cs$}

To control $\cs$ and $\ds$ we will use the notation
\begin{equation}\label{notation} \begin{array}{cc}
     m^c(k)=m^a(k)-m^a(k+1) & m^d(k)=m^a(k)-m^a(k+2) \\
    m^e(k)=m^b(k)-m^b(k+1) & m^f(k)=m^b(k)-m^b(k+2) \;. \\
  \end{array}
\end{equation}
Since the second $k$-derivative of $m$ is given by (see (\ref{mprime}) for the first derivative)
$$m(k)^{\prime\prime}=\left\{
         \begin{array}{ll}
           -1/(4\sqrt{k^3N}), & \hbox{for $k\geq N^{1-2\xi}$;} \\
          0, & \hbox{else.}
         \end{array}
       \right.$$
it is easy to verify that
\begin{equation}\label{estcdef}\|\widehat{m}_j^x\|_{\text{op}} \leq C N^{-2+3\xi} \text{ for }x\in\{c,d,e,f\}\;.\end{equation}
Recall that \begin{align*}\cs (\Psi,\phi)=&2N(N-1)(N-2)\Im\left(\laa\Psi ,g_{\beta}(x_{1}-x_{2})
\left[V_N(x_1-x_3),
\widehat{r}\right] \Psi\raa\right)
\\&N(N-1)(N-2)\Im\left(\laa\Psi ,g_{\beta}(x_{1}-x_{2})
\left[4 \pi|\phi|^2(x_3),
\widehat{r}\right] \Psi\raa\right)\;.\end{align*}
Since $p_j+q_j=1$, we can rewrite $\widehat{r}$ as
 $$
 \widehat{r}=\widehat{m}^bp_{1}p_{2}+\widehat{m}^a(p_{1}q_{2}+q_1p_2)=(\widehat{m}^b-2\widehat{m}^a)p_{1}p_{2}+\widehat{m}^a(p_{1}+p_2)\;.$$
Thus,
\begin{align}\nonumber|\cs (\Psi,\phi)|\leq &C N^3\left|\laa\Psi ,g_{\beta}(x_{1}-x_{2})
\left[V_N(x_1-x_3),
(\widehat{m}^b-2\widehat{m}^a)p_{1}p_{2}+\widehat{m}^a(p_{1}+p_2)\right] \Psi\raa\right|
\\&\nonumber+C N^3\left|\laa\Psi ,g_{\beta}(x_{1}-x_{2})
\left[4\pi|\phi|^2(x_3),
\widehat{r}\right] \Psi\raa\right|
\\\label{gleichung3}\leq&C N^3\left|\laa\Psi ,g_{\beta}(x_{1}-x_{2})p_2
\left[V_N(x_1-x_3),
\widehat{m}^a\right] \Psi\raa\right|
\\\label{gleichung1}&+C N^3\left|\laa\Psi ,g_{\beta}(x_{1}-x_{2})
V_N(x_1-x_3)
(\widehat{m}^b-2\widehat{m}^a)p_{1}p_{2}\Psi\raa\right|
\\\label{gleichung2}&+C N^3\left|\laa\Psi ,g_{\beta}(x_{1}-x_{2})
(\widehat{m}^b-2\widehat{m}^a)p_{1}p_{2}V_N(x_1-x_3) \Psi\raa\right|
\\\label{gleichung4}&+C N^3\left|\laa\Psi ,g_{\beta}(x_{1}-x_{2})
\widehat{m}^ap_{1} V_N(x_1-x_3)\Psi\raa\right|
\\\label{gleichung5}&+C N^3\left|\laa\Psi ,g_{\beta}(x_{1}-x_{2})V_N(x_1-x_3)
\widehat{m}^ap_{1} \Psi\raa\right|
\\\label{gleichung6}&+C N^3\left|\laa\Psi ,g_{\beta}(x_{1}-x_{2})
\left[4 \pi |\phi|^2(x_3),
\widehat{r}\right] \Psi\raa\right|
\;.
\end{align}
Using Lemma \ref{kombinatorik} (d),
we obtain the following estimate:
\begin{align*}
(\ref{gleichung3})=&
C N^3\left|\laa\Psi ,g_{\beta}(x_{1}-x_{2})p_2
\left[V_N(x_1-x_3),
p_1p_3\widehat{m}^d+p_1q_3\widehat{m}^c+q_1p_3\widehat{m}^c\right] \Psi\raa\right|
\\\leq&
C N^3\big|\laa\Psi ,V_N(x_1-x_3)g_{\beta}(x_{1}-x_{2})p_2\mathds{1}_{\text{supp}(V_N)}(x_1-x_3)
\\&\hspace{4cm}\left(p_1p_3\widehat{m}^d+p_1q_3\widehat{m}^c+q_1p_3\widehat{m}^c\right) \Psi\raa\big|
\\&+
C N^3\left|\laa\Psi ,g_{\beta}(x_{1}-x_{2})p_2\left(
p_1p_3\widehat{m}^d+p_1q_3\widehat{m}^c+q_1p_3\widehat{m}^c\right)V_N(x_1-x_3)\Psi\raa\right|
\;.
\end{align*}
Both lines are bounded by
\begin{align*}
&C N^3\|V_N(x_1-x_3)\Psi\|\;\|g_{\beta}(x_{1}-x_{2})p_2\|_{\text{op}}
\\&\left(2\|\mathds{1}_{\text{supp}(V_N)}(x_1-x_3)p_1\|_{\text{op}}+\|\mathds{1}_{\text{supp}(V_N)}(x_1-x_3)p_3\|_{\text{op}}\right)
\left(\|\widehat{m}^d\|_{\text{op}}+\|\widehat{m}^c\|_{\text{op}}\right)
\;.
\end{align*}
In view of Lemma \ref{kombinatorik} (e) with Lemma \ref{defAlemma} (i),  $\|g_{\beta}(x_{1}-x_{2})p_2\|_{\text{op}} \leq \|\phi\|_\infty \| g_\beta \| \leq C  \|\phi\|_\infty  N^{-1-\beta} \ln(N) $.
Using (\ref{estcdef}), together with 
$ \|\mathds{1}_{\text{supp}(V_N)}(x_1-x_3)p_1\|_{\text{op}}  \|V_N(x_1-x_3)\Psi\|\leq  \mathcal{K}(\phi, A_t) N^{-1/2}$, 
 we obtain, using $\xi <1/2$,
$$(\ref{gleichung3})\leq   \mathcal{K}(\phi,A_t) N^{-1/2+3\xi-\beta}\ln(N)
\leq   \mathcal{K}(\phi,A_t) N^{1/2+\xi-\beta}\ln(N)
\;.$$
We continue with
\begin{align*}
&(\ref{gleichung1})+(\ref{gleichung2})+(\ref{gleichung4})\\\leq&
C N^3\|V_N(x_1-x_3)\Psi\|\|g_{\beta}(x_{1}-x_{2})p_2\|_{\text{op}}
\\ & \times \|\mathds{1}_{\text{supp}(V_N)}(x_1-x_3)p_1\|_{\text{op}}\|
(\widehat{m}^b-2\widehat{m}^a)\|_{\text{op}}
\\&+C N^3\|g_{\beta}(x_{1}-x_{2})p_2\|_{\text{op}}
\|\widehat{m}^b-2\widehat{m}^a\|_{\text{op}}\|p_{1}V_N(x_1-x_3) \Psi\|
\\&+C N^3\|g_{\beta}(x_{1}-x_{2})p_1\|_{\text{op}}\|
\widehat{m}^a\|_{\text{op}}\|p_{1} V_N(x_1-x_3)\Psi\|
\\\leq &   \mathcal{K}(\phi,A_t) N^{1/2+\xi-\beta} \ln(N)
\;.
\end{align*}

Next, we estimate \eqref{gleichung5}.
The support of the function $g_{\beta}(x_{1}-x_{2})V_N(x_1-x_3)$ is such that
$|x_1-x_2| \leq C N^ {-\beta}$, as well as $|x_1-x_3| \leq C e^ {-N}$. Therefore,
$g_{\beta}(x_{1}-x_{2})V_N(x_1-x_3) \neq 0$ implies $ |x_2-x_3| \leq C N^ {- \beta} $. We estimate
\begin{align*}
\eqref{gleichung5} =&C N^3\left|\laa\Psi ,g_{\beta}(x_{1}-x_{2})V_N(x_1-x_3)p_{1}
\mathds{1}_{ B_{C N^{- \beta}}(0)}  (x_2-x_3)
\widehat{m}^a \Psi\raa\right|
\\
\leq &C N^3 \| p_{1} V_N(x_1-x_3) g_{\beta}(x_{1}-x_{2}) \Psi \| 
\|\mathds{1}_{ B_{C N^{- \beta}}(0)}  (x_2-x_3)
\widehat{m}^a \Psi \|
\\
\leq &C N^3
\|p_1 \mathds{1}_{\text{supp}(V_N)} (x_1-x_3)\|_{\text{op}}
 \|g_{\beta}(x_{1}-x_{2}) V_N(x_1-x_3)  \Psi \| 
\| \mathds{1}_{ B_{C N^{- \beta}}(0)}  (x_2-x_3)
\widehat{m}^a \Psi \|
\\
\leq & C N^{5/2} \| g_{\beta}\|_\infty
 \|\mathds{1}_{ B_{C N^{- \beta}}(0)}  \|^{\frac{1}{2}}_{\frac{p}{p-1}} 
\| \nabla_1 \widehat{m}^a \Psi \|^{  \frac{p-1}{ p}} \| \widehat{m}^a \Psi \|^{\frac{1}{p}}
\\
\leq &
 C N^{5/2} \| g_{\beta}\|_\infty N^{- \beta/2} \| \nabla_1 \widehat{m}^a \Psi \|^{1/2} \| \widehat{m}^a \Psi \|^{1/2}
 \\
\leq &
 C N^{3/2+\xi- \beta/2}
 \;.
\end{align*}
In the fourth line, we applied Sobolev inequality as in the proof of Lemma \ref{propo}, then setting $p=2$.
Furthermore, we used
$ \| \nabla_1 \widehat{m}^a \Psi \|^{1/2} \| \widehat{m}^a \Psi \|^{1/2} \leq C N^{-1 + \xi}$,
as well as $ \| g_{\beta}\|_\infty \leq C$, see Lemma \ref{defAlemma}.

Using Lemma \ref{kombinatorik} (d),
 (\ref{gleichung6}) can be bounded by
\begin{align*}
&C N^3\left|\laa\Psi ,g_{\beta}(x_{1}-x_{2})
\left[4 \pi|\phi|^2(x_3),
p_1p_2(\widehat{r}-\widehat{r}_2)+(p_1q_2+q_1p_2)(\widehat{r}-\widehat{r}_1)\right] \Psi\raa\right|
\\&\leq CN^3\|\phi\|_\infty^2 \left(\|\widehat{r}-\widehat{r}_2\|_{\text{op}}+\|\widehat{r}-\widehat{r}_1\|_{\text{op}}\right)
\|g_{\beta}(x_{1}-x_{2})p_2\|_{\text{op}}
\;.
\end{align*}
Note that $\|\widehat{r}-\widehat{r}_2\|_{\text{op}}+\|\widehat{r}-\widehat{r}_1\|_{\text{op}}\leq\sum_{j\in\{c,d,e,f\}}\|\widehat{m}^j\|_{\text{op}}
\leq C N^{-2+ 3 \xi}$ holds. 
With $\|g_{\beta}(x_{1}-x_{2})p_2\|_{\text{op}}\leq C N^{-1-\beta} \ln(N) $, it then follows that
$$|(\ref{gleichung6})|\leq C
\| \phi \|_\infty^2
N^{3\xi-\beta} \ln(N)\;.$$
In total, we obtain
\begin{align*}
|\gamma_d (\Psi,\phi)| \leq 
\mathcal{K}(\phi,A_t)
\left(
	N^{3/2+\xi-\beta/2}+N^{1/2+3\xi-\beta}\ln(N)
\right) \;.
\end{align*}


\paragraph{Control of $\ds$}
Recall that \begin{align*}&\ds (\Psi,\phi)=-\frac{1}{2}N(N-1)(N-2)(N-3)\\&\;\;\;\;\;\;\;\;\;\;\Im\left(\laa\Psi ,
g_{\beta}(x_{1}-x_{2})\left[V_N(x_3-x_4),
\widehat{r}\right] \Psi\raa\right)
\;.
\end{align*}
Using symmetry, Lemma \ref{kombinatorik} (d) and notation (\ref{notation}),
 $\ds$ is bounded by
\begin{align*}\ds (\Psi,\phi)\leq&N^4\big|\laa\Psi ,
g_{\beta}(x_{1}-x_{2})\big[V_N(x_3-x_4),
\widehat{m}^cp_1p_2p_3p_4+2\widehat{m}^dp_1p_2p_3q_4\\&\hspace{5cm}+
2\widehat{m}^ep_1q_2p_3p_4+4\widehat{m}^fp_1q_2p_3q_4\big] \Psi\raa\big|
\\\leq&4N^4\|V_N(x_3-x_4)\Psi\|\|\mathds{1}_{\text{supp}(V_N)}(x_3-x_4)p_3\|_{\text{op}}\|g_{\beta}(x_{1}-x_{2})p_1\|_{\text{op}}
\\&\hspace{3cm}(\|\widehat{m}^c\|_{\text{op}}+\|\widehat{m}^d\|_{\text{op}}+\|\widehat{m}^e\|_{\text{op}}+\|\widehat{m}^f\|_{\text{op}})
\;.
\end{align*}
We get with (\ref{estcdef}), Lemma \ref{defAlemma} and Lemma \ref{kombinatorik} that
$$ |\ds (\Psi,\phi)|\leq  \mathcal{K}(\phi,A_t) N^{1/2+3\xi- \beta} \ln(N)\;.$$

\paragraph{Control of $\gamma_f$}
Recall that 
\begin{align*}
\gamma_f(\Psi,\phi)
=
2
N(N-1)\frac{N-2}{N-1}\Im\left(\laa\Psi,g_{\beta}(x_{1}-x_{2})
\left[4 \pi |\phi|^2(x_1),
\widehat{r}\right] \Psi\raa\right)
\;.
\end{align*}
We obtain the estimate
\begin{align*}
|\gamma_f(\Psi,\phi)|
\leq
\mathcal{K}(\phi,A_t)
N^2
\|g_\beta\|
\| \widehat{r} \|_{\text{op}}
\leq
\mathcal{K}(\phi,A_t)
N^{\xi-\beta} \ln(N)
\;.
\end{align*}

Collecting all estimates, we get with $\xi <1/2$
\begin{align*}
|\gamma_c (\Psi,\phi)|+|\gamma_d (\Psi,\phi)|+|\gamma_e (\Psi,\phi)|+|\gamma_f(\Psi,\phi)|
\leq
\mathcal{K}(\phi,A_t)
N^{2-\beta/2} \ln(N)
\;.
\end{align*}
 Choosing $\beta$ sufficiently large, we obtain the desired decay and hence Lemma \ref{gammalemma fuer V}.
\subsection{Energy estimates}

\begin{lemma} \label{energylemma}
Let  $\Psi \in L^2_s( \mathbb{R}^{2N}, \mathbb{C})  \cap H^1( \mathbb{R}^{2N}, \mathbb{C}) ,\; \| \Psi \|=1$ with $ \|\nabla_1\Psi  \| \leq \mathcal{K}(\phi, A_t)$. Let 
$ \phi \in H^3(\mathbb{R}^2,\mathbb{C}) , \; \| \phi \|=1$. Define the sets
$ \mathcal{A}^{(d)}_{1},\overline{\mathcal{B}}^{(d)}_{1} $ as in Definition \ref{hdetail}.
Then, for $d$ large enough,
\begin{align*}
&\|\mathds{1}_{\mathcal{A}^{(d)}_{1}}\nabla_1q_1
\Psi \|^2 
+
\|\mathds{1}_{\overline{\mathcal{B}}^{(d)}_{1}}\nabla_1q_1
\Psi \|^2
\leq 
\mathcal{K}(\phi, A_t)
  \Big( \laa\Psi,\widehat{n}^{\phi}\Psi\raa+N^{-1/6} \ln(N)
\\
&+
\inf \left\lbrace 
\left|\mathcal{E}_{V_N}(\Psi)-\mathcal{E}_{4 \pi}^{GP}(\phi)\right|,
\left|\mathcal{E}_{W_\beta}(\Psi)-\mathcal{E}_{N \|W_\beta\|_1}^{GP}(\phi)\right|+ N^{-2 \beta} \ln(N)
\right\rbrace
\Big)
\;.
 \end{align*}
\end{lemma}

\begin{proof}
We start with expanding $\mathcal{E}_{W_\beta}(\Psi)-\mathcal{E}_{N \|W_\beta\|_1}^{GP}(\phi)$. This yields 
\begin{align*}
&\mathcal{E}_{W_\beta}(\Psi)-\mathcal{E}_{N \|W_\beta\|_1}^{GP}(\phi)
=
\|\nabla_1\Psi \|^2+\frac{N-1}{2}\|\sqrt{W_\beta}(x_1-x_2)\Psi\|^2
\nonumber
\\&
-\|\nabla\phi \|^2-\frac{1}{2} N\|W_\beta\|_1 \|\phi^2\|^2+
\laa \Psi, A_t (x_1) \Psi \raa - \langle \phi, A_t \phi \rangle
\nonumber
\\=&
 \|\mathds{1}_{\mathcal{A}^{(d)}_{1}}\nabla_1q_1\Psi \|^2+\|\mathds{1}_{\overline{\mathcal{B}}^{(d)}_{1}}
\mathds{1}_{\overline{\mathcal{A}}^{(d)}_{1}}
\nabla_1\Psi \|^2
+
M( \Psi, \phi)+Q_\beta( \Psi, \phi)
\;,
\end{align*}
where we have defined
\begin{align}
M( \Psi, \phi)=
\label{1 term in M}
&2 \Re\left(\laa\nabla_1q_1\Psi ,
\mathds{1}_{\mathcal{A}^{(d)}_{1}}
\nabla_1p_1\Psi \raa\right)
\\&
\label{2 term in M}
+ \|\mathds{1}_{\mathcal{A}^{(d)}_{1}}\nabla_1p_1\Psi \|^2-\|\nabla\phi \|^2
\\&
\label{3 term in M}
+\laa\Psi , A_t(x_1)\Psi\raa-\langle\phi , A_t\phi\rangle
\;,
\\
\nonumber
Q_\beta( \Psi, \phi)=&
\|\mathds{1}_{\mathcal{B}^{(d)}_{1}}\mathds{1}_{\overline{\mathcal{A}}^{(d)}_{1}}\nabla_1\Psi \|^2
\\
\nonumber
+&
\frac{N-1}{2}\laa\Psi ,(1-p_1p_2)W_{\beta}(x_1-x_2)(1-p_1p_2)\Psi\raa
\\
\nonumber
+&\frac{N-1}{2}\laa\Psi ,p_1p_2W_{\beta}(x_1-x_2)p_1p_2\Psi\raa -\frac{1}{2}N\|W_\beta\|_1\|\phi^2\|^2
\\
\nonumber
+&(N-1)\Re\laa\Psi ,(1-p_1p_2)W_{\beta}(x_1-x_2)p_1p_2\Psi\raa
\;.
\end{align}
Notice that the first two terms in $Q_\beta( \Psi, \phi)$ are nonnegative. This yields to the bound
\begin{align}
S_\beta( \Psi, \phi)=&
\label{1-pp term in Sbeta}
(N-1)|\laa\Psi ,(1-p_1p_2)W_{\beta}(x_1-x_2)p_1p_2\Psi\raa |
\\
+&
\label{pp pq term in Sbeta}
\left|
\frac{N-1}{2}\laa\Psi ,p_1p_2W_{\beta}(x_1-x_2)p_1p_2\Psi\raa -\frac{1}{2}N\|W_\beta\|_1\|\phi^2\|^2
\right|
\\
\nonumber
\geq&
-Q_\beta( \Psi, \phi)
\;.
\end{align}
We therefore obtain the following bound:
\begin{align}
\label{firsteneergybound}
 \|\mathds{1}_{\mathcal{A}^{(d)}_{1}}\nabla_1q_1\Psi \|^2
 +
 \|\mathds{1}_{\overline{\mathcal{B}}^{(d)}_{1}}
\mathds{1}_{\overline{\mathcal{A}}^{(d)}_{1}}
\nabla_1\Psi \|^2
\leq
&
\left|
\mathcal{E}_{W_\beta}(\Psi)-\mathcal{E}_{N \|W_\beta\|_1}^{GP}(\phi)
\right|
+
|M( \Psi, \phi)|+|S_\beta( \Psi, \phi)|
\;.
\end{align}

Next, we split up the energy difference $\mathcal{E}_{V_N}(\Psi)-\mathcal{E}_{4 \pi}^{GP}(\phi)$, 
\begin{align*}
\mathcal{E}_{V_N}(\Psi)-\mathcal{E}_{4 \pi}^{GP}(\phi)&=
\|\nabla_1\Psi \|^2+\frac{N-1}{2}\|\sqrt{V_N}(x_1-x_2)\Psi\|^2
-\|\nabla\phi \|^2
\\
&-2\pi\|\phi^2\|^2+
\laa \Psi, A_t (x_1) \Psi \raa - \langle \phi, A_t \phi \rangle
\;.
\end{align*}
In order to better estimate the terms corresponding to the two-particle interactions, we introduce, for $\mu>d$, the potential $M_{\mu} (x)$,
defined in Definition \ref{microscopic}, and continue with
\begin{align*}
\mathcal{E}_{V_N}(\Psi)-\mathcal{E}_{4 \pi}^{GP}(\phi)=&
\|\mathds{1}_{\mathcal{A}^{(d)}_{1}}\nabla_1\Psi\|^2
 +
  \|\mathds{1}_{\overline{\mathcal{B}}^{(d)}_{1}}
\mathds{1}_{\overline{\mathcal{A}}^{(d)}_{1}}
\nabla_1\Psi \|^2+
\|
\mathds{1}_{\mathcal{B}^{(d)}_{1}}
\mathds{1}_{\overline{\mathcal{A}}^{(d)}_{1}}
			\nabla_1\Psi \|^2 \nonumber\\&
+\frac{N-1}{2}\|\mathds{1}_{\overline{\mathcal{B}}^{(d)}_{1}}\sqrt{V_N}(x_1-x_2)\Psi \|^2
\\&+
\frac{1}{2}
\laa\Psi ,\sum_{j\neq
1}\mathds{1}_{\mathcal{B}^{(d)}_{1}}\left(V_N-M_{\mu}\right)(x_1-x_j)\Psi \raa
\\& +
\frac{1}{2}
\laa\Psi ,\sum_{j\neq
1}\mathds{1}_{\mathcal{B}^{(d)}_{1}}M_{\mu}(x_1-x_j)\Psi \raa
-\|\nabla\phi \|^2-2\pi\|\phi^2\|^2
\\&+\laa\Psi , A_t(x_1)\Psi\raa-\langle\phi , A_t\phi\rangle
\;.
\end{align*}
Using that $q_1=1-p_1$ and symmetry gives (after reordering)
\begin{align*} &
\mathcal{E}_{V_N}(\Psi)-\mathcal{E}_{4 \pi}^{GP}(\phi)
\\
=&
 \|\mathds{1}_{\mathcal{A}^{(d)}_{1}}\nabla_1q_1\Psi \|^2+
 \|\mathds{1}_{\overline{\mathcal{B}}^{(d)}_{1}}
\mathds{1}_{\overline{\mathcal{A}}^{(d)}_{1}}
\nabla_1\Psi \|^2
+
\frac{N-1}{2}\|\mathds{1}_{\overline{\mathcal{B}}^{(d)}_{1}}\sqrt{V_N}(x_1-x_2)\Psi \|^2
\\&
+\frac{N-1}{2}\laa\Psi ,\mathds{1}_{\mathcal{B}^{(d)}_{1}}(1-p_1p_2)M_{\mu}(x_1-x_2)(1-p_1p_2)\mathds{1}_{\mathcal{B}^{(d)}_{1}}\Psi\raa
\\&+ \|\mathds{1}_{\mathcal{B}^{(d)}_{1}}\mathds{1}_{\overline{\mathcal{A}_1}^{(d)}}\nabla_1\Psi \|^2
+
\frac{1}{2}
\laa\Psi ,\sum_{j\neq
1}\mathds{1}_{\mathcal{B}^{(d)}_{1}}\left(V_N-M_{\mu}\right)(x_1-x_j)\Psi \raa
\\&+\frac{N-1}{2}\laa\Psi ,\mathds{1}_{\mathcal{B}^{(d)}_{1}}p_1p_2M_{\mu}(x_1-x_2)p_1p_2\mathds{1}_{\mathcal{B}^{(d)}_{1}}\Psi\raa -2\pi\|\phi^2\|^2
\\&+2 \Re\left(\laa\nabla_1q_1\Psi ,
\mathds{1}_{\mathcal{A}^{(d)}_{1}}
\nabla_1p_1\Psi \raa\right)
\\&
+(N-1)\Re\laa\Psi ,\mathds{1}_{\mathcal{B}^{(d)}_{1}}(1-p_1p_2)M_{\mu}(x_1-x_2)p_1p_2\mathds{1}_{\mathcal{B}^{(d)}_{1}}\Psi\raa
\\&
+ \|\mathds{1}_{\mathcal{A}^{(d)}_{1}}\nabla_1p_1\Psi \|^2-\|\nabla\phi \|^2
\\&+\laa\Psi, A_t(x_1)\Psi\raa-\langle\phi , A_t\phi\rangle
\\
=&
 \|\mathds{1}_{\mathcal{A}^{(d)}_{1}}\nabla_1q_1\Psi \|^2+
  \|\mathds{1}_{\overline{\mathcal{B}}^{(d)}_{1}}
\mathds{1}_{\overline{\mathcal{A}}^{(d)}_{1}}
\nabla_1\Psi \|^2
+
M( \Psi, \phi)+\tilde{Q}_\mu( \Psi, \phi)
\;.
\end{align*}
with
\begin{align}
&\tilde{Q}_\mu( \Psi, \phi)=
\frac{N-1}{2}\laa\Psi ,\mathds{1}_{\mathcal{B}^{(d)}_{1}}(1-p_1p_2)M_{\mu}(x_1-x_2)(1-p_1p_2)\mathds{1}_{\mathcal{B}^{(d)}_{1}}\Psi\raa
\nonumber
\\
&+
\frac{N-1}{2}\|\mathds{1}_{\overline{\mathcal{B}}^{(d)}_{1}}\sqrt{V_N}(x_1-x_2)\Psi \|^2
\nonumber
\\
\label{vielezeilenb}&
+ \|\mathds{1}_{\mathcal{B}^{(d)}_{1}}\mathds{1}_{\overline{\mathcal{A}}^{(d)}_{1}}\nabla_1\Psi \|^2
+
\frac{1}{2}
\laa\Psi ,\sum_{j\neq
1}\mathds{1}_{\mathcal{B}^{(d)}_{1}}\left(V_N-M_{\mu}\right)(x_1-x_j)\Psi \raa
\\&
\nonumber
+(N-1)\Re\laa\Psi ,\mathds{1}_{\mathcal{B}^{(d)}_{1}}(1-p_1p_2)M_{\mu}(x_1-x_2)p_1p_2\mathds{1}_{\mathcal{B}^{(d)}_{1}}\Psi\raa
\\
&+\frac{N-1}{2}\laa\Psi ,\mathds{1}_{\mathcal{B}^{(d)}_{1}}p_1p_2M_{\mu}(x_1-x_2)p_1p_2\mathds{1}_{\mathcal{B}^{(d)}_{1}}\Psi\raa -2\pi\|\phi^2\|^2
\;.
\nonumber
\end{align}
The first two terms in $\tilde{Q}_\mu( \Psi, \phi)$ are nonnegative.
For $\mu >d$ Lemma \ref{positiv V-w} below shows that \eqref{vielezeilenb} is also nonnegative. Thus, for $\mu >d$, we obtain the bound 
\begin{align}
\label{vielezeilenc2}
\tilde{S}_\mu( \Psi, \phi)=
&(N-1)\left|
\laa\Psi ,\mathds{1}_{\mathcal{B}^{(d)}_{1}}(1-p_1p_2)M_{\mu}(x_1-x_2)p_1p_2\mathds{1}_{\mathcal{B}^{(d)}_{1}}\Psi\raa
\right|
\\
\label{vielezeilenc}
+&
\left|
\frac{N-1}{2}\laa\Psi ,\mathds{1}_{\mathcal{B}^{(d)}_{1}}p_1p_2M_{\mu}(x_1-x_2)p_1p_2\mathds{1}_{\mathcal{B}^{(d)}_{1}}\Psi\raa -2\pi\|\phi^2\|^2
\right|
\\
\nonumber
\geq& -\tilde{Q}_\mu (\Psi, \phi)
 \;.
\end{align}
In total, we obtain
\begin{align}
\label{secondenergybound}
 \|\mathds{1}_{\mathcal{A}^{(d)}_{1}}\nabla_1q_1\Psi \|^2
 +
 \|\mathds{1}_{\overline{\mathcal{B}}^{(d)}_{1}}
\mathds{1}_{\overline{\mathcal{A}}^{(d)}_{1}}
\nabla_1\Psi \|^2
\leq
|M( \Psi, \phi)|+\tilde{S}_\mu( \Psi, \phi)+ \left|\mathcal{E}_{V_N}(\Psi)-\mathcal{E}_{4 \pi}^{GP}(\phi)\right|
\;.
 \end{align}
Next, we will estimate $M( \Psi, \phi),S_\beta( \Psi, \phi)$ and $\tilde{S}_\mu( \Psi, \phi)$.

\begin{itemize}
\item Estimate of $S_\beta( \Psi, \phi)$ and $\tilde{S}_\mu( \Psi, \phi)$. \\
We first estimate \eqref{vielezeilenc}, using the same estimate as in \eqref{roemisch I potetnialestimate}.
Note that
$$
\laa\Psi ,\mathds{1}_{\mathcal{B}^{(d)}_{1}}p_1p_2M_{\mu}(x_1-x_2)p_1p_2\mathds{1}_{\mathcal{B}^{(d)}_{1}}\Psi\raa=
\langle\phi, M_{\mu} \star|\phi|^2\phi\rangle\laa\Psi ,\mathds{1}_{\mathcal{B}^{(d)}_{1}}p_1p_2\mathds{1}_{\mathcal{B}^{(d)}_{1}}\Psi\raa
\;.
$$
Using $\|\mathds{1}_{\overline{\mathcal{B}}^{(d)}_{1}} \Psi\|\leq C N^{1-d+ \epsilon}$, for any $\epsilon>0$, (see Lemma \ref{hdetail}) 
we obtain, together with $\|p_1p_2\Psi\|^2= 1+ 2 \|p_1 q_2\Psi\|^2+ \|q_1 q_2\Psi\|^2$
\begin{align*}
|\eqref{vielezeilenc}|
\leq &
3\| q_1 \Psi \|^2
+
 C
 \left(
 N^{1-d+ \epsilon}
+
N^{2-2d+ 2 \epsilon}
\right)
+
\frac{1}{2}
|N\langle\phi, M_{\mu} \star|\phi|^2\phi\rangle
- N \|M_\mu \|_1 \| \varphi^2 \|^2
|
\\
+&
\frac{1}{2}
| 4 \pi- N \|M_\mu \|_1 |\| \varphi^2 \|^2
+
\frac{1}{2}
\langle\phi, M_{\mu} \star|\phi|^2\phi\rangle
\;.
\end{align*}
Note that, using Young's inequality and
 \eqref{roemisch I potetnialestimate}
\begin{align*}
&|
\langle\phi, N M_{\mu} \star|\phi|^2\phi\rangle
- N \|M_\mu \|_1\| \varphi^2 \|^2|
=
\left|
\int_{\mathbb{R}^2} d^2 x
|\phi(x)|^2
\left(
N( M_{\mu} \star|\phi|^2)(x)
-N \|M_\mu \|_1|\phi(x)|^2
\right)
\right|
\\
\leq &
 \| \phi \|_\infty^2
 \|N(M_\mu\star|\phi|^2)-\|NM_\mu\|_1|\phi|^2\|_1
  \leq
  C
   \| \phi \|_\infty^2
   \|\Delta|\phi|^2\|_1
   N^{-2\mu} \ln(N) 
\\   
   \leq  &
   \mathcal{K}(\phi, A_t)
   N^{-2\mu} \ln(N) 
   \;.
\end{align*}
Since $|N \|M_{\mu}\|_1- 4 \pi | \leq C \frac{\ln(N)}{N} $ (see Lemma \ref{defAlemma})
and
$\langle\phi, M_{\mu} \star|\phi|^2\phi\rangle \leq \|\phi\|_\infty^4 \|M_\mu \|_1 \leq
C \|\phi\|_\infty^4 N^{-1}
 $,
it follows that
\begin{align}
\left|(\ref{vielezeilenc})\right|\leq & \mathcal{K}(\phi,A_t) \left(
\laa\Psi,\widehat{n}^{\phi}\Psi\raa+
 N^{1-d+ \epsilon}+N^{2-2d+2 \epsilon}+ 
	 	 N^{-2\mu} \ln(N) + N^{-1} \ln(N)
 \right)
 \nonumber
 \\
 \leq & \mathcal{K}(\phi,A_t)
  \left(\laa\Psi,\widehat{n}^{\phi}\Psi\raa+
 N^{-1} \ln(N)
\right)
\;,
\end{align}
where the last inequality holds for $d$ large enough (recall that we chose $\mu>d$).
\\
Using the same estimates, we obtain
\begin{align*}
\eqref{pp pq term in Sbeta}
\leq
\mathcal{K}(\phi,A_t)
 \left(\laa\Psi,\widehat{n}^{\phi}\Psi\raa+
N^{-2\beta} \ln(N) 
+
N^{-1} \ln(N) 
\right)
\;.
\end{align*}

Line~\eqref{vielezeilenc2} and line \eqref{1-pp term in Sbeta} are controlled by Lemma \ref{energie(1-pp)pp}, which is stated below.
\begin{align*}
\eqref{1-pp term in Sbeta},
\eqref{vielezeilenc2}
\leq
\mathcal{K}(\phi,A_t)( \laa \Psi, \widehat{n} \Psi \raa + N^{-1/6} \ln(N) )
\;.
\end{align*}

In total, we obtain, for any $\mu> d \geq 1$, the bound 
\begin{align*}
S_\beta (\Psi,\phi)
\leq &
\mathcal{K}(\phi,A_t) \left(\laa\Psi,\widehat{n}\Psi\raa+N^{-2\beta } \ln(N) +N^{-1/6 } \ln(N) 
\right)
\\
\tilde{S}_\mu (\Psi,\phi)
 \leq &
\mathcal{K}(\phi,A_t)\left( \laa\Psi,\widehat{n}\Psi\raa+N^{-1/6} \ln(N) \right)
\;.
\end{align*}

\item  Estimate of $M(\Psi, \phi)$.\\
First, we estimate \eqref{1 term in M}.
\begin{align*}
|\eqref{1 term in M}|\leq& 2|\laa\nabla_1q_1\Psi ,
\mathds{1}_{\overline{\mathcal{A}}^{(d)}_{1}}
\nabla_1p_1\Psi \raa|+2|\laa\nabla_1q_1\Psi ,
\nabla_1p_1\Psi \raa|
\\&\leq2\|\nabla_1q_1\Psi\|\;\|\mathds{1}_{\overline{\mathcal{A}}^{(d)}_{1}}\nabla_1p_1\|_{\text{op}}
+2|\laa\widehat{n}^{-1/2}q_1\Psi ,
\Delta_1 p_1\widehat{n}_{1}^{1/2}\Psi \raa|
\;.
\end{align*}
By Lemma \ref{hdetail}, we obtain 
$\|\mathds{1}_{\overline{\mathcal{A}}^{(d)}_{1}}\nabla_1p_1\|_{\text{op}} \leq 
C 
\| \nabla \phi \|_\infty
N^{1/2-d}$.
Furthermore, we use
$
\| \nabla_1 q_1\Psi\| \leq \| \nabla_1 \Psi\|+\| \nabla_1 p_1\Psi\|\leq \mathcal{K}(\phi, A_t)
$ (see also Lemma \ref{kinenergyboundedlemma}) and
$
|\laa\widehat{n}^{-1/2}q_1\Psi ,
\Delta_1 p_1\widehat{n}_{1}^{1/2}\Psi \raa|
\leq
\mathcal{K}(\phi, A_t)
\|\widehat{n}_{1}^{1/2}\Psi  \|
\|\widehat{n}^{1/2}\Psi  \|
\leq
\mathcal{K}(\phi, A_t)
(
\laa\Psi,\widehat{n}\Psi\raa+ N^{-1}
)
$.
Hence, for $d$ large enough,
\begin{align*}
|\eqref{1 term in M}|&\leq \mathcal{K}(\phi, A_t) (\laa\Psi,\widehat{n}\Psi\raa+N^{\frac{1}{2} -d} + N^{-1})
\leq
\mathcal{K}(\phi,A_t) (\laa\Psi,\widehat{n}\Psi\raa+ N^{-1})
\;.
\end{align*}

Line~\eqref{2 term in M} is estimated for $d$ large enough, noting that
$ \| \nabla_1 p_1 \Psi  \|^2=
\|\nabla\phi \|^2  \|  p_1 \Psi  \|^2$, by
 \begin{align*}\eqref{2 term in M}=&
 \|\mathds{1}_{\mathcal{A}^{(d)}_{1}}\nabla_1p_1\Psi  \|^2- \|\nabla\phi \|^2 \\
 \leq& |  \| \nabla_1 p_1 \Psi  \|^2- \|\nabla\phi \|^2 |
 +  \|\mathds{1}_{\overline{\mathcal{A}}^{(d)}_{1}}\nabla_1p_1\Psi  \|^2  \\
 \leq& C \left(  \| \nabla \varphi \|^2 \laa \Psi , q_1\Psi \raa +
\| \nabla \phi \|_\infty^2  N^{1 - 2d}  \right) 
 \\
 \leq &
 \mathcal{K}(\phi,A_t) \laa \Psi \widehat{n} , \Psi \raa
 \;.
\end{align*}
For line \eqref{3 term in M}, we use Lemma \ref{opdiff} to obtain
\begin{align*}
\eqref{3 term in M} \leq C\| A_t\|_{\infty}
\left(
\laa \Psi, \widehat{n} \Psi \raa + N^{-1/2}
\right)
\;.
\end{align*}
In total, we obtain
\begin{align*}
M(\Psi, \phi)
\leq
\mathcal{K}(\phi,A_t)
\left(
\laa \Psi, \widehat{n} \Psi \raa + N^{-1/2}
\right)
\;.
\end{align*}
\end{itemize}

Note that
\begin{align*}
 \|\mathds{1}_{\overline{\mathcal{B}}^{(d)}_{1}}
 \nabla_1q_1\Psi \|^2
 =&
  \|
  \mathds{1}_{\overline{\mathcal{A}}^{(d)}_{1}}
  \mathds{1}_{\overline{\mathcal{B}}^{(d)}_{1}}
 \nabla_1q_1\Psi \|^2
 + 
 \|
\mathds{1}_{\mathcal{A}^{(d)}_{1}} 
 \mathds{1}_{\overline{\mathcal{B}}^{(d)}_{1}}
 \nabla_1q_1\Psi \|^2
 \\
 \leq &
   \|
  \mathds{1}_{\overline{\mathcal{A}}^{(d)}_{1}}
  \mathds{1}_{\overline{\mathcal{B}}^{(d)}_{1}}
 \nabla_1
( 1-p_1)
 \Psi \|^2
 + 
 \|
\mathds{1}_{\mathcal{A}^{(d)}_{1}} 
 \nabla_1q_1\Psi \|^2
 \\
 \leq &
2    \|
  \mathds{1}_{\overline{\mathcal{A}}^{(d)}_{1}}
  \mathds{1}_{\overline{\mathcal{B}}^{(d)}_{1}}
 \nabla_1
 \Psi \|^2
+
2
\|
  \mathds{1}_{\overline{\mathcal{A}}^{(d)}_{1}}
  \mathds{1}_{\overline{\mathcal{B}}^{(d)}_{1}}
 \nabla_1 p_1
 \Psi \|^2
 + 
  \|
\mathds{1}_{\mathcal{A}^{(d)}_{1}} 
 \nabla_1q_1\Psi \|^2.
\end{align*}
Using
$
\|\mathds{1}_{\overline{\mathcal{B}}^{(d)}_{1}} \nabla_1 p_1\|_{\text{op}}\leq 
N \|\mathds{1}_{\overline{\mathcal{A}}^{(d)}_{1}}\nabla_1 p_1\|_{\text{op}}\leq 
 C \|\nabla \phi\|_{\infty}N^{3/2-d}\;,
$
\eqref{firsteneergybound}, \eqref{secondenergybound} and the bounds for
 $M( \Psi, \phi),S_\beta( \Psi, \phi)$ and $\tilde{S}_\mu( \Psi, \phi)$,
we then obtain the Lemma for $d$ large enough.

\end{proof}

\begin{lemma}\label{positiv V-w} 
$ $\\
\begin{itemize}
\item[(a)]
Let $R_\beta$ and $M_\beta$ be defined as in Lemma \ref{microscopic}.
Then, for  any $\Psi\in H^1(\mathbb{R}^{2N}, \mathbb{C})$
$$\|\mathds{1}_{|x_1-x_2|\leq  R_{\beta}}\nabla_1\Psi\|^2+\frac{1}{2}\laa\Psi,
(V_{N}-M_{\beta})(x_1-x_2)\Psi\raa\geq0\;.$$
\item[(b)] 
Let $M_\beta$ be defined as in Lemma \ref{microscopic}. 
Let  $\Psi\in
L_s^2(\mathbb{R}^{2N}, \mathbb{C}) \cap H^1(\mathbb{R}^{2N}, \mathbb{C})$.
Then, for sufficiently large $N$ and for $\beta>d$,
\begin{align*}
\|\mathds{1}_{\mathcal{B}^{(d)}_{1}}\mathds{1}_{\overline{\mathcal{A}}^{(d)}_{1}}\nabla_1\Psi \|^2
+
\frac{1}{2}
\laa\Psi ,\sum_{j\neq
1}\mathds{1}_{\mathcal{B}^{(d)}_{1}}\left(V_N-M_{\beta}\right)(x_1-x_j)\Psi \raa
\geq 0
\;.
\end{align*}
\end{itemize}

\end{lemma}
\begin{proof}
\begin{itemize}
\item[(a)]
We first show nonnegativity of the one-particle operator 
$H^{Z_n} : H^2 (\mathbb{R}^2, \mathbb{C}) \rightarrow L^ 2( \mathbb{R}^2, \mathbb{C})$  given by
$$H^{Z_n}=-\Delta+ \frac{1}{2} \sum_{z_k\in Z_n} (V_{N}(\cdot-z_k)-M_{\beta}(\cdot-z_k))$$
for any $n\in \mathbb{N}$ and any $n$-elemental subset $Z_n\subset\mathbb{R}^2$ which is
 such that the supports of the potentials
$M_{\beta}(\cdot-z_k)$ are pairwise disjoint for any two $z_k\in
Z_n$.

Since $f_{\beta}(\cdot-z_k)$ is the the zero energy scattering state of the potential
$1/2 V_{N}(\cdot-z_k)- 1/2 W_{\beta}(\cdot-z_k)$, it follows that
$$F^{Z_n}_{\beta}=\prod_{z_k\in Z_n}f_{\beta}(\cdot-z_k)\;.$$ fulfills
$H^{Z_n}  F^{Z_n}_{\beta}=0$
for any such $Z_n$.
By construction $f_{\beta}$ is a positive function, so is $F^{Z_n}_{\beta}$.
Since $\frac{1}{2} \sum_{z_k\in Z_n} (V_{N}(\cdot-z_k)-M_{\beta}(\cdot-z_k)) \in L^\infty (\mathbb{R}^2,\mathbb{C})$, this potential is a infinitesimal perturbation of $-\Delta$, thus 
$\sigma_{\text{ess}}( H^{Z_n} ) =[0, \infty )$. 
Assume now that $H^{Z_n}$ is not nonnegative. Then, there exists a ground state $\Psi_G\in H^2( \mathbb{R}^2,\mathbb{C})$ of $H^{Z_n}$ of
negative energy $E<0$. The phase of the ground state can be chosen such that  the ground state is real and positive (see e.g. \cite{teschl}, Theorem 10.12.).
 Since such a ground state of negative energy decays exponentially, that is
$\Psi_G (x) \leq C_1 e^{-C_2 |x|}, C_1,C_2 >0$ , 
the following scalar product is well defined (although $F^{Z_n}_{\beta} \notin  L^2( \mathbb{R}^2, \mathbb{C})$).
\begin{align}
\label{contra}\langle F^{Z_n}_{\beta},H^{Z_n}\Psi_G\rangle=\langle F^{Z_n}_{\beta},E\Psi_G\rangle<0
\;.
\end{align}

On the other hand we have since $F^{X_n}_{\beta_1,\beta}$ is the zero energy scattering state 
$$\langle
F^{Z_n}_{\beta},H^{Z_n}\Psi_G\rangle=\langle H^{Z_n}F^{Z_n}_{\beta},\Psi_G\rangle=0\;.$$
This contradicts (\ref{contra}) and the nonnegativity of $H^{Z_n}$ follows.

Now, assume that there exists a $\psi\in H^2 (\mathbb{R}^2,\mathbb{C}) $ such that the quadratic form
$$
Q (\psi)=
\|\mathds{1}_{|\cdot|\leq  R_{\beta}}\nabla\psi\|^2+
\frac{1}{2}\langle\psi,(V_{N}(\cdot)-M_{\beta}(\cdot))\psi\rangle<0\;.$$ 
Since $V_{\beta_1}$ and $M_{\beta_1}$ are spherically symmetric we can assume that $\psi$ is spherically symmetric. Subsituting $ \psi \rightarrow a \psi ,\; a \in \mathbb{R}$ , we can furthermore assume that, for all $|x| = R_\beta$, $\psi(x) =
1- \epsilon$ for $\epsilon>0$.

Define $\tilde{\psi}$  such that
$ \tilde{\psi}(x)=\psi(x) \text{ for } |x| \leq R_{\beta}$ and $\tilde{\psi}(x)=1$ for $|x|> R_{\beta}+ \epsilon$ and $\epsilon>0$. Furthermore, $\tilde{\psi}$ can be constructed such that
$ \| \mathds{1}_{|\cdot| \geq R_\beta} \nabla \tilde{\psi}\|^2 \leq C (\epsilon+ \epsilon^ 2)$.

Then $Q ( \tilde{\psi})= Q (\psi)<0$ holds, because the operator associated with the quadratic form is supported inside the ball $B_0 (R_{\beta})$.

Using $\tilde{\psi}$, we can construct a set of points $Z_n$ and a $\chi\in H^2(\mathbb{R}^2, \mathbb{C})$ such that $\langle\chi,H^{Z_n}\chi\rangle<0$, contradicting to nonnegativity of $H^{Z_n}$.

For $R>1$ let
$$\xi_R(x)=\left\{
         \begin{array}{ll}
           R^2/x^2, & \hbox{for $|x|>R$;} \\
           1, & \hbox{else.}
         \end{array}
       \right.
$$
Let now $Z_n$ be a subset $Z_n\subset\mathbb{R}^2$ with $|Z_n|=n$ which is such that the supports of the potentials
$M_{\beta}(\cdot-z_k)$ lie within the Ball around zero with radius $R$ and are pairwise disjoint for any two $z_k\in Z_n$. Since we are in two dimensions we can choose a $n$ which is of order $R^2$.

Let now $\chi_R(x)=\xi_R (x) \prod_{z_k\in Z_n} \tilde{\psi}(x-z_k)$. By construction, there exists a $D = \mathcal{O}(1)$ such that
$ \chi_R(x)= \tilde{\psi}(x-z_k)$ for $ |x-z_k| \leq D $.
From this, we obtain 
\begin{align*}
\langle\chi_R,H^{Z_n}\chi_R\rangle
=&
\| \nabla \chi_R \|^2+
n\frac{1}{2}\langle\psi,(V_{N}(\cdot)-M_{\beta}(\cdot))\psi\rangle
\\
=&
n Q(\psi)+
\sum_{z_k \in Z_n}
\| \mathds{1}_{|\cdot-z_k|\geq R_\beta} \nabla \chi_R \|^2
\\
\leq &
n Q(\psi) +C n (\epsilon +\epsilon^2) + \| \nabla \xi_R  \|^2
\\
=&
n Q(\psi) +C n (\epsilon+ \epsilon^2) +C
\;.
\end{align*}
Choosing $R$ and hence $n$ large enough and $\epsilon$ small, we can find a $Z_n$
such that
$\langle \chi_R,H^{Z_n}\chi_R\rangle$ is negative, contradicting nonnegativity of $H^{Z_n}$.
\\
Now, we can prove that
\begin{align}
\label{poslemmac}
\|\mathds{1}_{|x_1-x_2|\leq  R_{\beta_1}}\nabla_1\Psi\|^2+\frac{1}{2}\laa\Psi,
(V_{N}-M_{\beta})(x_1-x_2)\Psi\raa\geq 0
\;.
\end{align}

holds for  any $\Psi\in H^2(\mathbb{R}^{2N},\mathbb{C})$.
Using the coordinate transformation $\tilde{x}_1= x_1-x_2 \; , \tilde{x}_i=x_i \; \forall i \geq 2$, we have $\nabla_{x_1}= \nabla_{\tilde{x}_1}$. Thus \eqref{poslemmac} is equivalent to $
\|\mathds{1}_{|x_1|\leq  R_{\beta_1}}\nabla_1\Psi\|^2+\frac{1}{2}\laa\Psi,
(V_{N}-M_{\beta})(x_1)\Psi\raa
 \geq 0$ $\forall \Psi \in H^2(\mathbb{R}^{2N},\mathbb{C})$
which follows directly from $Q(\psi) \geq 0$ for all $\psi\in H^2(\mathbb{R}^{2},\mathbb{C})$.
By a standard density argument, we can conclude that $Q(\Psi) \geq 0$ $\forall \Psi \in H^1(\mathbb{R}^{2N},\mathbb{C})$.
\item[(b)]
Define 
$c_k= \lbrace
(x_1,\dots,x_N) \in \mathbb{R}^{2N}| |x_1-x_k|\leq R_\beta \rbrace$ and 
$\mathcal{C}_1= \cup_{k=2}^N c_k$.
For  $(x_1, \dots, x_N) \in \mathcal{B}^{(d)}_{1}$ it holds that
 $|x_i-x_j| \geq  N^{-d}$ for $2 \leq i,j \leq N$.
Let $  \beta >d$. Assume that $N^{-d} > 2 R_\beta$, which hold for $N$ sufficiently large, since $R_\beta \leq CN^{-\beta}$.
Then, it follows that, for $i \neq j$, 
 $ \left( c_i \cap \mathcal{B}^{(d)}_{1} \right)
  \cap
  \left( c_j \cap \mathcal{B}^{(d)}_{1} \right)= \emptyset$.
 Under the same conditions, we also have $\mathds{1}_{\overline{\mathcal{A}}^{(d)}_{1}} \geq \mathds{1}_{\mathcal{C}_{1}}$.   
 Therefore
 \begin{align*}
& 
\mathds{1}_{\overline{\mathcal{A}}^{(d)}_{1}} \mathds{1}_{\mathcal{B}^{(d)}_{1}}
\geq 
\mathds{1}_{\mathcal{C}_{1}} \mathds{1}_{\mathcal{B}^{(d)}_{1}}
 = \mathds{1}_{\mathcal{C}_{1} \cap \mathcal{B}^{(d)}_{1} }
 = \mathds{1}_{ \cup_{k=2}^N  \left( c_k \cap \mathcal{B}^{(d)}_{1} \right) }
 = \sum_{k=2}^N \mathds{1}_{c_k \cap \mathcal{B}^{(d)}_{1} }
 =\mathds{1}_{ \mathcal{B}^{(d)}_{1} } 
 \sum_{k=2}^N \mathds{1}_{c_k  }
 \;.
 \end{align*}
 Note that $\mathds{1}_{\mathcal{B}^{(d)}_{1}}$  depends only on $x_2, \dots, x_N$. By this
 \begin{align*}
\| \mathds{1}_{\overline{\mathcal{A}}^{(d)}_{1}} \mathds{1}_{\mathcal{B}^{(d)}_{1}}
\nabla_1 \Psi \|^2
\geq
\sum_{k=2}^N
\|
 \mathds{1}_{c_k  }
\nabla_1 \mathds{1}_{ \mathcal{B}^{(d)}_{1} } \Psi \|^2
=
(N-1)
\|
 \mathds{1}_{|x_1-x_2| \leq  R_\beta }
\nabla_1 \mathds{1}_{ \mathcal{B}^{(d)}_{1} } \Psi \|^2
\;.
 \end{align*}
This yields
 \begin{align*}
 \eqref{vielezeilenb}
 \geq
 (N-1)
 \left(
 \|
 \mathds{1}_{|x_1-x_2| \leq  R_\beta }
\nabla_1 \mathds{1}_{ \mathcal{B}^{(d)}_{1} } \Psi \|^2
+
\frac{1}{2}\laa \mathds{1}_{ \mathcal{B}^{(d)}_{1} } \Psi,
(V_{N}-M_{\beta})(x_1-x_2)\mathds{1}_{ \mathcal{B}^{(d)}_{1} } \Psi\raa
 \right)
 \geq 0
 \;.
 \end{align*}
  where the last inequality follows from (a)
\end{itemize}
\end{proof}

\begin{lemma}\label{energie(1-pp)pp}
Let $W_\beta \in \mathcal{V}_\beta$. 
Let 
$
\Psi \in 
 L^2_{s}(\mathbb{R}^{2N}, \mathbb{C})
 \cap
  H^1(\mathbb{R}^{2N}, \mathbb{C})
 $
 and $\|\nabla_1\Psi\|$ be bounded uniformly in $N$.
Let $d $ in Definition \ref{hdetail} of $\mathds{1}_{{\mathcal{B}_1^{(d)}}} $ sufficiently large.
Let $\Gamma \in \lbrace \Psi, \mathds{1}_{{\mathcal{B}_1^{(d)}}}  \Psi \rbrace$.
Then, for all $\beta >0$,
\begin{enumerate}
\item   
\begin{align*}
&N\left|\laa \Gamma, q_1p_2
W_\beta(x_1-x_2) p_1p_2 \Gamma\raa\right|\leq   C \|\phi \|_\infty^2
 \laa \Psi, \hat{n} \Psi \raa \;.
\end{align*}

\item  \begin{align*}
&N|\laa \Gamma ,
p_1p_2 W_\beta(x_1-x_2) q_1q_2 \Gamma\raa|  \leq
\mathcal{K}(\phi,A_t)
 \left(
\laa\Psi, \widehat{n} \Psi \raa
+
N^{-1/6} \ln(N)
\right)
\;.
\end{align*}

\item[(c)]
 \begin{align*}
&N|\laa \Gamma ,
(1-p_1p_2) W_\beta(x_1-x_2) p_1p_2 \Gamma\raa|  
\leq
\mathcal{K}(\phi,A_t)
 \left(
\laa\Psi, \widehat{n} \Psi \raa
+
N^{-1/6} \ln(N)
\right)
\;.
\end{align*}
\end{enumerate}
\end{lemma}

\begin{proof}

\begin{enumerate}
\item
Let first $\Gamma= \mathds{1}_{{\mathcal{B}_1^{(d)}}}  \Psi$.
Then,
\begin{align}
&N\left|\laa\mathds{1}_{{\mathcal{B}_1^{(d)}}}  \Psi, q_1p_2
W_\beta(x_1-x_2) p_1p_2 \mathds{1}_{{\mathcal{B}_1^{(d)}}}  \Psi\raa\right| 
\nonumber
 \\
\leq& 
\label{muchopequeno}
 N\left|\laa \mathds{1}_{{\overline{\mathcal{B}}^{(d)}_{1}}} \Psi, q_1p_2
W_\beta(x_1-x_2) p_1p_2 \mathds{1}_{{\mathcal{B}_1^{(d)}}}  \Psi\raa\right|  \\
\label{pequeno}
+& N\left|\laa \Psi, q_1p_2
W_\beta(x_1-x_2) p_1p_2 \mathds{1}_{{\mathcal{B}_1^{(d)}}}  \Psi\raa\right|  
\;.
\end{align}
Using Lemma \ref{propo}
together with $\|p_2 W_\beta (x_1-x_2) p_2 \|_{\text{op}} \leq \|\phi\|_\infty^2 \|W_\beta\|_1$,
 the first line can be bounded, for any $\epsilon>0$, by
\begin{align}
&
\eqref{muchopequeno}
\leq  \mathcal{K}(\phi, A_t) N  \| \mathds{1}_{{\overline{\mathcal{B}}^{(d)}_{1}}} \Psi \|
\| W_{\beta} \|_{1} \leq \mathcal{K}(\phi, A_t) N^{1- d + \epsilon}.
\end{align}
The second term is bounded by

\begin{align*}
\eqref{pequeno}
=&  N\left|\laa \sqrt{W_\beta(x_1-x_2)} q_1p_2 (\hat{n})^{-\frac{1}{2}}   \Psi, 
\sqrt{W_\beta(x_1-x_2)} p_1p_2 \hat{n}_{1}^{\frac{1}{2}} \mathds{1}_{{\mathcal{B}^{(d)}}_{1}} \Psi\raa\right|  \\
\leq& CN \|\sqrt{W_\beta(x_1-x_2)} p_2 \|_{\text{op}}^2 
\left(  \|q_1 (\hat{n})^{-\frac{1}{2}}  \Psi \|^2 
+ \| \hat{n}_1^{\frac{1}{2}} \mathds{1}_{{\mathcal{B}^{(d)}}_{1}} \Psi \|^2
\right)  \\
\leq& CN \|\sqrt{W_\beta(x_1-x_2)} p_2 \|_{\text{op}}^2 
\left(  \laa \Psi, \hat{n} \Psi \raa 
+ \| \hat{n}_1^{\frac{1}{2}} \Psi \|^2
+ \| \hat{n}_1^{\frac{1}{2}} \mathds{1}_{{\overline{\mathcal{B}}^{(d)}_{1}}} \Psi \|^2
\right)  \\
\leq& CN \| W_\beta \|_1 \| \phi \|_\infty^2 
\left( \laa \Psi, \hat{n} \Psi \raa + \|  \mathds{1}_{{\overline{\mathcal{B}}^{(d)}_{1}}} \Psi \|^2
\right)  \\
\leq& C \| \phi \|_\infty^2  \left( \laa \Psi, \hat{n} \Psi \raa + N^{1 - d + \epsilon}  \right)
.
\end{align*}
Choosing $d$ large enough, $N^{1 - d + \epsilon}$ is smaller than $\laa \Psi, \hat{n} \Psi \raa$. This yields  (a) in the case
$\Gamma= \mathds{1}_{{\mathcal{B}_1^{(d)}}}  \Psi$.
The inequality (a) can be proven  analogously for $\Gamma= \Psi$.
\item
Let $\Gamma= \mathds{1}_{{\mathcal{B}_1^{(d)}}}  \Psi$.
 We first consider (b) for potentials with $\beta< 1/4$.
We have to estimate 
\begin{align}
&N|\laa \mathds{1}_{{\mathcal{B}_1^{(d)}}} \Psi ,p_1 p_2 W_\beta(x_1-x_2) q_1 q_2
\mathds{1}_{{\mathcal{B}^{(d)}_{1}}}\Psi
\raa|
\leq
N|\laa\Psi ,p_1 p_2 W_\beta(x_1-x_2) q_1 q_2
\Psi\raa|
\nonumber \\
+&
N|\laa \mathds{1}_{\overline{\mathcal{B}}^{(d)}_{1}}\Psi ,p_1 p_2 W_\beta(x_1-x_2) q_1 q_2
\Psi\raa|
+
N|\laa\Psi ,p_1 p_2 W_\beta(x_1-x_2) q_1 q_2
\mathds{1}_{\overline{\mathcal{B}^{(d)}_{1}}}\Psi\raa|
\nonumber \\
+&
\label{betaklein}
N|\laa \mathds{1}_{\overline{\mathcal{B}}^{(d)}_{1}}\Psi ,p_1 p_2 W_\beta(x_1-x_2) q_1 q_2
\mathds{1}_{\overline{\mathcal{B}^{(d)}_{1}}}\Psi\raa|
\nonumber
\\
\leq & 
N|\laa \Psi ,p_1 p_2 W_\beta(x_1-x_2) q_1 q_2
\Psi\raa|
\\
+&
\label{muchomuchoklein}
C N \|\mathds{1}_{\overline{\mathcal{B}}^{(d)}_{1}}\Psi\| \|W_\beta \|_\infty
\;.
\end{align}
The last term is bounded, for any $\epsilon>0$, by
\begin{align*}
\eqref{muchomuchoklein} \leq C  N N^{1-d+ \epsilon} N^{-1+2 \beta}
\leq
N^{-2} \; ,
\end{align*}
where the last inequality holds choosing $d$ large enough.

Using Lemma \ref{kombinatorik} (c) and Lemma \ref{trick} with $O_{1,2}=q_2
W_\beta(x_1-x_2) p_2$, $\Omega=N^{-1/2}q_1\Psi$ and $\chi=N^{1/2}p_1\Psi$ we get
 \begin{align*}
 \eqref{betaklein}
&\leq \left\|q_1\Psi\right\|^2+
N^2\big|\laa q_2\,\Psi,p_1\sqrt{W_\beta}(x_1-x_2)
p_3
\sqrt{W_\beta}(x_1-x_3)\\&\hspace{3.7cm}\sqrt{W_\beta}(x_1-x_2)p_2\sqrt{W_\beta}(x_1-x_3)p_1
q_3\,\Psi\raa\big|
\\&+N^2(N-1)^{-1}\|q_2
W_\beta(x_1-x_2) p_2p_1\Psi\|^2&
\\&\leq \left\|q_1\Psi\right\|^2+N^2\|\sqrt{W_\beta}(x_1-x_2)p_1\|_{\text{op}}^4\;\|
q_2\,\Psi\|^2
\\&+C N\|
W_\beta(x_1-x_2) p_2\|_{\text{op}}^2
\;.
 \end{align*}
 With  Lemma \ref{kombinatorik} (e) we get the bound
 \begin{align*}
&
 \eqref{betaklein}
 \leq  \|q_1\Psi\|^2+
 N^2\|\phi  \|_\infty^4\|W_\beta\|_1^2\;\|
q_1\Psi\|^2
\\&+C N\|W_\beta\|^2\|\phi  \|_\infty^2
\;.
 \end{align*}
Note, that $\|W_\beta\|_1\leq CN^{-1}$, $\|W_\beta\|^2\leq CN^{-2+2\beta}$
Hence
\begin{align*}
\eqref{betaklein}\leq C\left(\laa\Psi,q_1\Psi\raa+
\mathcal{K}(\phi)
N^{-1+2\beta}\right)
\;.
\end{align*}
Note that, for $\beta<1/4$, $ N^{-1+2\beta} \leq N^ {-1/6} \ln(N)$.
Using the same bounds for $\Gamma= \Psi$, we obtain (b) for the case $\beta<1/4$.

\item[b)] for $1/4\leq \beta$:\\
We use $U_{\beta_1,\beta}$ from Definition \ref{udef} for some $0<\beta_1<1/4$.

 $\potdiff^\phi_\beta(x_1,x_2)-W_{\beta}+U_{\beta_1,\beta}$ has the form of $\potdiff^\phi_{\beta_1}(x_1,x_2)$ which has been controlled above. It is left to control $$
 N\left|\laa \mathds{1}_{{\mathcal{B}_1^{(d)}}} \Psi, p_1p_2\left(W_{\beta}(x_1-x_2)-U_{\beta_1,\beta}(x_1-x_2)\right)q_1q_2 \mathds{1}_{{\mathcal{B}_1^{(d)}}} \Psi\raa\right|\;.
 $$
Let  $\Delta h_{\beta_1,\beta}=W_{\beta}-U_{\beta_1,\beta}$. Integrating by parts and using that\\ $\nabla_1 h_{\beta_1,\beta}(x_1-x_2)=-\nabla_2 h_{\beta_1,\beta}(x_1-x_2)$ gives
 \begin{align}
\nonumber
&N\left|\laa \mathds{1}_{{\mathcal{B}_1^{(d)}}} \Psi, p_1p_2\left(W_{\beta}(x_1-x_2)-U_{\beta_1,\beta}(x_1-x_2)\right)q_1q_2  \mathds{1}_{{\mathcal{B}_1^{(d)}}} \Psi\raa\right|
\\\label{split1}&\hspace{2cm}=N\left|\laa\nabla_1p_1 \mathds{1}_{{\mathcal{B}_1^{(d)}}} \Psi, p_2\nabla_2 h_{\beta_1,\beta}(x_1-x_2)q_1q_2 \mathds{1}_{{\mathcal{B}_1^{(d)}}} \Psi\raa\right|
\\\label{split2}
&\hspace{2.4cm}+N\left|\laa \mathds{1}_{{\mathcal{B}_1^{(d)}}} \Psi, p_1p_2\nabla_2 h_{\beta_1,\beta}(x_1-x_2)\nabla_1q_1q_2 \mathds{1}_{{\mathcal{B}_1^{(d)}}} \Psi\raa\right|
\;.
\end{align}
Let $ (a_1,b_1)=(q_1,\nabla p_1)$ or $ (a_1,b_1)=( \nabla q_1,p_1)$. Then, both terms can be estimated as follows:
\\
We use Lemma \ref{trick} with $\Omega=N^{-\eta/2} a_1\mathds{1}_{{\mathcal{B}_1^{(d)}}} \Psi$,
 $O_{1,2}=N^{1+\eta/2}q_2\nabla_2 h_{\beta_1,\beta}(x_1-x_2)p_2$ and $\chi= b_1 \mathds{1}_{{\mathcal{B}_1^{(d)}}} \Psi$. We choose $ \eta < 2 \beta_1$.

\begin{align}\label{nulltesumme}
&N\left|\laa \mathds{1}_{{\mathcal{B}_1^{(d)}}} \Psi, a_1p_2\nabla_2 h_{\beta_1,\beta}(x_1-x_2)b_1q_2 \mathds{1}_{{\mathcal{B}_1^{(d)}}} \Psi\raa\right|
\nonumber
\\
&\leq N^{-\eta}\|a_1 \mathds{1}_{{\mathcal{B}_1^{(d)}}} \Psi\|^2
\\&+
\label{erstesumme}\frac{N^{2+\eta}}{N-1}\|q_2\nabla_2 h_{\beta_1,\beta}(x_1-x_2)b_1 p_2 \mathds{1}_{{\mathcal{B}_1^{(d)}}} \Psi\|^2
 \\\label{zweitesumme}&+N^{2+\eta}\left|\laa \mathds{1}_{{\mathcal{B}_1^{(d)}}} \Psi, b_1 p_2 q_3\nabla_2 h_{\beta_1,\beta}(x_1-x_2)\nabla_3 h_{\beta_1,\beta}
(x_1-x_3)b_1 q_2p_3\mathds{1}_{{\mathcal{B}_1^{(d)}}} \Psi\raa\right|^{1/2}
 \;.
 \end{align}
We obtain (note that $\mathds{1}_{{\mathcal{B}_1^{(d)}}} $ does not depend on $x_1$)
$$(\ref{nulltesumme})\leq N^{-\eta}\|a_1 \mathds{1}_{{\mathcal{B}_1^{(d)}}} \Psi\|^2
=
 N^{-\eta}\| \mathds{1}_{{\mathcal{B}_1^{(d)}}} a_1\Psi\|^2
\leq  
\mathcal{K}(\phi, A_t)
N^{-\eta}\;.$$
since both $ \| \nabla q_1 \Psi\|$ and $ \| q_1 \Psi\|$ are bounded uniformly in $N$.
Since $q_2$ is a projector it follows that
\begin{align*}
 (\ref{erstesumme})
 \leq&\frac{N^{2+\eta}}{N-1}\|\nabla_2 h_{\beta_1,\beta}(x_1-x_2)p_2\|_{\text{op}}^2
 \|b_1 \mathds{1}_{{\mathcal{B}_1^{(d)}}} \Psi\|^2
\leq 
C
\frac{N^{2+\eta}}{N-1}
\|\phi\|_\infty ^2
\|\nabla h_{\beta_1,\beta}\|^2
\|b_1 \mathds{1}_{{\mathcal{B}_1^{(d)}}} \Psi\|^2
\\
\leq&
\mathcal{K}(\phi, A_t)
N^{\eta-1}\ln(N)
\|\phi\|_\infty ^2
 \;,
 \end{align*}
where we used Lemma \ref{ulemma} in the last step.

Next, we estimate 
\begin{align}
(\ref{zweitesumme})\leq& 
N^{2+\eta}\| p_2
\nabla_2 h_{\beta_1,\beta}(x_1-x_2)b_1 q_2
\mathds{1}_{{\mathcal{B}_1^{(d)}}} \Psi\|^2\
\nonumber
\\
\leq&
\label{ersterterm}
2 N^{2+\eta}\| p_2
\nabla_2 h_{\beta_1,\beta}(x_1-x_2)b_1 q_2
\mathds{1}_{\overline{\mathcal{B}}^{(d)}_{1}}\Psi\|^2\
\\
+&
\label{zweiterterm}
2 N^{2+\eta}\| p_2
\nabla_2 h_{\beta_1,\beta}(x_1-x_2)b_1 q_2
\Psi\|^2
\;.
\end{align}
The first term can be estimated as 
\begin{align*}
\eqref{ersterterm} 
\leq &
C N^{2+ \eta} \|\nabla_2 h_{\beta_1,\beta}(x_1-x_2)b_1\|_{\text{op}}^2 
\|\mathds{1}_{\overline{\mathcal{B}}^{(d)}_{1}}\Psi\|^2
\\
\leq &
C N^{2+ \eta} \|\nabla_2 h_{\beta_1,\beta}\|^2 (\| \phi \|^2_\infty+\| \nabla \phi \|^2_\infty)
\|\mathds{1}_{\overline{\mathcal{B}}^{(d)}_{1}}\Psi\|^2
\\
\leq &
\mathcal{K}(\phi, A_t)
N^{2+ \eta} N^{-2} \ln(N) N^{2-2d+ 2 \epsilon}
=
\mathcal{K}(\phi, A_t)
 N^{2-2d+2 \epsilon + \eta} \ln(N)
 \;,
\end{align*}
for any $\epsilon >0$.
For $d$ large enough, this term is subleading.
The last term can be estimated as
\begin{align*}
\eqref{zweiterterm}
\leq&
2N^{2+\eta}\| p_2
 h_{\beta_1,\beta}(x_1-x_2)b_1 \nabla_2q_2
\Psi\|^2
\nonumber
\\+&
2N^{2+\eta}\| |\varphi(x_2)\rangle \langle \nabla \varphi (x_2)|
 h_{\beta_1,\beta}(x_1-x_2)b_1 q_2
\Psi\|^2\
\nonumber
\\
\leq &
C N^{2+\eta}\| p_2
 h_{\beta_1,\beta}(x_1-x_2)
 \|_{\text{op}}^2
 \| b_1\nabla_2q_2
\Psi\|^2\
\nonumber
\\
+&
C N^{2+\eta}
\| |\varphi(x_2)\rangle \langle \nabla \varphi(x_2)|
 h_{\beta_1,\beta}(x_1-x_2) \|_{\text{op}}^2
  \| b_1 q_2
\Psi\|^2\
\nonumber
\\
\leq &
C N^{2+\eta} \left( \|\nabla \varphi \|_\infty^2+ \| \varphi \|_\infty^2 \right)
\| h_{\beta_1,\beta} \|^2 (1+ \| \nabla \phi \|^2)
\nonumber
\\ \leq&
\mathcal{K}(\phi, A_t)
N^{\eta- 2\beta_1} \ln(N)^2
\;.
\end{align*}
Combining both estimates  we obtain, for any $\beta>1$,
\begin{align*}
&N\left|\laa \mathds{1}_{{\mathcal{B}_1^{(d)}}} \Psi, p_1p_2W_{\beta}(x_1-x_2)q_1q_2  \mathds{1}_{{\mathcal{B}^{(d)}}_{1}}\Psi\raa\right|
\\
&\leq
\inf_{ \eta>0}
\inf_{ 0<\mu <1/4}
\left(
 \mathcal{K}(\phi, A_t)
 \left(
\laa\Psi, \widehat{n} \Psi \raa
+
N^{-1 + 2 \mu}
+
N^{- \eta}
+
N^{\eta-1} \ln(N)
+
N^{\eta -2 \mu} \ln(N)
\right)
\right)
\\
&\leq
 \mathcal{K}(\phi, A_t)
 \left(
\laa\Psi, \widehat{n} \Psi \raa
+
N^{-1/6} \ln(N)
\right)
\;.
\end{align*}
where the last inequality comes from choosing $\eta=1/3$ and $\mu=1/4$.
For $\Gamma=  \Psi$, (b) can be estimated the same way, yielding the same bound.
\item[(c)]
This follows from (a) and (b), using that $1-p_1p_2= q_1q_2+p_1q_2+q_1p_2$.
\end{enumerate}
\end{proof}

\section*{Acknowledgments}
We are grateful to David Mitrouskas for many valuable discussions and would like to thank Serena Cenatiempo for helpful discussions.
M.J. gratefully acknowledges financial support by the German National Academic Foundation.
N.L. gratefully acknowledges financial support by the Cusanuswerk and the European Research Council (ERC) under the European Union's Horizon 2020 research and innovation programme (grant agreement No 694227).


\begin{thebibliography}{}

\bibitem{Bagnato}
V.  Bagnato and D.Kleppner
\emph{Bose-Einstein condensation in low-dimensional traps},
Phys. Rev. A 44, 7439 (1991).


\bibitem{abdallah}
N. Ben Abdallah, F. M\'ehats, and O. Pinaud,
\emph{
  Adiabatic
approximation of
the Schr\"odinger Poisson system with a partial confinement},
SIAM Journal on
Mathematical Analysis, 36(3):986â1013 (2005).




\bibitem{SchleinNorm}
C. Boccato, S. Cenatiempo and B. Schlein,
\emph{Quantum many-body fluctuations around nonlinear Schr\"odinger dynamics},
arXiv:1509.03837 (2015).


\bibitem{benedikter}  N. Benedikter, G. De Oliveira and B. Schlein, \emph{Quantitative derivation of the Gross-Pitaevskii equation}, Comm. Pur. Appl. Math.  08 (2012)

\bibitem{carles}
R. Carles and J. Drumond Silva,
\emph{
Large time behavior in nonlinear Schrodinger equation with time dependent potential}, Communications in Mathematical Sciences, International Press, 2015, 13 (2), pp.443-460.


\bibitem{chen2d}
X. Chen and J. Holmer, \emph{ The Rigorous Derivation of the 2D Cubic Focusing NLS from Quantum Many-body Evolution}, arXiv:1508.07675 (2015).


\bibitem{cherny}
A. Yu. Cherny and A. A. Shanenko
\emph{Dilute Bose gas in two dimensions: density expansions and the Gross-Pitaevskii equation
}, PhysRevE.64.027105 (2001).


\bibitem{erdos1} L. Erd\"os, B. Schlein and  H.-T. Yau, \emph{Derivation of the Gross-Pitaevskii Hierarchy for the Dynamics
of Bose-Einstein Condensate}, Comm.\ Pure Appl.\ Math.\ \textbf{59}
, no. 12, 1659--1741 (2006).

\bibitem{erdos2} L. Erd\"os, B. Schlein and  H.-T. Yau, \emph{ Derivation of the cubic non-linear Schr\"odinger equation from
quantum dynamics of many-body systems}, Invent. Math. 167
, 515--614 (2007).


\bibitem{erdos3} L. Erd\"os, B. Schlein and  H.-T. Yau, \emph{ Derivation of the Gross-Pitaevskii  equation for the dynamics of Bose-Einstein condensate}, Ann. of Math. (2) 172
, no. 1, 291--370 (2010).



\bibitem{erdos4} L. Erd\"os, B. Schlein and  H.-T. Yau, \emph{Rigorous derivation of the Gross-Pitaevskii  equation with a larger interaction potential}, J. Amer. Math. Soc. 22
, no. 4, 1099--1156 (2009).


\bibitem{experiment}
A. GÃ¶rlitz, J. M. Vogels, A. E. Leanhardt, C. Raman, T. L. Gustavson, J. R. Abo-Shaeer, A. P. Chikkatur, S. Gupta, S. Inouye, T. Rosenband, and W. Ketterle
\emph{Realization of Bose-Einstein Condensates in Lower Dimensions},
Phys. Rev. Lett. 87, 130402 (2001)


\bibitem{schlein2d}
K. Kirkpatrick, B. Schlein and Gigliola Staffilani, \emph{Derivation of the two-dimensional nonlinear Schr\"odinger equation from many body quantum dynamics}, American Journal of Mathematics 133, no. 1 (2011): 91-130.

\bibitem{keler}
J. v. Keler,
\emph{ Mean Field Limits in Strongly Confined Systems},
arXiv:1412.3437 (2014)

\bibitem{teufel}
J. v. Keler and S. Teufel,
\emph{The NLS Limit for Bosons in a Quantum Waveguide},
Annales Henri Poincar{\'e}, p.1-40 (2016).

\bibitem{ketterle}
W. Ketterle, \emph{Nobel lecture: When atoms behave as waves: Bose-Einstein condensation and the atom laser},
Rev. Mod. Phys. 74, no.~4, 1131--1151 (2002).








\bibitem{knowles}A. Knowles and P.~Pickl, \emph{Mean-Field Dynamics: Singular Potentials and Rate of Convergence},  Comm. Math. Phys.  298, 101-139 (2010).



\bibitem{lewin}
M. Lewin, Phan Th\'anh Nam, N. Rougerie, 
\emph{A note on 2D focusing many-boson systems
},  arXiv:1509.09045 (2015)

\bibitem{liebanalysis}
E. Lieb and M. Loss
\emph{Analysis}, Graduate studies in mathematics, American Mathematical Society (2010)

\bibitem{lieb100bec}
E. Lieb and R.Seiringer
\emph{Proof of Bose-Einstein condensation for dilute trapped gases.}, Phys Rev Lett. vol. 88, 170409 (2002)


\bibitem{lieb}
E. Lieb and R.Seiringer,
\emph{The Stability of Matter in Quantum Mechanics
}, Cambridge University Press, Cambridge (2010).


\bibitem{lssy}
E.H Lieb, R. Seiringer, J.P. Solovej and J. Yngvason, \emph{The
mathematics of the Bose gas and its condensation}, Oberwolfach
Seminars, {\bf 34} Birkhauser Verlag, Basel, (2005).

\bibitem{lsy}
E.H Lieb, R. Seiringer and J. Yngvason, \emph{ A Rigorous Derivation of the Gross-Pitaevskii Energy Functional for a Two-Dimensional Bose Gas
},
Commun. Math. Phys. 224, 17 (2001).

\bibitem{ls}
E.H. Lieb and J. Yngvason, 
\emph{ The Ground State Energy of a Dilute Two- dimensional Bose Gas}, J. Stat. Phys. 103, 509 (2001).







\bibitem{michelangeli}
A. Michelangeli, \emph{Equivalent definitions of asymptotic 100\% BEC}, Nuovo Cimento Sec. B.,  123, 181--192 (2008).

\bibitem{marcin1}
Phan Th\`anh Nam and M. Napi\'orkowski,
\emph{A note on the validity of Bogoliubov correction to mean-field dynamics},
arXiv:1604.05240 (2016).

\bibitem{marcin2}
Phan Th\`anh Nam and Marcin Napi\'orkowski,
\emph{Bogoliubov correction to the mean-field dynamics of interacting bosons},
arXiv:1509.04631 (2016).

\bibitem{picklnorm}
S. Petrat, D. Mitrouskas and P. Pickl
\emph{Bogoliubov corrections and trace norm convergence for
the Hartree dynamics}, arXiv:1609.06264 (2016).

\bibitem{ptertov}
D. S. Petrov and G. V. Shlyapnikov
\emph{ Interatomic collisions in a tightly confined Bose gas}, 
PhysRevA.64.012706 (2000).


\bibitem{pickl2}
P.~Pickl, \emph{Derivation of the time dependent Gross-Pitaevskii equation without positivity condition on the interaction}, J. Stat. Phys. 140, 76--89 (2010).


\bibitem{picklgp3d}
P.~Pickl, \emph{Derivation of the time dependent Gross-Pitaevskii equation with external fields},  arXiv:1001.4894 Rev. Math. Phys., 27, 1550003 (2015) .



\bibitem{pickl1}
P.~Pickl, \emph{A simple  derivation of mean field limits for quantum systems}, Lett. Math. Phys. 97, 151--164 (2011).




\bibitem{rodnianskischlein}
I.~Rodnianski and B.~Schlein, \emph{Quantum fluctuations and rate of
  convergence towards mean field dynamics}, Comm. Math. Phys.  291, no 1, 31--61 (2009).

\bibitem{seiringerpos}
R. Seiriner \emph{Absence of bound states implies non-negativity of the scattering length
}, J. Spectr.Theory 2, 321â328 (2012)

\bibitem{teschl}
G. Teschl \emph{Mathematical Methods in Quantum Mechanics
With Applications to Schr\"odinger Operators},
Graduate Studies in Mathematics, Volume 157, Amer. Math. Soc., Providence (2014).




\end{thebibliography}
\end{document}